\documentclass[11pt]{book}

\usepackage{etoolbox}
\makeatletter
\patchcmd{\chapter}{\if@openright\cleardoublepage\else\clearpage\fi}{}{}{}
\makeatother

\usepackage{titlesec}

\titlespacing*{\chapter}{0pt}{0pt}{40pt}

\usepackage[ansinew]{inputenc}
\usepackage{latexsym,tensor}
\usepackage{amsmath}
\usepackage{mathtools}
\usepackage{amsthm}
\usepackage{amsfonts}
\usepackage{amssymb}
\usepackage{textcomp}
\usepackage{cancel}
\usepackage{verbatim}
\usepackage{bm}
\usepackage{float} %imagenes donde se quiera
\usepackage[pdftex]{hyperref}
\usepackage{graphicx}
\usepackage{tensor}
\usepackage{enumerate}

\usepackage[centering,vcentering,marginratio=1:1,hscale=0.85,vscale=0.82]{geometry}
\numberwithin{equation}{section}

\usepackage{fancyhdr}
\makeatletter
\let\ps@plain\ps@fancy
\makeatother

\makeatletter \def\cleardoublepage{\clearpage\if@twoside \ifodd\c@page\else
   \thispagestyle{empty}
   \newpage
   \if@twocolumn\hbox{}\newpage\fi\fi\fi}
\makeatother

\def\openone{\leavevmode\hbox{\small$1$\normalsize\kern-.33em$1$}}

\newcommand{\ket}[1]{| {#1} \rangle}
\newcommand{\bra}[1]{\langle {#1} |}

\newcommand{\ematriz}[3]{\langle {#1} |{#2}|{#3}\rangle}	
\newcommand{\braket}[2]{\langle {#1}|{#2}\rangle}
\newcommand{\proj}[2]{\left| {#1} \right\rangle\!\left\langle {#2} \right|}
\newcommand{\ii}{\mathrm{i}}

\newcommand{\qu}[1]{\mathsf{#1}}
\newcommand{\qud}[1]{\dot{\mathsf{#1}}}
\newcommand{\sym}{\boldsymbol{\Omega}}
\newcommand{\bs}[1]{\boldsymbol{#1}}
\newcommand{\rr}[1]{\left(#1\right)}
\newcommand{\be}{\begin{equation}}
\newcommand{\ee}{\end{equation}}
\newcommand{\tq}{\hat{\tilde q}}
\newcommand{\tp}{\hat{\tilde{p}}}
\newcommand{\q}{\hat{\Theta}}
\newcommand{\bS}{\boldsymbol{S}}
\newcommand{\bF}{\boldsymbol{F}}
\newcommand{\bO}{\boldsymbol{\Omega}^{-1}}
\newcommand{\bOr}{\boldsymbol{\Omega}}
\newcommand{\bSi}{\boldsymbol{\sigma}}

\newcommand{\bXi}{\hat{\boldsymbol{\Xi}}}

\def\slashchar#1{\setbox0=\hbox{$#1$} % set a box for #1
\dimen0=\wd0 % and get its size
\setbox1=\hbox{/} \dimen1=\wd1 % get size of /
\ifdim\dimen0>\dimen1 % #1 is bigger
\rlap{\hbox to \dimen0{\hfil/\hfil}} % so center / in box
#1 % and print #1
\else % / is bigger
\rlap{\hbox to \dimen1{\hfil$#1$\hfil}} % so center #1
/ % and print /
\fi}

\renewcommand{\d}{\text{d}}
\newcommand{\tr}{\operatorname{Tr}}

\newcommand{\bel}[1]{\begin{equation}\label{#1}}

\theoremstyle{plain}

\theoremstyle{definition}

\theoremstyle{remark}

\author{Eduardo Martin-Martinez}
\date{}

\begin{document}

\chapter*{Quantum Mechanics in Phase Space}
\tableofcontents
\newpage

\vspace*{-2.4cm}
\chapter*{Preface}
\vspace*{-1cm}

These are my personal lecture notes that I prepared to teach the course quantum theory III at the University of Waterloo. I believe that there is some value on posting these notes on arXiv since they may perhaps be helpful to people first approaching the formulation of quantum mechanics in phase space and Gaussian quantum mechanics. 

However these are just my personal notes and are not meant to be a review of any kind nor substitute any good textbook by any means. Rather, they are written as an informal starter on these topics for those interested. There is a substantial amount of home-brewed content in these notes at points where I could not find textbooks that cover the content with the desired optics so please be always observant, and use these notes under your own discretion. In case of doubt please go to well-established reviews and textbook materials. 

The lecture notes/informal character of these notes resulted in a lack of explicit references at the pertinent places in the text. However, here are a series of fantastic materials that you can go to for further exploration. I have used some of them as reference material to write these notes:

\begin{itemize}
    \item[-] J. V. Jos\'e, E. J. Saletan, ``Classical dynamics: A contemporary approach''.\\ Cambridge University Press, 1998.
    \item[-] Luis J. Garay \href{https://uc2ef3ca7f56d7e650af9028e330.dl.dropboxusercontent.com/cd/0/inline2/BrOAJnhIqFQ9VzpCBWDAQZMaDTh0PNwtdwebfh5QWP9ybYahV18F4IsiCyfKpgr6k6fZXT96b5ORrOgP7mPFR4UGJMX9IqB_beUFtobeR4fFufYqfe9hR4N5GNYeDN0HWTXXd3FpK8mDutzSGwqisMATkxlnec2mk8f_64Ao0qblalzwNl3bLKdD7TbdYKg3znno5Vje3Z6MXGs_akc4kQp3HgXXaF0tMlr7s4oximcsClonOgAVuAuATVbiruHkw3-F-A_u72rSPllUmb5w9jFz5grTW6oLrLIg9UHZdpfilnCdLfgQSr3XdNYAUZgg1mOGQp7iRdIAeP62tcJAjq8W8giuP4UXw9vJr_y3cxBsp5D8HX09VQ92uzCigop4SfsZsUIFPw8zECj8GyGDledqJBSkQwZDNbLZXKN4sO_IqQ/file}{Notes on classical mechanics} (in Spanish)
    \item[-] W. P. Schleich, ``Quantum Optics in Phase Space''. Wiley-VCH Verlag Berlin, 2001.
    \item[-] M. O. Scully, M. S. Zubairy, ``Quantum optics''. Cambridge University Press, 2012.
    \item[-] K. E. Cahill and R. J. Glauber, \href{https://journals.aps.org/pr/abstract/10.1103/PhysRev.177.1882}{Phys. Rev. 177, 1882 (1969)}.
    \item[-] G. Adesso , ``Entanglement of Gaussian States". Ph.D. Thesis, U. of Salerno \href{https://arxiv.org/abs/quant-ph/0702069}{arXiv:quant-ph/0702069}
      \item[-] C. Weedbrook, S. Pirandola, R. Garc\'ia-Patr\'on, N. J. Cerf, T. C. Ralph, J. H. Shapiro, and S. Lloyd, \href{https://journals.aps.org/rmp/abstract/10.1103/RevModPhys.84.621}{Rev. Mod. Phys. \textbf{84}, 621 (2012)}.
      \item[-] G. Adesso, S. Ragy, and A. R. Lee, \href{https://doi.org/10.1142/S1230161214400010}{Open Syst. Inf. Dyn. \textbf{21}, 1440001 (2014)}.
      \item[-] A. Serafini,  ``Quantum Continuous Variables''. CRC Press, 2017
      \item[-] L. Lami, A. Serafini, G. Adesso, \href{https://iopscience.iop.org/article/10.1088/1367-2630/aaa654}{New J. Phys. 20, 023030 (2018)}
      \item[-] T. F. Demaire,``Pedagogical introduction to the entropy of entanglement for Gaussian states"\\ \href{https://arxiv.org/abs/1209.2748}{arXiv:1209.2748}
\end{itemize}

Additionally the course assumes some basic knowledge of differential geometry at some points. There are many good texts that deal with these notions at different depths. I can recommend the following two:
\begin{itemize}
 \item[-] B. F. Schutz, ``Geometrical Methods of Mathematical Phsyics''. Cambridge University Press, 1980.
 \item[-]J. M. Lee, ``Manifolds and Differential Geometry'' American Mathematical Society, 2009
\end{itemize}

I also have some recorded lectures on differential geometry  that are tailored to the differential geometry basis required here. These are videos 1, 2, 14, 15, 16, 17 and 18 of  \href{https://youtube.com/playlist?list=PLeoh1MW56PeLn-tYxepNXBnfTMdbBemfJ}{this YouTube playlist}. The playlist is a course on General Relativity but the seven lectures indicated are an introduction to the Differential Geometry topics assumed to be familiar to the reader of these notes.

\newpage

\chapter*{Phase Space Quantum Mechanics}
\quad\\[-25mm]

\section{Crash-Course on classical phase-space mechanics}

\subsection{Configuration space and its tangent fibre bundle}

Consider a physical system with $n$ degrees of freedom, described by a configuration space $\mathcal{Q}$ of $n$ dimensions. Let $q^a\in \mathbb{R}$, $a=1,\dots,n$ be some generalized coordinates of such space. A point $\mathsf{q}\in\mathcal{Q}$ can be represented by the value of its generalized coordinates $q^a$. The map that takes $\mathcal{Q}$ to the multiplet $q^1,\dots,q^n$ (the coordinates of $\mathsf{q}$) is therefore a map between  $\mathsf{q}$ and $\mathbb{R}^n$. If there is a compatible collection of such invertible and bicontinuous maps\footnote{The maps are continuous and their inverses are continuous too. This is what we call homeomorphism, or topological isomorphism.} mapping an open set\footnote{We are assuming that there is a natural topology on $\mathcal{Q}$, and therefore a notion of `neighbourhood', and furthermore we are entering the realms of differential geometry and the theory of differentiable manifolds, which is indeed extremely important for understanding classical mechanics. However, I am only grazing the surface here. For more details check a text on Differential Geometry (e.g., John Lee's "Introduction to Smooth Manifolds").} around all points, $\mathsf{q}\in\mathcal{Q}$  to open sets of $\mathbb{R}^n$, the configuration space has a structure of differentiable manifold. 

For example, a system of two penduli restricted to move on a plane is described with two angular coordinates $q_1,q_2 \in[0,2\pi)$, therefore the system's configuration space is a torus, $\mathcal{Q}=T^2=S^1\times S^1$.

The trajectory of the system (how it moves as a function of some evolution parameter, e.g., time) is determined by a parametric curve $\mathsf{q}(s)$ in $Q$, where $s \in \mathbb{R}$. One can think, for example as the trajectory as a function of time if we identify the parameter $s$ with time. In each point $\mathsf{q}$ we can build the tangent vector space $T_q\mathcal{Q}$ generated by all the possible vectors $\dot{\mathsf{q}}$ which are tangent to every curve that pass through $\mathsf{q}$. These vectors, whose components we denote as $\dot q^a$, represent all the possible `generalized velocities' that one can have at the point $\mathsf{q}$. The dot symbolizes a derivative with respect to the parameter $s$, i.e., $\dot{\mathsf{q}}=\text{d}\mathsf{q}/\text{d}s$.

We call the \textit{velocity phase space} $\mathsf{T}\mathcal{Q}$ (also called, the tangent fibre bundle to the configuration space $\mathcal{Q}$) the set of all the points $\mathsf{q}\in\mathcal{Q}$ together with all their corresponding tangent spaces $T_q\mathcal{Q}$. $\mathsf{T}\mathcal{Q}$ is also a differentiable manifold in its own right.

As examples, the configuration space of a pendulum restricted to move on a plane is the circle $\mathcal{Q}=S^1$. At each point, the possible values of its velocity are not limited, therefore the tangent space at each point of its trajectory is the real line $T_q\mathcal{Q}=\mathbb{R}$. The tangent fibre bundle (velocity phase space) is the cylinder $\mathsf{T}\mathcal{Q}=S^1\times\mathbb{R}$.

\subsection{The Legendre transform}

Let $V$ be a vector space, and let $l:V\to\mathbb{R}$ a convex function\footnote{i.e., the Hessian $\left|\frac{\partial^2 l}{\partial v^a\partial v^b} \right|>0$, where $v^a$ are coordinates on $V$. For advanced readers: In the notation of my differential geometry notes, the second derivative right above would not be of $l$, but $\bar l$, the function that takes coordinates and maps them to $\mathbb{R}$, but we will be omitting the bar in these notes as it is common everywhere for simplicity.}. We define the \textit{Legendre Transform} of $l$ as a new function $h:V^*\to\mathbb{R}$ defined over the dual vector space $V^*$ as follows. Let $\mathsf{p}\in V^*$, then
\begin{equation}
h(\mathsf{p})\coloneqq \underset{\mathsf{v}\in V}{\text{max}}\, f(\mathsf{p},\mathsf{v}), \qquad\text{where}\quad f(\mathsf{p},\mathsf{v})\coloneqq p_av^a-l(\mathsf{v}) \;.
\end{equation}
The values of $\mathsf{v}$ for which $f(\mathsf{p},\mathsf{v})$ is stationary of course satisfy that
\begin{equation}
0=\frac{\partial f }{\partial v^a}=p_a-\frac{\partial l}{\partial v^a} \;,
\end{equation}
that is to say, for the values of $\mathsf{v}$ such that 
\begin{equation}
\frac{\partial l}{\partial v^a}=p_a \;.
\end{equation}

The reason to demand that the function is convex is to guarantee that the critical point is a maximum and that the relationship between $\mathsf{p}$ and $\mathsf{v}$ is invertible. We can then obtain the value $\mathsf{v}(\mathsf{p})$ for which $f$ is maximum, and therefore we get that the Legendre transformed function $h$ is
\begin{equation}
h(\mathsf{p})=f\big[\mathsf{p},\mathsf{v}(\mathsf{p})\big] \;.
\end{equation}

The Legendre transform is involutive, which means that the transform of the transform is the original function. Also note that the Legendre transform's domain does not necessarily have to be the whole $V^*$ but instead an open subset of the dual space. For example, consider the function $l(v)=e^v$. For that one $h(p)=p\log p -p$ which is only defined for values of the coordinate $p>0$.

\subsection{Hamilton equations}

Let us consider a system whose Lagrangian  $L(\mathsf{q},\dot{\mathsf{q}},t)$ is non-singular (i.e., that $D_{ab}=\partial L/\partial \dot q^a\partial \dot q^b$ is positive definite). We define the Hamiltonian of the system as the Legendre transform of the Lagrangian with respect to the (generalized) velocities $\qud{q}$:
\begin{equation}
H(\qu{q},\qu{p},t)=\underset{\qud{q}}{\text{max}}\big[p_a\dot q^a-L(\qu{q},\qud{q},t)\big]=\big[p_a\dot q^a-L(\qu{q},\qud{q},t)\big]_{\qud{q}\to\qud{q}(\qu{q},\qu{p})} \;,
\end{equation}
where the canonical momentum $\qu{p}$ to the configuration variable $\qu{q}$ is 
\begin{equation}
p_a=\frac{\partial L}{\partial \dot q^a} \;.
\end{equation}

It is easy to compute (making use of the definition of the Hamiltonian as the Legendre transform of the Lagrangian, and the Euler-Lagrange equations) that on the one hand
\begin{equation}
\frac{\partial H}{\partial t}=-\frac{\partial L}{\partial t} \;,
\end{equation}
and that
\begin{equation}
\frac{\partial H}{\partial q^a}=-\dot p_a,\qquad \frac{\partial H}{\partial p_a}=\dot q^a.
\label{Hamilton}
\end{equation}
These are called the Hamilton equations and are completely equivalent to the Euler-Lagrange equations.

The Hamiltonian has a very interesting property: its time dependence is completely determined by its explicit dependence on $t$, that is
\begin{equation}
\dot H\coloneqq\frac{\text{d} H}{\text{d} t}=\frac{\partial H}{\partial q^a}\dot q^a+\frac{\partial H}{\partial p_a}\dot p_a+\frac{\partial H}{\partial t}=\frac{\partial H}{\partial t},
\end{equation}
since the two first summands are zero for the trajectories satisfying Hamilton's equations \eqref{Hamilton}. The fact that $\dot H=\partial_t H$ tells us that if the Hamiltonian is not explicitly time dependent, it is conserved along trajectories satisfying Hamilton's equations (that is, classical solutions to the equations of motion). 

\textbf{Exercise}: Consider a system whose Lagrangian is quadratic on the velocities, and without a linear term, that is
\begin{equation}
L=\frac12 g_{ab}\dot q^a\dot q^b-V,
\end{equation}
where $g_{ab}=g_{ab}(\qu{q},t)$, $V=V(\qu{q},t)$ are arbitrary functions of the positions and time. Then show that its Hamiltonian is
\begin{equation}
H=\frac12 g^{ab} p_a  p_b+V,
\end{equation}
where $g^{ab}=g^{ab}(\qu{q},t)$ is the inverse matrix to $g_{ab}$. That is, that if the quadratic form is the kinetic energy, $V$ can be considered a potential energy and the Hamiltonian is the mechanical-energy. Notice that if the potential energy $V$ is not explicitly time dependent then the mechanical energy is conserved. This is not true, for example, for a charged particle in an electromagnetic field (remember there is a term $\bm p\cdot\bm A$).

\subsection{Phase space}

In the Lagrangian description of a dynamical system the dynamical variables are functions in what we called \textit{velocity phase space}, that is, in the tangent fibre bundle $\mathsf{T}\mathcal{Q}$, and the physical trajectories can be interpreted as curves $(\qu{q}(t),\qud{q}(t))$ in that space.

In the Hamiltonian formulation of mechanics the dynamical variables are instead functions of the (generalized) positions $\qu{q}$ and the momenta $\qu{p}$. Notice then that at each point of configuration space $q\in\mathcal{Q}$ we have introduced two different vector spaces. On the one hand we have the tangent vector space $T_q\mathcal{Q}$ to $\mathcal{Q}$ at the point $\qu{q}$, and on the other hand we have arrived (through a Legendre transform)  to the vector space whose elements are momenta. We have seen that the momentum $\mathsf{p}$ are forms, elements of the dual space to the vector space of velocities\footnote{The generalized momentum $\qu{p}$ is a linear map that takes an element of the tangent space $\qud{q}$ and returns a scalar through its contraction $p_a\dot q^a$. That is the definition of a liner functional, or one-form.}. The dual vector space where the momenta $\qu{p}$ live is known as the cotangent vector space $T_q^*\mathcal{Q}$ to the configuration space $\mathcal{Q}$ at the point $\mathsf{q}$. The set of all the points $\qu{q}\in \mathcal{Q}$ and all the respective cotangent vector spaces $T_q^*\mathcal{Q}$ form a differentiable manifold called the cotangent fibre bundle $\mathsf{T}^*\mathcal{Q}$, or \textit{phase space}. We call \textit{dynamical variable} to any function defined over phase space.

As examples, the configuration space of a pendulum restricted to move on a plane is the circle $\mathcal{Q}=S^1$. At each point $\theta$ in configuration space, the possible values of its momentum are not limited, therefore the cotangent space at each point of its trajectory is the real line $T_q^*\mathcal{Q}=\mathbb{R}$. The cotangent fibre bundle (phase space) is the cylinder $\mathsf{T}^*\mathcal{Q}=S^1\times\mathbb{R}$.

The physical trajectories on the cotangent fibre bundle $\mathsf{T}^*\mathcal{Q}$ (i.e., the physical classical trajectories on phase space) are curves on  $\mathsf{T}^*\mathcal{Q}$ whose tangent vector is determined by the Hamiltonian according to Hamilton's equations.

In the Hamiltonian formalism, configuration variables $q^a$ and momenta $p_a$ are treated on equal footing, so we are going to introduce a notation that addresses them as such in a convenient way. Let us introduce the $2n$ variables $\xi^\alpha$ such that $(\xi^1,\dots,\xi^{2n})=(q^1,\dots,q^n,p_1,\dots,p_n)\in \mathsf{T}^*\mathcal{Q}$, which we call canonical variables. Note that they mix variables of the configuration space $q^a$ with the momenta $p_a$.

In terms of these canonical variables, Hamilton equations can therefore be written compactly as
\begin{equation}
\tilde{\Omega}_{\alpha\beta}\dot \xi^\beta=\partial_\alpha H \qquad \text{or} \qquad  \dot \xi^\alpha=\tilde{\Omega}^{\alpha\beta}\partial_\beta H,
\label{compactHam}
\end{equation}
where $\partial_\beta\coloneqq\partial/\partial\xi^\beta$, $\tilde{\Omega}_{\alpha\beta}$ are the components of the \textit{symplectic matrix} of dimension $2n\times 2n$:
\begin{equation}
(\tilde{\Omega}_{\alpha\beta})=\begin{pmatrix}
0_n& -\openone_n\\
\openone_n & 0_n
\end{pmatrix},
\label{sympmat}
\end{equation}
and $\tilde{\Omega}^{\alpha\beta}$ are the components of the inverse matrix to $\tilde{\Omega}$, that is $\tilde{\Omega}^{\alpha\beta}\tilde{\Omega}_{\beta\gamma}=\delta^\alpha_\gamma$, or in matrix form
\begin{equation}
(\tilde{\Omega}^{\alpha\beta})=\begin{pmatrix}
0_n& \openone_n\\
-\openone_n & 0_n
\end{pmatrix},
\end{equation}

The symplectic matrix satisfies therefore that it is antisymmetric $\tilde{\Omega}_{\alpha\beta}=-\tilde{\Omega}_{\beta\alpha}$, and also that $\tilde{\Omega}^2=-\openone$, that is $\tilde{\Omega}^{-1}=-\tilde{\Omega}=\tilde{\Omega}^\intercal$, or in components $\tilde{\Omega}^{\alpha\beta}=-\tilde{\Omega}_{\alpha\beta}$.

\subsection{Poisson Brackets}

We define the Poisson bracket of two dynamical variables $f(\xi^\alpha), g(\xi^\alpha)$ (that is, two functions of generalized positions and momenta) as
\begin{equation}
\big\{f,g\big\}\coloneqq\partial_\alpha f \tilde{\Omega}^{\alpha\beta}\partial_\beta g.
\end{equation}
 In terms of the $\mathsf{q}$ and $\mathsf{p}$ variables, the Poisson bracket takes the form
\begin{equation}
\big\{f,g\big\}=\frac{\partial f}{\partial q^a}\frac{\partial g}{\partial p_a}-\frac{\partial g}{\partial q^a}\frac{\partial f}{\partial p_a}.
\end{equation}
In particular, the Poisson bracket between phase space variables is
\begin{equation}
\big\{\xi^\alpha,\xi^\beta\big\}= \tilde{\Omega}^{\alpha\beta},
\end{equation}
which in terms of  $\mathsf{q}$ and $\mathsf{p}$ variables is
\begin{equation}
\big\{q^a,q^b\big\}= 0, \qquad \big\{q^a,p_b\big\}= \delta^a_b,\qquad \big\{p_a,p_b\}=0.
\end{equation}

The Poisson bracket has the following properties (exercise: prove it!). Let $f,g,h$ be elements of $\mathcal{F}(\mathsf{T}^*\mathcal{Q})$ (functions on the cotangent fibre bundle to $Q$, i.e., dynamical variables):
\begin{enumerate}
\item Anti-symmetry: $\big\{f,g\big\}=-\big\{g,f\big\}$,
\item Jacobi Identity: $\big\{f,\big\{g,h\big\}\big\}+\big\{g,\big\{h,f\big\}\big\}+\big\{h,\big\{f,g\big\}\big\}=0$,
\item Linearity: $\big\{af+bg,h\big\}=a\big\{f,h\big\}+b \big\{g,h\big\},\quad a,b\in\mathbb{R}$,
\item Leibniz Rule: $\big\{fg,h\big\}=f\big\{g,h\big\}+g\big\{f,h\big\}$.
\end{enumerate}

There are two extremely important things to learn from this. First, properties 1 to 3 indicate that the space $\mathcal{F}(\mathsf{T}^*\mathcal{Q})$ of dynamical variables is a Lie Algebra under the Poisson bracket. Second, properties 3 and 4 tell us that given a dynamical variable $g$, we can consider the operation `take the Poisson bracket with $g$', $\big\{g,\circ\big\}$, as a derivative of the argument $\circ$, that is, the Poisson bracket with $g$ is a differential operator\footnote{This makes the space of dynamical variables a Poisson Algebra.}. In particular, we can consider the differential operator $\big\{\xi^\alpha,\circ\big\}$, that to each dynamical variable $f$ assigns the vector
\begin{equation}
X^\alpha_f\coloneqq\big\{\xi^\alpha,f\big\}=\tilde{\Omega}^{\alpha\beta}\partial_\beta f.
\label{Hamilflow}
\end{equation}
Now, from Hamilton equations \eqref{compactHam}, we know that for the particular case where $f$ is the Hamiltonian, $X^\alpha_H=\dot\xi^\alpha$. Using \eqref{Hamilflow} we see that Hamilton equations can be written in terms of the Poisson bracket as
\begin{equation}
\dot\xi^\alpha=\big\{\xi^\alpha,H\} \;,
\label{motioneq}
\end{equation}
or in other words
\begin{equation}
\dot q^a=\big\{q^a,H\},\qquad \dot p_a=\big\{p_a,H\} \;.
\label{motioneqpq}
\end{equation}

In more general terms, considering the Poisson brackets of general dynamical variables, that can possibly depend explicitly on time, we get that  (exercise, prove the last two steps using Hamilton equations!) 
\begin{equation}
\dot f=\dot\xi^\alpha\partial_\alpha f +\partial_t f=X_H^\alpha\partial_\alpha f+\partial_t f=\big\{f,H\big\}+\partial_tf.
\label{fdot}
\end{equation}
For any two arbitrary, possibly time dependent dynamical variables $f(\xi^\alpha,t)$, $g(\xi^\alpha,t)$ it is satisfied that
\begin{equation}
\frac{\text{d}}{\text{d}t}{\big\{f,g\big\}}=\big\{\dot f,g\big\}+\big\{f,\dot g\big\}.
\label{bracketeering}
\end{equation}
This is easy to prove using \eqref{fdot}, Jacobi's identity and the fact  (also easy to prove) that
\begin{equation}
\partial_t{\big\{f,g\big\}}=\big\{\partial_t f,g\big\}+\big\{f,\partial_t g\big\}.
\end{equation}

Summarizing this last result: if a curve in phase space is generated by a Hamiltonian evolution (by the vector $X^\alpha_H$), then the Poisson bracket of two arbitrary dynamical variables evolves along this curve according to equation \eqref{bracketeering}. The opposite is also true: if the evolution of the Poisson bracket of two dynamical variables along a curve in phase space satisfies \eqref{bracketeering}, then there exists (locally) a Hamiltonian that generates this evolution.

The proof is not very involved: if we assume that equation \eqref{bracketeering} is satisfied for any pair of dynamical variables $f$ and $g$ along a curve of tangent vector $\dot\xi^\alpha=X^\alpha(\boldsymbol{\xi})$, then it is satisfied as well for the particular choice $f=\xi^\alpha$ and $g=\xi^\beta$. Since the Poisson bracket between canonical variables is constant, the equation \eqref{bracketeering} tells us that
\begin{equation}
0=\big\{\dot \xi^\alpha,\xi^\beta\big\}+\big\{ \xi^\alpha,\dot \xi^\beta\big\}=\big\{X^\alpha,\xi^\beta\big\}+\big\{ \xi^\alpha,X^\beta\big\}=-\partial_\gamma(\tilde{\Omega}^{\beta\gamma}X^\alpha-\tilde{\Omega}^{\alpha\gamma}X^\beta),
\end{equation}
where in the last identity we have used \eqref{Hamilflow} and that $\partial_\alpha\xi^\beta=\delta_{\alpha}^\beta$. Now contracting this with $\tilde{\Omega}_{\alpha\sigma}\tilde{\Omega}_{\beta\nu}$ we get
\begin{equation}
\partial_\nu(\tilde{\Omega}_{\alpha\sigma}X^\alpha)-\partial_\sigma(\tilde{\Omega}_{\beta\nu}X^\beta)=0,
\end{equation}
and this is indeed the (local) integrability condition (coincidence of cross-derivatives) for the differential equation $\partial_\alpha H=\tilde{\Omega}_{\alpha\beta}X^\beta$, that is, Hamilton equations in their form \eqref{compactHam}. In other words, the necessary and sufficient condition for this equation to have a local solution. That means that assuming \eqref{bracketeering}  for any two dynamical variables is sufficent for a function $H$ to exist such that $\dot\xi=X^\alpha=\tilde{\Omega}^{\alpha\beta}\partial_\beta H$ and therefore for the evolution to be generated by a Hamiltonian. 

A corollary of these results is that the Poisson bracket of two conserved quantities is also a  conserved quantity, and the set of all conserved quantities, which is a finite dimensional vector space, is also a Lie algebra (in fact, a subalgebra of the Lie algebra of dynamical variables $\mathcal{F}(\mathsf{T}^*\mathcal{Q})$ with the Poisson bracket).

\subsection{The Symplectic form}

In the previous sections we have considered only a particular choice of coordinates for the phase space $\mathsf{T}^*\mathcal{Q}$: the canonical coordinates $\xi^\alpha=(q_a,p^a)$, which come directly from the Legendre transform of the velocities. In these coordinates we have identified the symplectic matrix $\tilde{\Omega}_{\alpha\beta}$ given by \eqref{sympmat}. We can, of course, write the abstract object $\boldsymbol{\Omega}$, whose coordinates in the canonical coordinate basis are $\tilde{\Omega}_{\alpha\beta}$. This object is a 2-form which has the following expression
\begin{equation}
\boldsymbol{\Omega}=\frac12\tilde{\Omega}_{\alpha\beta}\,\d\xi^\alpha \wedge \d\xi^\beta=\frac12\d p_a\wedge \d q^a-\frac12\d q^a\wedge \d p_a=\d p_a\wedge \d q^a.
\end{equation}
We call\footnote{Note the subtleties of the symplectic manifold notation: $\boldsymbol{\Omega}$ is a two-form in the symplectic manifold, but since the cotantgent fibre bundle takes objects of configuration space and of each cotangent space the lower and upper indices in positions and momenta still refer to that structure to notate covariance/contravariance. In any case $\boldsymbol{\Omega}$ is a two-form, hence twice covariant tensor, in the symplectic space.} $\boldsymbol{\Omega}$ the \textit{symplectic form}. The symplectic form is coordinate independent, but its components are certainly coordinate dependent. 

Consider a coordinate change $\xi^\alpha\to\zeta^\alpha$ , the symplectic form can be written in either coordinate basis as 
\begin{equation}
\boldsymbol{\Omega}=\frac12\tilde{\Omega}_{\alpha\beta}\,\d\xi^\alpha \wedge \d\xi^\beta=\frac12 \underbrace{\tensor{\Lambda}{_\alpha^\gamma}\tensor{\Lambda}{_\beta^\delta}\tilde{\Omega}_{\gamma\delta}}_{{\Omega}_{\alpha\beta}} \,\d\zeta^\alpha \wedge \d\zeta^\beta \;,
\end{equation}
where $\tensor{\Lambda}{_\alpha^\gamma}=\frac{\partial\xi^\gamma}{\partial\zeta^{\delta}}$ is the change of basis matrix, and ${\Omega}_{\alpha\beta}$ are the coefficients of the symplectic form in the new basis (which are different from \eqref{sympmat}). We will use $\tilde\Omega_{\alpha\beta}$ for the components of the symplectic form  in the canonical basis, and we will use $\Omega_{\alpha\beta}$ for any other basis or for an arbitrary basis. We know that the symplectic matrix is non-degenerate because $\tilde\Omega_{\alpha\beta}$  is invertible. This means that the symplectic form can be used much in the same way that we used the metric tensor in relativity: we can use it to build an isomorphism between vectors and one-forms through contraction. In particular, we can build a map\footnote{Just to recap, this is a map from the tangent vector space to the $\overbrace{\text{cotangent fibre bundle to configuration space}}^{\text{phase space}}$ to the cotantgent vector space to the cotangent fibre bundle to configuration space (phase space).} $i\boldsymbol{\Omega}:T(\mathsf{T}^*\mathcal{Q})\to T^*(\mathsf{T}^*\mathcal{Q})$ that  assigns  to each vector $\mathsf{X}$  the one-form
\begin{equation}
i_\mathsf{X}\boldsymbol{\Omega}=\Omega_{\alpha\beta}X^\alpha\d\zeta^\beta.
\label{contraction}
\end{equation} 
and recall that $\{\d\zeta^\beta\}$ is an arbitrary coordinate basis of $T^*(\mathsf{T}^*\mathcal{Q})$, dual to $\{\partial_{\zeta^\alpha}\}$, which is a coordinate basis of  $T(\mathsf{T}^*\mathcal{Q})$, such that the contraction $(\partial_{\zeta^\alpha},\d\zeta_\beta)=\delta^\alpha_\beta$. 

(\textbf{Exercise}: check that $i_{\partial_{q^a}}\boldsymbol{\Omega}=-\d p_a$, $i_{\partial_{p_a}}\boldsymbol{\Omega}=\d q^a$) 

It is easy to check that the two-form $\boldsymbol{\Omega}$ is a closed form, which means $\d\boldsymbol{\Omega}=0$. Since it is non-singular, this implies that it is also exact and therefore there exists (locally) a one-form $\boldsymbol\theta$ such that  $\boldsymbol{\Omega}=\d \boldsymbol\theta$. The one-form  $\boldsymbol\theta$ is called the canonical form. It is a good exercise to prove that in canonical coordinates, the canonical form is $\boldsymbol\theta=p_a\d q^a$.

\subsection{Coordinate independent Hamilton equations}

Hamilton equations, as written\footnote{Notice that we are being slightly more careful with the partial derivative notation. In the past  we notated $\partial_\alpha\coloneqq \partial/\partial\xi^\alpha$, but now we are considering also other arbitrary basis and not only the canonical coordinate ones, so when we refer to the canonical coordinates we will specify it in the notation as written here.} in \eqref{compactHam}, are: 
\begin{equation}
\tilde\Omega_{\alpha\beta}X^\beta=\partial_{\xi^\alpha}H,
\label{reham}
\end{equation}
 where $X^\beta=\dot\xi^\beta$. If the system is Hamiltonian, Hamilton equations are satisfied and solving Hamilton equations consists of finding the vectors $\mathsf{X}$ that satisfy \eqref{reham}, and then find their integral to find the trajectory in phase space. 

Comparing with \eqref{contraction}, we see that the left-hand side of \eqref{reham} has the components of the contraction $i_\mathsf{X}\boldsymbol{\Omega}$ of the symplectic form with the dynamical vector $\mathsf{X}$ (with opposite sign, remembering the symplectic form is anti-symmetric). We also see that the right-hand side has the components of the one-form $\d H$ in the basis $\{\d \xi^\alpha\}$. This means that we can write Hamilton equations in a coordinate independent way as
\begin{equation}
i_\mathsf{X}\boldsymbol{\Omega}=-\d H.
\label{Hamiltontruth}
\end{equation}
This gives the most general formulation of the equation on $\mathsf{X}$ whose integral curves are the classical trajectories. In an analogy with general relativity, where physicists use the metric to lower and raise indices, you can think of the differential of the Hamiltonian as the co-vector obtained `lowering the index' of the dynamical vector with the symplectic form.

This is an elegant and revealing form for time evolution, and showcases how each Hamiltonian determines univocally the dynamical vector, and reciprocally, how a given dynamical vector (provided the 1-form on the l.h.s. is exact, and therefore the evolution is Hamiltonian) determines the Hamiltonian except for an additive constant. This is the ambiguity of the `origin of potential'.

\subsection{Flow of a vector field}

Let us define the \textit{flow of a vector field}. Let $\mathsf{X}$ be a vector field on phase space, and let us consider the map $\varphi_s^\mathsf{X}$ over phase space that assigns to each point $\boldsymbol{\zeta}\in \mathsf{T}^*\mathcal{Q}$ the point $\boldsymbol{\zeta}_s\in \mathsf{T}^*\mathcal{Q}$ that is at a parametric distance $s$ over the integral curve of $\mathsf{X}$ (in simpler words, the solution to the equation $\d\boldsymbol{\zeta}_s/\d s=\mathsf{X}$ that goes through $\boldsymbol\zeta$). This sounds a lot like `translations' along a trajectory, and indeed the set $\varphi^\mathsf{X}=\{\varphi_{s}^\mathsf{X},s\in\mathbb{R}\}$ of all such transformations has group structure under the composition law $\varphi_{s_1}^\mathsf{X}\circ\varphi_{s_2}^\mathsf{X}=\varphi_{s_1+s_2}^\mathsf{X}$. The set $\varphi^{\mathsf{X}}$ is called the flow of the vector field $\mathsf{X}$.

\textbf{Example}: Consider the vector field $X^\alpha=(v^a,\bm 0)$ in canonical coordinates, i.e., $\mathsf{X}=v^a\partial_{q^a}$ . This vector generates a flow in phase space that consists of translations in configuration space. Indeed, the  integral curves of $\xi^\alpha=\int  \d s \, X^\alpha$ are 
\begin{equation}
q^a(s)=q_0^a+v^a\, s,\qquad  p_a(s)=p_{a0}, 
\end{equation}
where $q_0^a$ and $p_{a0}$ are constants.

Let us consider now a small displacement along the flow of $\mathsf{X}$ by an infinitesimal amount $\delta s$. The evolution of any tensor (function, scalar, vector, form, etc.) $\mathsf{W}$ defined over phase space along the flow of $\mathsf{X}$ will be determined by its Lie derivative (and its possible explicit dependence on the parameter of the transformation\footnote{Notice that the flow is an operator on $\mathsf{T}^*\mathcal{Q}$, so technically what we are applying to arbitrary tensor is the push-forward induced by the flow, that is how tensors vary along the flow generated by $\mathsf{X}$. We use the same notation to not complicate it, but we need to keep this in mind.} $s$),
\begin{equation}
\varphi_{\delta s}^\mathsf{X} \mathsf{W}=\mathsf{W}+\delta s\mathcal{L}_\mathsf{X} \mathsf{W}+ \delta s \partial_s  \mathsf{W}.
\end{equation}
In other words, the infinitesimal transformation can be written as
\begin{equation}
\varphi_{\delta s}^\mathsf{X}=\openone +\delta s (\mathcal{L}_\mathsf{X} +\partial_s).
\end{equation}

We can now obtain the action of a finite transformation as follows: Any continuous finite transformation can be decomposed into an infinite composition of infinitesimal transformations. One way of seeing what this is is writing $s=k(s/k)$ and then taking the limit $k\to\infty$. Doing so,
\begin{equation}
\varphi_{s}^\mathsf{X}=\lim_{k\to\infty}\varphi_{k(s/k)}^\mathsf{X} =\lim_{k\to\infty}\left(\varphi_{s/k}^\mathsf{X}\right)^k= \lim_{k\to\infty}\left[\openone+\frac{s}{k}(\mathcal{L}_\mathsf{X} +\partial_s) \right]^k=e^{s(\mathcal{L}_\mathsf{X} +\partial_s)}=\openone+\sum_{k=1}^{\infty}\frac{s^k}{k!}(\mathcal{L}_\mathsf{X} +\partial_s)^k,
\end{equation}
therefore showing that the flow acts as an operator on tensors in phase space (and in particular functions, dynamical variables) as the exponential of the Lie derivative with respect to the vector defining the flow.

\subsection{Hamiltonian vector fields and Hamiltonian flows}

Given a dynamical variable $f$ (a phase space scalar, a function on  $\mathsf{T}^*\mathcal{Q}$), we define the Hamiltonian vector field $\mathsf{X}_f$ associated to $f$ to be the only vector that satisfies
\begin{equation}
i_{\mathsf{X}}\boldsymbol{\Omega}=-\d f,
\label{unamas}
\end{equation}
and therefore can also be written\footnote{Remembering from \eqref{contraction} that $i_\mathsf{X}\boldsymbol{\Omega}=\Omega_{\alpha\beta}X^\alpha\d\zeta^\beta
$, then multiplying on both sides by the $\Omega^{\alpha\beta}$and writting $\d f$ in the coordinate basis $\partial_\alpha$.} as, $X^\alpha_f=\Omega^{\alpha\beta}\partial_\beta f $, or $\mathsf{X}_f=\Omega^{\alpha\beta}\partial_\beta f \partial_\alpha$.

Not all vector fields are Hamiltonian: It is obvious from \eqref{unamas} that a vector field $\mathsf{X}$ is a Hamiltonian vector field iff the one-form $i_{\mathsf{X}}\boldsymbol{\Omega}$ is exact (which in this case is implied by it being closed, i.e., $\d i_{\mathsf{X}}\boldsymbol{\Omega}=0$).

\textbf{Example}: Consider the vector field that generates phase space translations ($X^\alpha=(v^a,\bm 0)$ in canonical coordinates, i.e., $\mathsf{X}=v^a\partial_{q^a}$)  from the previous example. This vector field is Hamiltonian  since $i_{\mathsf{X}}\boldsymbol{\Omega}=-v^a\d p_a$ and therefore $\d i_{\mathsf{X}}\boldsymbol{\Omega}=0$. In fact, its associated dynamical variable is $f=v^ap_a$.

Each dynamical variable $f$ determines uniquely a Hamiltonian vector field $\mathsf{X}_f$ and hence a Hamiltonian flow 
\begin{equation}
\varphi^f\coloneqq\varphi^{\mathsf{X}_f}=e^{s(\mathcal{L}_{\mathsf{X}_f} +\partial_s)}.
\end{equation}
The variable $f$ is called the \textit{generating function} of the flow $\varphi^f$.

Now, importantly, let us note that on dynamical variables (scalar fields over $\mathsf{T}^*\mathcal{Q}$) the Lie derivative is particularly easy. In particular, for a scalar function $g$, $\mathcal{L}_{\mathsf{X}_f}\, g=\mathsf{X}_f\, g=\big\{g,f\big\}$, and therefore 
\begin{equation}
 \varphi^f g= e^{s(\mathsf{X}_f +\partial_s)}g=e^{s\left(\big\{\circ,f\big\} +\partial_s\right)}g \;.
\end{equation}

\textbf{Example}: Consider the vector field that generates phase space translations ($X^\alpha=(v^a,\bm 0)$ in canonical coordinates, i.e., $\mathsf{X}=v^a\partial_{q^a}$) from the previous examples. On dynamical variables $g(\mathsf{q},\mathsf{p})$, the Hamiltonian flow of translations would be
\begin{equation}
 \varphi^f g= e^{s\boldsymbol{X}} g=e^{sv^a\partial_{q^a}}g.
\end{equation}
Notice how cute: The projection of the canonical momentum on some direction of configuration space $v^a$ generates a (Hamiltonian flow of) translations in that direction by a magnitude $s$, and the map that implements it is an exponential map.

Time evolution is generated by the Hamiltonian flow, and in this (coordinate independent) language Hamilton equations reduce to finding the Hamiltonian vector field $\mathsf{X}_H$ associated to the Hamiltonian, and its flow. This is solving the problem of time evolution. As we saw in \eqref{fdot}, the classical time evolution of any dynamical variable $f$ is given by
\begin{equation}
\dot f=\mathsf{X}_H\, f+\partial_tf=\big\{f,H\big\}+\partial_tf.
\end{equation}
With our current knowledge we can generalize for the classical time evolution of any tensor $\mathsf{W}$ defined over phase space will be given by its derivative with respect to the Hamiltonian vector field (plus perhaps some explicit time dependence):
\begin{equation}
\dot{\mathsf{W}}=\mathcal{L}_{\mathsf{X}_H}\mathsf{W}+\partial_t \mathsf{W}.
\end{equation}

\textbf{Exercise:} Prove that the symplectic form is conserved under time evolution.

\subsection{Symplectic manifolds}

It is worth noting that these notions of phase space can be generalized to the so called symplectic manifolds. The construction is somewhat analogous to the construction of a spacetime in general relativity, where the role of the metric (a symmetric twice covariant tensor) is played by the symplectic form (an anti-symmetric twice covariant tensor, or two-form).

Let $\mathcal{M}$ be a differentiable manifold of dimension $2n$. We define a symplectic form to be any two-form $\boldsymbol{\Omega}$ that is closed and non-degenerate. We call the pair the pair $(\mathcal{M},\boldsymbol{\Omega})$ a symplectic manifold.

In this symplectic manifold, each scalar $H$ defines a Hamiltonian evolution where the parameter of evolution $t$ is a parameter of the integral curves of the Hamiltonian vector field $\mathsf{X}_H$ defined by the function $H$ and the symplectic form $\boldsymbol{\Omega}$ as in \eqref{unamas}.

The Poisson bracket between two scalars in the symplectic manifold is defined from the symplectic form as follows:
\begin{equation}
\big\{f,g\big\}\coloneqq \mathsf{X}_g f=-\boldsymbol{\Omega}(\mathsf{X}_f,\mathsf{X}_g)=\Omega^{\alpha\beta}\partial_\alpha f\,\partial_\beta \, g.  
\end{equation}
This Poisson bracket is trivially antisymmetric (because $\boldsymbol{\Omega}$ is a two-form), and  (\textbf{Exercise:} prove that) the fact that the symplectic form is closed implies the Jacobi identity for  the Poisson bracket.

There are infinite possible choices of closed and non-degenerate two-forms to pair with a given manifold to make the symplectic manifold. Different choices of the symplectic form give the same manifold $\mathcal{M}$ different symplectic structures.

\subsection{Symplectic transformations}

A symplectic transformation (sometimes called canonical transformation) is any transformation in phase space that preserves the symplectic form $\boldsymbol{\Omega}$. Think for example of a transformation that maps a vector $\boldsymbol{\zeta}$ on phase space to another one, $\boldsymbol{\zeta}'$. In this case, the condition that the symplectic form is invariant under the transformation means that the symplectic form evaluated at the point $\boldsymbol{\zeta}$ is the same that the symplectic form evaluated at the point $\boldsymbol{\zeta}'$, i.e., that this equality is satisfied
\begin{equation}
\boldsymbol{\Omega}[\boldsymbol{\zeta}'(\boldsymbol{\zeta})]-\boldsymbol{\Omega}(\boldsymbol{\zeta})=0.
\end{equation}
The expressions of $\sym$ at the points $\bs \zeta$ and $\bs\zeta'$ are
\begin{equation}
\boldsymbol{\Omega}(\boldsymbol{\zeta})=\frac12\Omega_{\alpha\beta}(\zeta)\d\zeta^\alpha\wedge \d\zeta^\beta,\qquad \boldsymbol{\Omega}(\boldsymbol{\zeta}')=\frac12\Omega_{\alpha\beta}(\zeta'){\d\zeta'}^\alpha\wedge {\d\zeta'}^\beta,
\end{equation}
therefore, 
\begin{equation}
\boldsymbol{\Omega}[\boldsymbol{\zeta}'(\boldsymbol{\zeta})]=\frac12\Omega_{\alpha\beta}[\boldsymbol{\zeta}'(\boldsymbol{\zeta})]\partial_\alpha{\zeta'}^\gamma\partial_\beta{\zeta'}^\delta \d \zeta^\alpha\wedge \d \zeta^\beta.
\end{equation}
Hence the  transformation $\boldsymbol{\zeta}\to\boldsymbol{\zeta}'$ is symplectic iff
\begin{equation}
\Omega_{\gamma\delta}(\bs\zeta')\partial_\alpha{\zeta'}^\gamma\partial_\beta{\zeta'}^\delta-\Omega_{\alpha\beta}(\bs\zeta)=0.
\end{equation}
The following important propositions are left to be proven as exercises:
\begin{enumerate}
\item A transformation generated by the vector $\qu{X}$ is symplectic iff the Lie derivative of the symplectic form along $\qu{X}$ is zero, that is,
\begin{equation}
\mathfrak{L}_{X} \bs\Omega=\frac{1}{2}\left(X^{\gamma} \partial_{\gamma} \Omega_{\alpha \beta}+\Omega_{\alpha \gamma} \partial_{\beta} X^{\gamma}+\Omega_{\gamma \beta} \partial_{\alpha} X^{\gamma}\right) \mathrm{d} \zeta^{\alpha} \wedge \mathrm{d} \zeta^{\beta}=0.
\label{Liezero}
\end{equation}
\item A transformation is symplectic iff its generating vector $\bs X$ is Hamiltonian, that is, iff there exists a function $f$ such that $i_{\bs X} \bs \Omega=-\mathrm{d} f$.
\item A transformation is symplectic iff it preserves Poisson brackets, that is, iff
\begin{equation}
\left\{\zeta^{\alpha}, \zeta^{\beta}\right\}_{\zeta}=\left\{\zeta^{\alpha}, \zeta^{\beta}\right\}_{\zeta'}.
\end{equation}
\item Hamiltonian evolution is a symplectic transformation (easy to prove assuming \eqref{Liezero}).
\end{enumerate}

\subsection{Liouville theorem}

The theorem says that  the  element of volume of phase space, which is given by the $2n$-form
\begin{equation}
\bs v=\bigwedge_{a=1}^{n} \mathrm{d} p_{a} \wedge \mathrm{d} q^{a},
\end{equation}
is invariant under symplectic transformations.

To prove it, it is enough to note that the volume of phase space is just a simple function of the symplectic form:
\begin{equation}
\bs v=\frac{1}{n !} \bs \Omega \wedge \cdots \wedge \bs\Omega=\frac{1}{n !} \bs\Omega^{\wedge n}.
\end{equation}
Therefore, the invariance of $\sym$ under symplectic transformations guarantees the invariance of $\bs v$.

Additionally, the theorem has another claim: The volume of a certain region $\mathcal{U}$ of a symplectic manifold does not change under a Hamiltonian flow, even if the shape of the region does change.

Proof: Let \(\varphi_{s}^{f}\) be an active Hamiltonian transformation generated by the dynamical variable \(f .\) Under this transformation, a certain \((2 n)\)-dimensional region  \(\mathcal{U}\) is mapped to \(\varphi_{s}^{f} \mathcal{U} .\) The volume of this new region is \(\int_{\varphi_{s}^{f}, \mathcal{U}} \bs v .\) This integral can be understood as a change of coordinates in \(\mathcal{U}\) that adapts the region, and therefore
\[
\int_{\varphi_{s}^{f} \mathcal{U}} \bs v=\int_{\mathcal{U}} \varphi_{-s}^{f} \bs v=\int_{\mathcal{U}} \bs v.
\]

\subsection{Statistical distributions in phase space}

Let us consider a system of $N$ particles with $n$ degrees of freedom where we are not interested in (or can't access) the information about every single particle, but its statistical behaviour. We can describe the system using a density distribution $\rho(\bs \xi, t)$ on phase space, where $\rho$ can be interpreted as the probability that the system's state is contained in an infinitesimal volume of phase space and satisfies that
\begin{equation}
\int \rho \bs v =\int \rho(\xi, t) \mathrm{d}^{2 n} \xi=1.
\end{equation}
One can prove\footnote{See for instance \href{https://www.oulu.fi/tf/statfys/lecture_2006/notes2.pdf}{https://www.oulu.fi/tf/statfys/lecture\_2006/notes2.pdf}.} that the probability integrated in any arbitrary region $\mathcal{U}$ does not change under Hamiltonian evolution, and therefore
\begin{equation}
0=\frac{\mathrm{d}}{\mathrm{d} t} \int_{\mathcal{U}} \rho v=\int_{\mathcal{U}} \frac{\mathrm{d} \rho}{\mathrm{d} t} v,
\end{equation}
where the last step is taken using that the volume element of phase space is invariant under Hamiltonian evolution. Since this is true for an arbitrary integration region then we conclude that $\dot \rho=0$. But we also know that, in the same fashion as any other dynamical variable,
\begin{equation}
\dot{\rho}=\big\{\rho, H\big\}+\partial_{t} \rho\Rightarrow \partial_{t} \rho=-\big\{\rho, H\big\},
\end{equation}
which is known as Liouville equation.

The expectation value of a dynamical variable in the state $\rho(q,p)$ is given by the statistical average over phase space of the function with the density $\rho(q,p)$, that is
\begin{equation}
\langle f( q, p)\rangle_{\rho}=\iint \d q\, \d p\, \rho(q, p) f(q, p).
\label{classstat}
\end{equation}

\section{Quantum Mechanics in Phase Space}

\subsection{Let's get quantum: Canonical Quantization and its issues}

We are going to discuss the process of canonical quantization\footnote{Canonical quantization is not the only process of quantization. However, other methods, such as path integral quantization, share the same drawbacks, albeit they manifest in different ways.}. This process takes a classical system and a set of its dynamical variables, and returns a quantum system and quantum observables. More specifically, a quantization of a classical system is a linear map---denoted by a hat ($\hat{\phantom{a}}$)---from (a subspace of) the space of dynamical observables (functions on phase space) to (a subspace of) the space of self-adjoint operators on a Hilbert space. This map is not arbitrary but must satisfy some properties.

In the first place it must represent the Poisson bracket structure. To represent that structure when the dynamical variables are linear operators on a Hilbert space, there is a Lie bracket kind of operation that naturally comes to mind: the commutator. Therefore, the quantization map must satisfy \([\hat{\circ}, \hat{\circ}]=\ii\hbar\, \widehat{\{\circ, \circ\}}\). The reason for the imaginary unit to be there is because the Poisson bracket of two dynamical variables is a dynamical variables, and, in contrast, the commutator of two self-adjoint operators is not self-adjoint (but anti-selfadjoint). Ideally, one would like to represent all functions $f$ on phase space (i.e. all physical classical observables) and their Poisson brackets as quantum operators.

Second, the representation must be irreducible, so that there exists no proper subspace of the Hilbert space invariant under the action of all the represented observables. If a representation were reducible, it would be the direct sum of irreducible representations that we could study separately. We would then need additional information to select one of them which would introduce additional ambiguities.

However, in general, these two conditions are not compatible and it is impossible to map all possible dynamical functions to quantum observables (Groenewold's theorem)\footnote{See Luis Garay lecture notes on QFT on curved spacetimes (\href{https://sites.google.com/site/luisjgaray/lecture-notes}{https://sites.google.com/site/luisjgaray/lecture-notes}) and for further details M.J. Gotay, Obstructions to Quantization, arXiv:math-ph/9809011v1; M.J. Gotay, J. Math. Phys. 40 (1999) 2107, arXiv:math-ph/9809015, On the Groenewold-VanHove problem for $\mathbb{R}^{2n}$.}.  There are very few cases where the two requisites are compatible so that a quantization of all the classical dynamical variables can be carried out. One such case is the torus, but not in \(\mathbb{R}^{2}\). Because of this incompatibility, what we do when we do quantum theory is select a complete subset of such classical observables\footnote{A set of classical observables is complete if the operators in the set close under the Poisson bracket, and if any other classical observable for which the poisson bracket is zero with all of them is just a constant. } and deal with the problem of representing other ones the best way possible. Notice that this may include (and often does) the Hamiltonian of the system itself. 

After quantization, we want to be sure that the representation is such that it is complete, i.e., that the only operator that commutes with the whole set is (a multiple of) the identity. Schur's lemma guarantees that this is indeed the case if the representation is irreducible.

An example of this is the canonical quantization of $\mathbb{R}^2$ (the phase space of a single particle). Let us consider canonical coordinates \((q, p) .\) If we choose as a complete set the Heisenberg algebra \(\{1, q, p\},\) then a quantization is given by the Schr\"odinger map:
\(\hat{\openone}:=1, \quad \hat{q}:=q, \quad \hat{p}:=-\ii \hbar \partial_{q}\). 

However, knowing a dynamical function does not imply knowing its quantization. Not only we would have powers with any polynomials of order higher than two. Additionally, we can always add factors that would not exist classically that generate non-trivial physics upon quantization.

\subsection{Mapping from classical to quantum}

Since the \(\hat{q}\) and \(\hat{p}\) variables do not commute, given a classical dynamical variable \(f({q}, {p})\) there is \textit{a priori} no unambiguously defined corresponding quantum dynamical variable, as there are factor ordering problems that make the map of classical to quantum \( f({q}, {p}) \rightarrow\) \(\hat A_f\coloneqq f(\hat{{q}}, \hat{{p}})\) ambiguous. We may think of several rules or prescriptions to resolve this ambiguity, but none is intrinsic to the transition from Classical to Quantum Mechanics. The earliest rule was proposed by Weyl in \(1927 .\) Taking \(n=1\) for simplicity, i.e., one \(\hat{q}\) and one \(\hat{p},\) this rule is to place \(\hat{q}\) and \(\hat{p}\) in all possible relative positions in any expression and then take the average. For monomials, exponentials and general functions, the rule is:
\begin{equation}
\left.q^{m} p^{n}=\frac{\partial^{m}}{\partial \lambda^{m}} \frac{\partial^{n}}{\partial \mu^{n}} \frac{(\lambda q+\mu p)^{m+n}}{(m+n) !} \right|_{\lambda=\mu=0}\longrightarrow \left.\frac{\partial^{m}}{\partial \lambda^{m}} \frac{\partial^{n}}{\partial \mu^{n}} \frac{(\lambda \hat q+\mu \hat p)^{m+n}}{(m+n) !} \right|_{\lambda=\mu=0}.
\label{Weylmap}
\end{equation}
That guarantees that the exponentials are quantized as  \(\exp [\ii(\lambda q+\mu p)] \rightarrow \exp [\ii(\lambda \widehat{q}+\mu \widehat{p})]\). The Weyl rule possess the property that if $f( q, p)\in\mathbb{R}$ then $\hat A_f=f(\hat q,\hat p)$ is Hermitian. Of course, for products the rule is non trivial, namely
\[f \rightarrow \widehat{A}_{f}, g \rightarrow \widehat{A}_{g} \not\Rightarrow f g \rightarrow \widehat{A}_{f} \widehat{A}_{g}.\]

Since now the dynamical functions are operators in a Hilbert space, so is the density function $\rho(\bm q,\bm p)$ that characterized the state of the system. It goes to the density operator $\hat \rho$. By construction, $\hat \rho$ is a probability distribution over possible states of the system, represented as eigenvectors on the Hilbert space on which the dynamical variables act. In the case of a single particle, this space is spanned by the generalized eigenvectors of the position operator $\hat q$ or the momentum operator $\hat p$, and the association is done following the standard rules from basic quantum mechanics (see, e.g., \href{https://sites.google.com/site/emmfis/teaching/amath-473}{QT2 lecture notes}). For classical states one would expect that $\rho(\bm q,\bm p)$ can contain the same information as $\hat \rho$ (so we didn't really need the whole quantization business). But we know that after canonical quantization $\hat \rho$ can contain in principle more information (superposition of classical states, for example), since it is now an operator on a Hilbert space rather than a function in Phase space. It is important to notice that quantum mechanics has inherited the whole symplectic structure of classical mechanics, only upgraded now by being represented by self-adjoint operators in a Hilbert space. However, it may be difficult to characterize the similarities unless we attempt to describe a quantum state in the same way as a classical state exactly, which is with a probability distribution on phase space. Can we do it? Can we capture the probabilistic nature of quantum theory with a probability distribution on phase space? If we can, then the theory, fancy as it is, will be nothing but a classical theory and we didn't really need to go the lengths of canonical quantization. 

We will follow  the steps that Wigner took to see if this is possible.

\subsection{The Wigner Function and the Weyl-Wigner correspondence}

The symmetric ordering rule by Weyl, which is the simplest but by far not the unique way of associating Hermitian operators to dynamical variables, can be used to describe the quantum dynamics of phase-space. It was Wigner in 1932 who found a way to represent quantum states by real phase-space distributions in the same way we did with density distributions in classical mechanics. Wigner actually provides the following expression (without justification!) in the context of quantum mechanical corrections to thermodynamic equillibrium:
\begin{equation}
W(q, p) \coloneqq \frac{1}{2 \pi \hbar} \int_{-\infty}^{\infty}\!\!\! \d x\, e^{-\frac{\ii}{\hbar} p x}\left\langle q+\tfrac{1}{2} x \right|\hat{\rho}\left| q-\tfrac{1}{2} x\right\rangle, \label{Wignerdef}
\end{equation}
where the matrix elements of the density operator are taken in the `eigenbasis' of the position operator $\hat{q}$. That is,
\begin{equation}
\hat \rho=\int \d q'' \int \d q'\rho(q'',q')\proj{q''}{q'}, \qquad \rho(q'',q')\coloneqq \left\langle q''\right|\hat{\rho}\left| q'\right\rangle.
\label{densitymat}
\end{equation}
Thinking about it, this is pretty intuitive: we have a distribution on configuration space (position representation) $\rho(q'',q')$. If we change the matrix elements on the right to the momentum representation we would get a phase space distribution from \eqref{densitymat},
\begin{equation}
\!\hat \rho\!=\!\int\!\!\d q''\!\!\!\int\!\!\d q' \!\!\!\int \!\!\d p \rho(q'',q')\ket{p}\!\!\underbrace{\braket{p}{q''}}_{\frac{e^{-\frac{\ii}{\hbar}p q''}}{\sqrt{2\pi\hbar}}}\!\langle{q'}|\!=\!\!\int\!\!\d q'\!\!\! \int \!\!\d p\! \underbrace{\frac{1}{\sqrt{2\pi\hbar}}\!\int\!\! \d q''\! e^{-\frac{\ii}{\hbar}p q''}\!\!\rho(q'',q')}_{\tilde\rho(p,q')}\ket{p}\langle{q'}|=\!\frac{1}{\sqrt{2\pi\hbar}}\int\!\!\d q' \!\!\!\int\!\! \d p\, {\tilde\rho(p,q')}\ket{p}\langle{q'}|,
\end{equation}
where we now found a phase space distribution that represents the density matrix in the hybrid $q$, $p$ representation, namely:
\begin{equation}
\tilde\rho(p,q')\coloneqq \frac{1}{\sqrt{2\pi\hbar}}\int\!\d q'' e^{-\frac{\ii}{\hbar}p q''}\!\!\rho(q'',q').
\label{nonsym}
\end{equation}
Et voil\`{a}! A phase space distribution. However, that's not the only one that can be thought of. Why are we transforming the second variable of the distribution only? Would it be the same as transforming the first? Or are there any other ways of associating such distribution? Indeed, in the spirit of Weyl, Wigner wanted to have a symmetric distribution in terms of what argument of the position distribution is moved to the momentum representation, so he referred the two different positions to a `centre' between the two an introduce a new variable $x$ as follows
\begin{equation}
q^{\prime}=q-\frac{1}{2} x \quad \text { and } \quad q^{\prime \prime}=q+\frac{1}{2} x,
\label{wigner1}
\end{equation}
so that now we can rewrite the density matrix in the position representation in \eqref{densitymat} as  
\begin{equation}
\hat \rho=\int \d q \int \d x\,\rho_\textsc{s}(q,x)\proj{q+\tfrac{1}{2} x}{q-\tfrac{1}{2} x}, \qquad \rho_\textsc{s}(q,x)\coloneqq \left\langle q+\tfrac{1}{2} x\right|\hat{\rho}\left| q-\tfrac{1}{2} x\right\rangle
\label{densitymat3}
\end{equation}
%\left\langle q+\tfrac{1}{2} x \right|\hat{\rho}\left| q-\tfrac{1}{2} x\right\rangle

And now we can switch to the momentum representation for the variable $x$ of the symmetric position representation distribution $\rho_\textsc{s}(q,x)$ through a Fourier transform, hence defining the Wigner function as a symmetric version of \eqref{nonsym}:
\begin{equation}
W(q, p) \coloneqq \frac{1}{2 \pi \hbar} \int_{-\infty}^{\infty}\!\!\! \d x\, e^{-\frac{\ii}{\hbar} p x}\left\langle q+\tfrac{1}{2} x \right|\hat{\rho}\left| q-\tfrac{1}{2} x\right\rangle=\frac{1}{2 \pi \hbar} \int_{-\infty}^{\infty} \!\!\!\d (q''-q^{\prime})  e^{-\frac{\ii}{\hbar} p\left(q^{\prime \prime}-q^{\prime}\right)}\left\langle q^{\prime \prime}|\hat{\rho}| q^{\prime}\right\rangle,
\label{Wigneryes}
\end{equation}
since $x=q''-q'$. The constant $(2\pi\hbar)^{-1}$ corresponds to a normalization of the distribution so that (as we will see in more detail later)
\begin{equation}
\int \! \d q \int\! \d p\, W(q, p)=1.
\label{norm}
\end{equation}
Summarizing, the Wigner function defined here consists of getting the density matrix in the position representation $\rho(q'',q')=\left\langle q''\right|\hat{\rho}\left| q'\right\rangle$, symmetrize the two variables through a symmetric translation in the two position variables, and transform one of the two position variables into the momentum representation, that is:
\begin{equation}
W(q, p)=\frac{1}{2 \pi \hbar} \int_{-\infty}^{\infty}\!\!\! \d x\, e^{-\frac{\ii}{\hbar} p x}\rho_\textsc{s}(q,x)
\label{Wignerdef2}\end{equation}

In classical mechanics we had that the expectation value of a dynamical function $f(q,p)$ in a state $\rho(q,p)$ could be found as
\begin{equation}
\langle f( q, p)\rangle_{\rho}=\iint \d q\, \d p\, \rho(q, p) f(q, p).
\label{classstat2}
\end{equation}
Can we build a quantum analogue for this? To answer this question,  let us represent a Weyl-quantized dynamical observable $\hat A_f$, which is the Weyl representation of the classical dynamical observable $f(q,p)$, in a way analogous to the way in which one represents the density matrix as a Wigner function. Namely, let us introduce the Wigner -Weyl representation of $\hat A_f$ as
\begin{equation}
A(q,p)\coloneqq   \int_{-\infty}^{\infty}\!\!\! \d x\, e^{-\frac{\ii}{\hbar} p x}\left\langle q+\tfrac{1}{2} x \right|\hat{A}_f\left| q-\tfrac{1}{2} x\right\rangle.
\end{equation}
Let's see whether this Wigner-Weyl representation behaves like the classical analogue of the dynamical function $f(q,p)$. For that, we know that the expectation value of $\hat A_f$ is
\begin{equation}
\langle \widehat{f(q,p)}\rangle_{\hat\rho}=\langle\hat A_f\rangle_{\hat\rho}=\operatorname{Tr}\big(\hat{\rho} \hat{A}_{f}\big)=\int \d q'' \bra{q''}\hat{\rho} \hat{A}_{f}\ket{q''}=\int \d q''\int \d q' \bra{q''}\hat{\rho} \ket{q'}\!\bra{q'}\hat{A}_{f}\ket{q''}.
\end{equation}
We now carry out the change of variable $q''=q+x/2$, $q'=q-x/2$, so that
\begin{align}
\langle\hat A_f\rangle_{\hat\rho}&=\int\!\! \d q\!\!\int\!\!\d x \, \bra{q+\tfrac{1}{2} x}\hat{\rho} \ket{q-\tfrac{1}{2} x}\bra{q-\tfrac{1}{2} x}\hat{A}_{f}\ket{q+\tfrac{1}{2} x} \nonumber \\
&=\int\!\! \d q\!\!\int\!\! \d x \!\!\int\!\d y\, \delta(y+x)\! \bra{q+\tfrac{1}{2} x}\hat{\rho} \ket{q-\tfrac{1}{2} x}\!\bra{q+\tfrac{1}{2} y}\hat{A}_{f}\ket{q-\tfrac{1}{2} y},
\end{align}
where we have added a delta integrated over $y$ so that the expressions are identical. Noting now that the delta can be written as  
\begin{equation}
\delta(x+y)= \frac{1}{2\pi\hbar}\int \d p\,  e^{\frac{-\ii}{\hbar} p(x+y)}
\end{equation}
we arrive to
\begin{equation}
\langle\hat A_f\rangle_{\hat\rho}=\int \d q \int \d p \underbrace{\frac{1}{2\pi\hbar}\int\!\!  \d x\, e^{\frac{-\ii}{\hbar} p x }  \bra{q+\tfrac{1}{2} x}\hat{\rho} \ket{q-\tfrac{1}{2} x}}_{W(q,p)}\underbrace{\int\!\d y\,e^{\frac{-\ii}{\hbar} p y }\bra{q+\tfrac{1}{2} y}\hat{A}_{f}\ket{q-\tfrac{1}{2} y}}_{A(q,p)}.
\label{prod1}
\end{equation}
Hence, we obtain an equation analogue to the classical version \eqref{classstat2}:
\begin{equation}
\langle\hat A_f\rangle_{\hat\rho}=\operatorname{Tr}\big(\hat{\rho} \hat{A}_{f}\big)=\iint \d q\, \d p\, W(q, p) A(q, p),
\label{product}
\end{equation}
where the Wigner function plays the role of the probability distribution on phase space, and $A(q,p)$ plays the role of the dynamical variable.

\noindent\textbf{Exercise:} Prove the duality between the Weyl ordering and the Weyl-Wigner representation of observables, in that the Weyl-Wigner representation of the Weyl-ordered quantization $\hat A_f$ associated to $f$, that is, $A(p,q)$, is equal to the dynamical function $f$ itself, i.e., $A(p ,q)=f(p,q)$.

This also tells us the normalization of the Wigner function: If we choose $f(q,p)=1$, then its Weyl ordering quantization is $\hat A_f=\openone$, and its Wigner-Weyl representation is $A(q,p)=1$, therefore we obtain that
\begin{equation}
\operatorname{Tr}\big(\hat{\rho}\big)=1=\iint \d q\, \d p\, W(q, p).
\end{equation}
So the integral of the Wigner function over phase space is 1.

\subsection{Properties of the Wigner Function}

\subsubsection{Marginals of the Wigner function}

As the previous section suggests, the Wigner function (which is actually a distribution over phase space) has indeed many of the nice properties of a probability distribution (like the classical $\rho(q,p)$ is). For example, the fact that it's normalized to 1 as  we saw above. Furthermore, same as a probability distribution over phase space, integrating over one of the variables one recovers the marginal probability distributions for $p$ and $q$. Indeed, let's recall the position and momentum distributions 

The position distribution is the probability density distribution of finding  the particle at position $q$. This is of course defined only as a distribution that has to be integrated over a finite length to yield a probability. This distribution is easy to obtain from Born's rule and the position representaiton for $\hat\rho$ shown in \eqref{densitymat}. Indeed,
\begin{equation}
\rho(q)=\tr(\proj{q}{q}\hat\rho)=\bra{q}\hat\rho\ket{q}=\int \d q'' \int \d q'\rho(q'',q')\braket{q}{q''}\braket{q'}{q} =\rho(q,q)
\end{equation}
where we used \eqref{densitymat} and the fact that $\braket{q}{q'}=\delta(q-q')$. Probabilities of finding a particle in a region $q\in[q_0,q_1]$ are given by integrals of this distribution: 
\begin{equation}
P(q_0\le q\le q_1)=\int_{q_0}^{q_1}\!\!\!\d q\, \rho(q).
\end{equation}
Now let us compare the probability distribution $\rho(q)=\rho(q,q)$ we just obtained with the marginal of the Wigner function after integrating momentum out:
When we integrate both sides of \eqref{Wigneryes} over \(p\) and interchange the integrations over \(x\) and \(p\) we find
\begin{equation}
\int_{-\infty}^{\infty} \d p\, W(q, p)=\int_{-\infty}^{\infty} \d x\, \left\langle q+\tfrac{1}{2} x \right|\hat{\rho}\left| q-\tfrac{1}{2} x\right\rangle \underbrace{\frac{1}{2 \pi \hbar} \int_{-\infty}^{\infty} \d p \exp \left(-\frac{\ii}{\hbar} p x\right)}_{\delta(x)}=\bra{q}\hat\rho\ket{q}=\rho(q),
\end{equation}
So indeed summing over all possible momenta (the position marginal) of the Wigner function gives the position probability distribution.

Analogously, we can prove that the momentum probability distribution (the probabilty of finding the particle with momentum $p$) is also a marginal (after integrating positions) of the Wigner function. Let's prove it. Born's rule tells us that the probability (density) of finding the particle with momentum $p$ is
\begin{equation}
\rho(p)\coloneqq \tr(\hat\rho\proj{p}{p})=\ematriz{p}{\hat\rho}{p}=\int \d q'' \int \d q' \braket{p}{q''}\ematriz{q''}{\hat\rho}{q'}\braket{q'}{p}.
\end{equation}
Now, recalling $\braket{p}{q''}=\dfrac{e^{-\frac{\ii}{\hbar}p q''}}{\sqrt{2\pi\hbar}}$:
\begin{equation}
\rho(p)=\ematriz{p}{\hat\rho}{p}=\frac{1}{2 \pi \hbar} \int_{-\infty}^{\infty} \!\!\!\d q''\int_{-\infty}^{\infty} \!\!\!\d q^{\prime}\,  e^{-\frac{\ii}{\hbar} p\left(q^{\prime \prime}-q^{\prime}\right)}\left\langle q^{\prime \prime}|\hat{\rho}| q^{\prime}\right\rangle.
\label{tocompare}
\end{equation}
And from \eqref{Wigneryes},
\begin{equation}
\int_\mathbb{R}\!\d q\, W(q, p) = \frac{1}{2 \pi \hbar} \int_{-\infty}^{\infty}\!\!\! \d q\int_{-\infty}^{\infty}\!\!\! \d x\, e^{-\frac{\ii}{\hbar} p x}\left\langle q+\tfrac{1}{2} x \right|\hat{\rho}\left| q-\tfrac{1}{2} x\right\rangle=\frac{1}{2 \pi \hbar} \int_{-\infty}^{\infty} \!\!\!\d q''\int_{-\infty}^{\infty} \!\!\!\d q^{\prime} e^{-\frac{\ii}{\hbar} p\left(q^{\prime \prime}-q^{\prime}\right)}\left\langle q^{\prime \prime}|\hat{\rho}| q^{\prime}\right\rangle,
\end{equation}
where in the last step we have undone the change of variables \eqref{wigner1}. Comparing with \eqref{tocompare}, we finally get what we expected:
\begin{equation}
\int_\mathbb{R}\!\d q\, W(q, p)=\ematriz{p}{\hat\rho}{p} =\rho(p),
\end{equation}
i.e., the momentum distribution is the marginal of the Wigner function after integrating out $q$.

\subsubsection{Inner product of two states and Wigner function bound}

The inner product between two general states $\hat\rho_1$ and $\hat \rho_2$, given by $\tr(\hat\rho_1\hat\rho_2)$, can be quickly computed in terms of the Wigner functions. It is enough to evaluate a calcualtion completely analogous to  \eqref{prod1} taking $\hat\rho\to\hat\rho_1$, $\hat A_f\to\hat\rho_2$, which yields
\begin{equation}
\tr(\hat\rho_1\hat\rho_2)=\int \d q \int \d p \underbrace{\frac{1}{2\pi\hbar}\int\!\!  \d x\, e^{\frac{-\ii}{\hbar} p x }  \bra{q+\tfrac{1}{2} x}\hat{\rho}_1 \ket{q-\tfrac{1}{2} x}}_{W_1(q,p)}\underbrace{\int\!\d y\,e^{\frac{-\ii}{\hbar} p y }\bra{q+\tfrac{1}{2} y}\hat\rho_2\ket{q-\tfrac{1}{2} y}}_{2\pi\hbar W_2(q,p)},
\label{firstexp}
\end{equation}
and therefore
\begin{equation}
\tr(\hat\rho_1\hat\rho_2)=2\pi\hbar \int \d q \int \d p\,  W_1(q,p) W_2(q,p),
\label{lamelaa}
\end{equation}
where $W_1(q,p)$ and $W_2(q,p)$ are  the Wigner functions representing $\hat\rho_1$ and $\hat\rho_2$ respectively.

Now if we take $\hat\rho_1=\hat\rho_2=\hat\rho$, we know that $\tr\hat\rho^2\le 1$, which introduces a bound for the integral of the square of the Wigner function. From \eqref{lamelaa} we get
\begin{equation}
\tr\hat\rho^2=2\pi\hbar \int \d q \int \d p\,  [W(q,p)]^2\le 1\Rightarrow \int \d q \int \d p\,  [W(q,p)]^2\le \frac{1}{2\pi\hbar}
\end{equation}
with the inequality saturating to an equality for pure states.

Additionally, we can find a bound for the Wigner function. Let us consider the Wigner function of a pure state $\hat\rho=\proj{\psi}{\psi}$. From \eqref{Wigneryes} we get the simple expression 
\begin{equation}
W(q, p) = \frac{1}{2 \pi \hbar} \int_{-\infty}^{\infty}\!\!\! \d x\, e^{-\frac{\ii}{\hbar} p x}\left\langle q+\tfrac{1}{2} x \right|\!\psi \rangle\!\langle\psi\! \left| q-\tfrac{1}{2} x\right\rangle=\frac{1}{2 \pi \hbar} \int_{-\infty}^{\infty}\!\!\! \d x\, e^{-\frac{\ii}{\hbar} p x}\psi\big(q+\tfrac{1}{2} x\big) \psi^*\big(q-\tfrac{1}{2} x\big)
\label{wignerpure}
\end{equation}
but this can be interpreted as the $L^2$ product of two $L^2$ norm 1 wavefunctions:
\begin{equation}
\phi_1(x)=\frac{1}{\sqrt{2}}e^{\frac{i}{\hbar} p x}\psi^*\big(q+\tfrac{1}{2} x\big),\qquad \phi_2(x)=\frac{1}{\sqrt{2}}\psi^*\big(q-\tfrac{1}{2} x\big),
\end{equation}
so that
\begin{equation}
W(q, p) = \frac{1}{ \pi \hbar} \int_{-\infty}^{\infty}\!\!\! \d x\, \phi_1^*(x)\phi_2(x)=\frac{1}{ \pi \hbar} \braket{\phi_1}{\phi_2}.
\end{equation}
The inner product of two states is bounded in modulus by the Cauchy-Schwarz inequality:
\begin{equation}
\left|\left\langle\phi_{1} | \phi_{2}\right\rangle\right|^{2} \leq\left\langle\phi_{1} | \phi_{1}\right\rangle \cdot\left\langle\phi_{2} | \phi_{2}\right\rangle= 1 \cdot 1\Rightarrow \left|\left\langle\phi_{1} | \phi_{2}\right\rangle\right|\le 1,
\end{equation}
which in turn gives the following bound for the Wigner function of two pure states:
\begin{equation}
|W(q, p)| \le \frac{1}{ \pi \hbar} 
\end{equation}
Now, since non-pure states are convex combinations of pure states (with positive, adding to one coefficients), we could easily prove that this bound holds for any state, pure or not. \textbf{Exercise: }prove it!

\subsubsection{The Wigner function can be negative!}

So far it looked like the Wigner function was a very good candidate for a probability distribution on phase space, however it fails in the most basic of the interpretations of probability densities: it is not always positive semidefinite. 

This is easy to prove again with expression \eqref{lamelaa}. Just take two orthogonal states $\hat \rho_1,\hat\rho_2$ with Wigner functions whose support is the whole phase space\footnote{For example, as we will see later, the ground state and all the Fock excitations of a harmonic oscillator.}. The orthogonality condition $\tr(\hat\rho_1\hat\rho_2)=0$ means that 
\begin{equation}
2\pi\hbar \int \d q \int \d p\,  W_1(q,p) W_2(q,p)=0,
\end{equation}
and since their supports are the whole phase space, this is only possible if at least one of them takes on negative values as well as positive. In fact, this equation establishes that the Wigner function cannot have any semidefinite signature. We will see, when we analyze the Wigner functions for systems of harmonic oscillators, that the energy eigenstates (the Fock states) have Wigner functions with negative values.

This condition means that it is impossible to interpret the Wigner function as a true probability distribution. In fact, there is no way, regardless of the operator ordering picked, to have a positive semidefinite phase space probability distribution in quantum theory. We will see that for classical states (those that can be explained without quantum theory) the Wigner function is indeed positive semidefinite, but there are states for which it is not. The negativity of the Wigner function is one of the many manifestations of genuinely quantum features of states and the theory.

\subsubsection{Other properties of the Wigner function}

Here is a list of other remarkable properties for you to prove as exercise or to do some further reading: 
\begin{enumerate}
\item $\hat \rho^{\dagger}=\hat \rho \Rightarrow W(q,p)\in\mathbb{R}$ (easy to prove from the definition).
\item Hudson theorem: For pure states, $W(q,p)\ge0, \forall q,p\in\mathbb{R}\Rightarrow W(q,p)$ is a Gaussian. Hence,  classical pure  states have Gaussian Wigner functions.
\item Folland and Sitaram theorem: $W(q,p)$ has compact support $\Rightarrow W(q,p)=0\quad\forall p,q\in\mathbb{R}$.  
\end{enumerate}

\subsection{Von Neumann Equation in Phase Space}

We know that the time evolution of the density matrix is given by the von Neumann equation
\begin{equation}
\frac{\partial \hat{\rho}}{\partial t}=-\frac{\ii}{\hbar}[\hat{H}, \hat{\rho}].
\end{equation}
This allows us to take the time derivative of the Wigner function. From \eqref{Wigneryes}
\begin{equation}
\frac{\partial W}{\partial t}= \frac{-\ii}{2 \pi \hbar^2} \int_{-\infty}^{\infty}\!\!\! \d x\, e^{-\frac{\ii}{\hbar} p x}\left\langle q+\tfrac{1}{2} x \right|[\hat H,\hat{\rho}]\left| q-\tfrac{1}{2} x\right\rangle.
\end{equation}
Let us now consider a Hamiltonian of the `Mechanical Energy' family:
\begin{equation}
\hat H=\frac{\hat p^2}{2m}+U(\hat q).
\end{equation}
We can then split the equation of motion for the Wigner function in two terms:
\begin{equation}
\frac{\partial W}{\partial t}=\mathcal{T}+\mathcal{U},
\end{equation}
where the `kinetic' and `potential' terms are
\begin{equation}
\mathcal{T}= \frac{-\ii}{4 m \pi \hbar^2} \int_{-\infty}^{\infty}\!\!\! \d x\, e^{-\frac{\ii}{\hbar} p x}\left\langle q+\tfrac{1}{2} x \right|[\hat p^2,\hat{\rho}]\left| q-\tfrac{1}{2} x\right\rangle,\qquad \mathcal{U}= \frac{-\ii}{2 \pi \hbar^2} \int_{-\infty}^{\infty}\!\!\! \d x\, e^{-\frac{\ii}{\hbar} p x}\left\langle q+\tfrac{1}{2} x \right|[ U(\hat q),\hat{\rho}]\left| q-\tfrac{1}{2} x\right\rangle.
\end{equation}
Both terms can be written in terms of the Wigner function itself. As an exercise, show that
\begin{equation}
\mathcal{T}=-\frac{p}{m} \frac{\partial}{\partial q} W(q,p),\qquad \mathcal{U}=\sum_{l=0}^{\infty} \frac{(\ii \hbar / 2)^{2 l}}{(2 l+1) !} \frac{\d^{2 l+1} U(q)}{\d q^{2 l+1}} \frac{\partial^{2 l+1}}{\partial p^{2 l+1}} W(q, p).
\end{equation}

\section{The Harmonic oscillator}

To get a better intuition on Wigner functions, let us consider the simple case of the harmonic oscillator. The Hamiltonian of the harmonic oscillator is given by
\begin{equation}\hat{H}=\frac{\hat{p}^{2}}{2 m}+\frac{m}{2} \omega^{2}\hat{q}^{2}=\hbar\omega\left(\hat a^\dagger \hat a+\frac{1}{2}\right),\end{equation}
where for the last step we have defined the annihilation and creation operators as
\begin{equation} \hat a =\sqrt{\frac{m \omega}{2 \hbar}}\left(\hat{q}+\frac{\ii}{m \omega} \hat{p}\right), \qquad \hat a^{\dagger} =\sqrt{\frac{m \omega}{2 \hbar}}\left(\hat{q}-\frac{\ii}{m \omega} \hat{p}\right), \label{defcreationannihilation} \end{equation}
or equivalently
\begin{equation}\hat{q}=\sqrt{\frac{\hbar}{2} \frac{1}{m \omega}}\left(\hat a^{\dagger}+\hat a\right),\qquad \hat{p}=\ii \sqrt{\frac{\hbar}{2} m \omega}\left(\hat a^{\dagger}-\hat a\right).\end{equation}
Notice that these annihilation and creation operator are not born within quantum theory. Rather, they are the Weyl quantization of the classical variables
\begin{equation}  \alpha =\sqrt{\frac{m \omega}{2 \hbar}}\left({q}+\frac{\ii}{m \omega} {p}\right), \qquad  \alpha^* =\sqrt{\frac{m \omega}{2 \hbar}}\left({q}-\frac{\ii}{m \omega} {p}\right). 
\label{changvar}
\end{equation}
We could choose to describe the system in terms of canonical variables $q$ and $p$ or in terms of the annihilation and creation variables $\alpha$ and $\alpha^*$. Finally, note that $\left[a, a^{\dagger}\right]=\openone$.

\subsection{Wigner function of the ground state}

From basic quantum mechanics we know the position representation wavefunction for the ground state of the harmonic oscillator:
\begin{equation}
\psi_0(q)=\braket{q}{0}=\frac{1}{\left(2 \pi a_{0}^{2}\right)^{1 / 4}} \mathrm{e}^{-q^{2} /\left(4 a_{0}^{2}\right)}, \qquad a_{0}=\sqrt{\frac{\hbar}{2 m \omega}},
\end{equation}
which directly substituting into \eqref{wignerpure} yields
\begin{equation}W(q, p)=\frac{1}{2\pi\hbar}\int \d x \frac{1}{\sqrt{2 \pi a_{0}^{2}}} \exp \left[-\frac{(q+x / 2)^{2}}{4 a_{0}^{2}}-\frac{(q-x / 2)^{2}}{4 a_{0}^{2}}\right] \mathrm{e}^{i p x / \hbar}=\frac{1}{\pi\hbar} \exp \left[-\frac{q^{2}}{2 a_{0}^{2}}-\frac{p^{2}}{2 p_{0}^{2}}\right],
\label{ground}
\end{equation}
where $p_{0}=\hbar/(2 a_{0})$. The Wigner function of the ground state is a (spherical) Gaussian distribution. If we look at the product of the two uncertainties we get that
\begin{equation}
p_0a_0=\frac{\hbar}{2 a_{0}}a_0=\frac\hbar2=\Delta q\Delta p,
\end{equation}
which is precisely the saturation of Heisenberg's uncertainty principle.

\subsection{Coherent states}

A \textbf{coherent state} $\ket{\alpha}$ is defined to be the eigenstate of the annihilation operator (acting from the left)  with eigenvalue $\alpha$:
    \begin{equation}
        \hat a\ket{\alpha} = \alpha\ket\alpha\,.
    \end{equation}
    Now, consider the \textbf{displacement operator} 
    \begin{equation}
        \hat D(\alpha) \coloneqq e^{\alpha\hat a^\dagger - \alpha^*\hat a}=e^{-|\alpha|^{2} / 2} e^{\alpha a^{\dagger}} e^{-\alpha^{*} a}=e^{|\alpha|^{2} / 2} e^{-\alpha^{*} a} e^{\alpha a^{\dagger}},\qquad \alpha\in\mathbb{C}\,.
        \label{firstdisp}
    \end{equation}
   For the last equality we have used the BCH formulas for two operators \(\hat A\) and \(\hat B\) satisfying
$$
[[\hat A, \hat B], \hat A]=[[\hat A, \hat B], \hat B]=0
$$
which implies
$$
e^{\hat A+\hat B}=e^{-[\hat A, \hat B] / 2} e^{\hat A} e^{\hat B}.
$$
If we write \(\hat A=\alpha \hat a^{\dagger}, \hat B=-\alpha^{*} \hat a,\) and viceversa, one gets the last two steps of \eqref{firstdisp}.

We can also show that $\hat D(\alpha)\ket{0}$ is a normalized coherent state with eigenvalue $\alpha$, i.e., we can identify \mbox{$\ket\alpha = \hat D(\alpha)\ket{0}$}. Let us compute the action of the displacement operator on the ground state of the harmonic oscillator:  
    \begin{equation}
        \begin{split}
            \hat D(\alpha)\ket 0 &= e^{\alpha\hat a^\dagger}e^{-\alpha^*\hat a}e^{-|\alpha|^2/2}\ket 0\\
            &= e^{-|\alpha|^2/2}e^{\alpha\hat a^\dagger}\rr{\openone -\alpha^*\hat a+\dots}\ket 0\\
            &= e^{-|\alpha|^2/2}e^{\alpha\hat a^\dagger}\ket 0\\
            &= e^{-|\alpha|^2/2}\rr{\sum_{n=0}^\infty \frac{(\alpha\hat a^\dagger)^n}{n!}}\ket 0\\
            &= e^{-|\alpha|^2/2}\sum_{n=0}^\infty \frac{\alpha^n}{\sqrt{n!}}\ket n\,.
        \end{split}
        \label{momen}
    \end{equation}
 A displaced ground state is a coherent state.  To see that this is a coherent state, we apply $\hat a$:
    \begin{equation}
        \begin{split}
        \hat a\hat D(\alpha)\ket 0 
        &= e^{-|\alpha|^2/2}\sum_{n=1}^\infty \frac{\alpha^n}{\sqrt{n!}}\sqrt{n}\ket{n-1}\\
        &= e^{-|\alpha|^2/2}\sum_{n=1}^\infty \frac{\alpha^n}{\sqrt{(n-1)!}}\ket{n-1}\\
        &= \alpha e^{-|\alpha|^2/2}\sum_{n=1}^\infty \frac{\alpha^{n-1}}{\sqrt{(n-1)!}}\ket{n-1}\\
        &= \alpha e^{-|\alpha|^2/2}\sum_{n=0}^\infty \frac{\alpha^{n}}{\sqrt{(n)!}}\ket{n}\\
        &= \alpha \hat D(\alpha)\ket0\,.
        \end{split}
    \end{equation}
    The normalization follows because $\hat D(\alpha)$ is clearly unitary, so the norm is preserved:
    \begin{equation}
        \braket{\alpha}{\alpha} = \ematriz{0}{\hat D(\alpha)^\dagger\hat D(\alpha)}{0} = \braket{0}{0} = 1\,.
    \end{equation}
    
We can prove that there is no such from-the-left eigenstate for the creation operator by contradiction. Suppose such an eigenstate exists. Then 
    \begin{equation}
        \begin{split}
            \hat a^\dagger\ket{\alpha'} &=   \sum_{n=0}^\infty c_n\sqrt{n+1}\ket{n+1} = \alpha\sum_{n=0}^\infty c_n\ket{n}\,.
        \end{split}
    \end{equation}
    Comparing coefficients, we see that
    \begin{equation}
        c_0 = 0\,, c_1 = c_0\sqrt{n+1}\,, \dots \,, c_n = c_{n-1}\sqrt{n}\,.
    \end{equation}
    However, since $c_0 = 0$, we see that $c_1=0$, hence $c_n=0$ for all $n$. This implies that $\ket{\alpha'}=0$, contradicting the fact that it is an eigenvector of $\hat a^\dagger$. This is the reason why despite the ``symmetry" of the ladder operators, only annihilation operator has sensible eigenstates acting from the left. 
    
    Interestingly, unlike in finite-dimensional vector spaces, where over complex field $\mathbb{C}$ every linear operator has an eigenvalue and hence an eigenvector, this result shows that this expectation does not hold in infinite-dimensional Hilbert spaces.

The coherent states form a basis of the Hilbert space of a harmonic ocillator, but it is not an orthonormal basis.  Consider two normalized coherent states $\ket\alpha,\ket\beta$,   where $\alpha,\beta\in\mathbb{C}$. We will show that for any $\alpha,\beta\in \mathbb{C}$, these two coherent states \textit{cannot} be orthogonal. To see that we need to use BCH formulas again. In particular, it would be useful that if
\be
e^{\hat X} e^{\hat Y}=e^{\hat Z},
\ee
then
\be
\hat Z=\hat X+\hat Y+\frac{1}{2}[\hat X, \hat Y]+\frac{1}{12}[\hat X,[\hat X, \hat Y]]-\frac{1}{12}[\hat Y,[\hat X, \hat Y]]+\cdots.
\ee
In our case this implies that
\begin{equation}
        e^{\alpha\hat a^\dagger-\alpha^*\hat a}e^{\beta\hat a^\dagger-\beta^*\hat a} = e^{(\alpha+\beta)\hat a^\dagger-(\alpha^*+\beta^*)\hat a}e^{(\alpha\beta^*-\beta\alpha^*)/2}\,.
        \label{useful}
    \end{equation}
Playing a bit with the signs we get
\begin{equation}
        \begin{split}
            \braket{\alpha}{\beta} &= \bra{0}\hat D^\dagger(\alpha)\hat D(\beta)\ket{0}\\
            &= \bra0\hat D(-\alpha)\hat D(\beta)\ket{0}\\
            &= \bra{0}e^{(-\alpha\beta^*+\beta\alpha^*)/2}\hat D(\beta-\alpha)\ket{0}\\
            &= e^{(-\alpha\beta^*+\beta\alpha^*)/2} e^{-|\beta-\alpha|^2/2}\\
            &= e^{(-|\beta|^2-|\alpha|^2+2 \beta\alpha^*)/2}\neq\delta({\alpha-\beta})\,,
        \end{split}
        \label{cohproduct}
    \end{equation}
   where for the last steps (displacement expectation on vacuum) we used \eqref{momen}. As an overcomplete basis, the resolution of the identity can be written in terms of coherent states as
\begin{equation}
\openone=\sum_{n=0}^{\infty}\proj{n}{n}=\frac1\pi\int\!\d^2\alpha\, \proj{\alpha}{\alpha}=\frac1\pi\int\!\d(\text{Re}\,\alpha)\d(\text{Im}\,\alpha)\, \proj{\alpha}{\alpha}
\end{equation}
We can prove this relationship, for example, by substituting \eqref{momen}:
\begin{equation}
\int \!\d^{2} \alpha\proj{\alpha}{\alpha}=\int \!\d^{2} \alpha e^{-|\alpha|^{2}} \sum_{m, n} \frac{\left(\alpha^{*}\right)^{n} \alpha^{m}}{\sqrt{n ! m !}}| m\rangle\langle n|,
\label{intermed}
\end{equation}
where the measure \(\d^{2} \alpha\) means ``summing" over all complex values of \(\alpha,\) i.e., integrating over the whole complex plane. Now, writing \(\alpha\) in polar form:
\begin{equation}
\alpha=r e^{i \phi} \Rightarrow \d^{2} \alpha=r\d r \d \phi\,, 
\end{equation}
we get
\begin{equation}
\int \!\d^{2} \alpha\, e^{-|\alpha|^{2}}\!\left(\alpha^{*}\right)^{n}  \alpha^{m}=\!\int_{0}^{\infty}\!\!\! \d r\, r e^{-r^{2}} r^{m+n} \underbrace{\int_{0}^{2 \pi}\! \!\!\d \phi\, e^{i(m-n) \phi}}_{2 \pi \delta_{mn}} =2 \pi \delta_{mn} \frac{1}{2} \int_{0}^{\infty}\!\!\! \d (r^{2})\left(r^{2}\right)^{m} e^{-r^{2}}=\pi m ! \, \delta_{mn}.
\label{orthogonalcohconj}
\end{equation}
Using this in \eqref{intermed} we finally get
\begin{equation}
\int \d^{2}\, \alpha|\alpha\rangle\!\bra{\alpha}=\pi \sum_{n} \ket{n}\langle n|=\pi\openone \qquad\Longleftrightarrow\qquad \frac{1}{\pi}\int \d^2\alpha\,|\alpha\rangle\langle\alpha|=\openone.
\end{equation}

There is a very useful identity for coherent states that can be readily proven using the BCH formulas:
the displacement property, which can be proved by noticing from \eqref{firstdisp} that
\begin{equation}
\hat D(\alpha) =e^{|\alpha|^{2} / 2} e^{-\alpha^{*} \hat a} e^{\alpha \hat a^{\dagger}},\qquad
 \hat D(\alpha)=e^{-|\alpha|^{2} / 2} e^{\alpha \hat a^{\dagger}} e^{-\alpha^{*} \hat a}.
\end{equation}
Therefore, using the first expression for  \(\hat D^{\dagger}(\alpha)\) and the second expression for \(\hat D(\alpha) ,\) we get
\bel{belcanto}
\hat D^{\dagger}(\alpha) \hat a \hat D(\alpha)=e^{\alpha^{*} \hat a} e^{-\alpha \hat a^{\dagger}} \hat a e^{\alpha \hat a^{\dagger}} e^{-\alpha^{*} \hat a}.
\ee
Now, using the BCH formula, 
\be
e^{-\alpha \hat A} \hat B e^{\alpha \hat A}=\hat B-\alpha[\hat A, \hat B]+\frac{\alpha^{2}}{2 !}[\hat A,[\hat A, \hat B]]+\ldots
\ee
For \(A=a^{\dagger}, B=a,\) this becomes
\be
e^{-\alpha \hat  a^{\dagger}} \hat a e^{\alpha \hat a^{\dagger}}=\hat a+\alpha \Rightarrow \hat D^{\dagger}(\alpha) \hat a \hat D(\alpha)=\hat a+\alpha.
\label{displacementopa}
\ee

\subsection{The coherent state representation of the Wigner function}\label{CoherentRep}

Although we will revisit this in a much more powerful way in Section \ref{CoherentRepRev}, we can already see that we can write the Wigner function in the representation given by the coherent states. To do so, we first need to express the probability density $\rho_{s}(q,x)$ given by \eqref{densitymat3} in this basis, which we can do introducing identities,
\be
\bra{q+\tfrac{x}{2}} \hat{\rho} \ket{q-\tfrac{x}{2}}=\frac{1}{\pi^{2}}\iint \d^{2}\lambda \, \d^{2}\mu \; \braket{q+\tfrac{x}{2}}{\lambda}\bra{\lambda}\hat{\rho}\ket{\mu}\braket{\mu}{q-\tfrac{x}{2}}  \;.
\ee
Now, we can use the position representation of the Fock states for the Harmonic oscillator
\be
\braket{y}{n}=\frac{1}{\sqrt{2^{n}n!}}\left( \frac{m\omega}{\pi \hbar} \right)^{1/4} e^{-\frac{m\omega}{2\hbar}y^{2}}H_{n}\left( \sqrt{\frac{m\omega}{\hbar}}y \right)
\ee
where $H_{n}$ stands for the $n$-th Hermite polynomial. Recalling \eqref{momen} and defining $z=\sqrt{\frac{m\omega}{\hbar}}y$,  we get
\be
\braket{y}{\lambda}=e^{-\tfrac{|\lambda|}{2}^2}\left(\frac{m\omega}{\pi \hbar} \right)^{1/4} e^{-\tfrac{z}{2}^{2}} \, \sum_{n=0}^{\infty}{\frac{1}{n!}\left( \frac{\lambda}{\sqrt{2}} \right)^n H_{n}(z)} \;.
\ee
Using a property of Hermite polynomials that tells us that
\be
\sum_{n=0}^{\infty}{\frac{t^{n}}{n!}H_{n}(y)}=e^{2yt-t^{2}},
\ee
we conclude that we can write
\be
\braket{y}{\lambda}= \left( \frac{m\omega}{\pi \hbar} \right)^{1/4} e^{(z^{2}-|\lambda|^{2})/2} \, e^{-(z-\lambda/\sqrt{2})^{2}} \;.
\ee
If we apply the change of variables
\be
\beta=\frac{\lambda+\mu}{2}, \qquad \eta=\mu-\lambda,
\ee 
we have to deal with the following brakets
\be \label{braket1}
\braket{q+\tfrac{x}{2}}{\beta-\tfrac{\eta}{2}}=\left( \frac{m\omega}{\pi \hbar} \right)^{1/4} e^{\frac{1}{2}\left(\frac{m\omega}{\hbar}\left(q+\frac{x}{2} \right)^2-\left| \beta-\frac{\eta}{2} \right|^2 \right)} \, e^{-\left( \sqrt{\frac{m\omega}{\hbar}}\left(q+\frac{x}{2} \right) - \frac{1}{\sqrt{2}}\left(\beta-\frac{\eta}{2} \right) \right)^2},
\ee
\be \label{braket2}
\braket{\beta+\tfrac{\eta}{2}}{q-\tfrac{x}{2}}=\left( \frac{m\omega}{\pi \hbar} \right)^{1/4} e^{\frac{1}{2}\left(\frac{m\omega}{\hbar}\left(q-\frac{x}{2} \right)^2-\left| \beta+\frac{\eta}{2} \right|^2 \right)} \, e^{-\left( \sqrt{\frac{m\omega}{\hbar}}\left(q-\frac{x}{2} \right) - \frac{1}{\sqrt{2}}\left(\beta^*-\frac{\eta^*}{2} \right) \right)^2}.
\ee
Hence, starting with the Wigner function in terms of position and momentum \eqref{Wignerdef}, we need to integrate the terms involving $x$,
\be
\begin{split}
\int_{-\infty}^{\infty} & \d x \; e^{-\frac{\ii}{\hbar}px+\frac{m\omega}{2\hbar}\left(\left( q+\frac{x}{2} \right)^{2} + \left( q-\frac{x}{2} \right)^{2}\right)-\left( \sqrt{\frac{m\omega}{\hbar}}\left( q+\frac{x}{2} \right) -\frac{1}{\sqrt{2}} \left( \beta -\frac{\eta}{2} \right) \right)^2-\left( \sqrt{\frac{m\omega}{\hbar}} \left( q-\frac{x}{2} \right) - \frac{1}{\sqrt{2}} \left( \beta^*+\frac{\eta^*}{2} \right) \right)^2}\\
&=\sqrt{\frac{4\pi \hbar}{m\omega}} \, e^{-\left( \beta-\frac{\eta}{2} \right)\left( \beta^*+\frac{\eta^*}{2} \right)-i\left( \beta-\frac{\eta}{2} \right) \sqrt{\frac{2}{\hbar m \omega}}p+\left(\beta - \frac{\eta}{2} \right) \sqrt{\frac{2m\omega}{\hbar}}q+i\left( \beta^*+\frac{\eta^*}{2} \right) \sqrt{\frac{2}{\hbar m \omega}}p+\left( \beta^*+\frac{\eta^*}{2} \right)\sqrt{\frac{2m\omega}{\hbar}}q-\frac{1}{\hbar m \omega}p^2-\frac{m\omega}{\hbar}q^2}.\\
\end{split}
\ee
With the change variables \eqref{changvar}, and taking into account those terms in \eqref{braket1} and \eqref{braket2} that don't involve $x$, upon some (rather tedious) arithmetics, we get
\be \label{Wignercoherent}
W(\alpha,\alpha^*)=\frac{1}{\pi^3 \hbar} \iint \d^2\eta \,\d^2\beta \; e^{-2|\alpha-\beta|^{2}}e^{(\alpha\eta^*-\alpha^*\eta)}e^{(\eta \beta^*-\eta^* \beta)/2}\bra{\beta-\tfrac{\eta}{2}}\hat{\rho}\ket{\beta+\tfrac{\eta}{2}} \;.
\ee
Notice the fact that the element of volume\footnote{Notice the distinction between the  volume form of phase space $\text{d}q\wedge\text{d}p$, the volume form of the complex plane $\text{d}^2\alpha=\text{d}(\text{Re}\,\alpha)\wedge\text{d}(\text{Im}\,\alpha)$ and its expression in terms of products of the differentials of $\alpha$ and $\alpha^*$. It is satisfied that $\text{d}q\wedge\text{d}p=\ii\hbar\, \d\alpha\wedge \d\alpha^*=2\hbar\,\text{d}^2\alpha$.} $\d q\wedge\d p=\ii\hbar\, \d\alpha\wedge \d\alpha^*$ gives the correct normalization,
\be
\ii\hbar \iint \d \alpha \, \d \alpha^* \; W(\alpha,\alpha^*)=1 \;.
\ee
Analogously, we can build the coherent state representation of the operator $\hat{A}_{f}$ as follows:
\be
A(\alpha,\alpha^*)=\frac{2}{\pi^2} \iint \d^2\eta \,\d^2\beta \; e^{-2|\alpha-\beta|^{2}}e^{(\alpha\eta^*-\alpha^*\eta)}e^{(\eta \beta^*-\eta^* \beta)/2}\bra{\beta-\tfrac{\eta}{2}}\hat{A}_{f}\ket{\beta+\tfrac{\eta}{2}} \;,
\ee
so that
\be
\langle \hat{A}_{f} \rangle_{\hat{\rho}}= \ii\hbar \iint \d \alpha \, \d \alpha^* \; W(\alpha,\alpha^*) A(\alpha,\alpha^*) \;.
\ee

\subsubsection{Revisiting the Wigner function of the ground state of the harmonic oscillator}

Using the coherent state representation of the Wigner function it becomes easier to evaluate many Wigner functions. For example let's revisit the ground state $\hat{\rho}=\ket{0}\bra{0}$. From equation \eqref{momen} we know that
\begin{equation}
\braket{0}{\alpha}= e^{-|\alpha|^2/2}
\end{equation}
Therefore,
%\begin{equation}
%\begin{split}
%W_{\ket{0}}(\alpha, \alpha^{*})=\frac1\pi\iint \d \eta^{*} \d \eta\left\langle \alpha-\tfrac{\eta}{2}|0\rangle\langle 0| \alpha+\tfrac{\eta}{2}\right\rangle \mathrm{e}^{\frac{1}{2}\left(\eta^{*} \alpha-\eta \alpha^{*}\right)}=\frac1\pi\iint \d \eta^{*} \d \eta\, e^{-|\alpha|^2-|\eta|^2} \mathrm{e}^{\frac{1}{2}\left(\eta^{*} \alpha-\eta \alpha^{*}\right)}=\frac{2}{\pi}e^{-2|\alpha|^2}
%\end{split}
%\label{coherentground}
%\end{equation}
\begin{equation}
\begin{split}
W_{\ket{0}}(\alpha, \alpha^{*})&=\frac{1}{\pi^3 \hbar} \iint \d^{2} \beta \, \d^{2} \eta \; e^{-2|\alpha-\beta|^{2}}e^{(\alpha \eta^*-\alpha^*\eta)}e^{(\eta \beta^*-\eta^* \beta)/2} \braket{\beta-\tfrac{\eta}{2}}{0}\braket{0}{\beta+\tfrac{\eta}{2}} \\
&= \frac{1}{\pi^3 \hbar} \iint \d^{2}\beta \, \d^{2}\eta \; e^{-2|\alpha|^2-2|\beta|^2+2(\alpha \beta^*+\alpha^* \beta)}e^{-|\beta|^{2}-|\eta|^{2}/4}e^{(\alpha \eta^*-\alpha^* \eta)}e^{(\eta \beta^* - \eta^* \beta)/2} \\
&= \frac{4}{\pi^2 \hbar} \, e^{-6|\alpha|^{2}} \int \d^{2}\beta \; e^{-4|\beta|^{2}+4(\alpha \beta^*+\alpha^* \beta)}=\frac{1}{\pi \hbar} \, e^{-2|\alpha|^{2}},
\end{split}
\label{coherentground}
\end{equation}
performing the integrals upon writing the complex variables in terms of its real and imaginary parts. It is easy to check that
\begin{equation}
\ii\hbar \int \d\alpha\d\alpha^*\,W_{\ket{0}}(\alpha, \alpha^{*})=\frac{2}{\pi} \int \d (\mathrm{Re}\,\alpha) \; e^{-2(\mathrm{Re}\,\alpha)^2} \;\int \d (\mathrm{Im}\,\alpha) \; e^{-2(\mathrm{Im}\,\alpha)^2}=1,
\end{equation}
where we used that $\text{d}\alpha\wedge\text{d}\alpha^*=-2\ii\text{d}^2\alpha=-2\ii\, \text{d}(\text{Re}\, \alpha)\wedge \text{d}(\text{Im}\, \alpha).$
We can also check that \eqref{coherentground} is equal to \eqref{ground} upon substitution \eqref{changvar}. This Wigner function is indeed a circular Gaussian function centred at zero and with minimal uncertainty.

\subsubsection{Wigner function of a coherent state}

To compute the Wigner funciton of a coherent state $\ket\lambda$ we just need to use the result we derived in \eqref{cohproduct}:
\begin{equation}
\braket{\alpha}{\lambda}=e^{(-|\lambda|^2-|\alpha|^2+2 \lambda\alpha^*)/2},
\end{equation}
therefore
\begin{equation}
\begin{split}
W_{\ket{\lambda}}(\alpha, \alpha^{*})&=\frac{1}{\pi^3 \hbar} \iint \d^{2} \beta \, \d^{2} \eta \; e^{-2|\alpha-\beta|^{2}}e^{(\alpha \eta^*-\alpha^*\eta)}e^{(\eta \beta^*-\eta^* \beta)/2} \braket{\beta-\tfrac{\eta}{2}}{\lambda}\braket{\lambda}{\beta+\tfrac{\eta}{2}} \\
&= \frac{1}{\pi^3 \hbar} \iint \d^{2}\beta \, \d^{2}\eta \; e^{-2|\alpha-\beta|^{2}}e^{-|\beta-\lambda|^{2}-|\eta|^{2}/4}e^{(\lambda\eta^*+\lambda^*\eta)/2}e^{(\alpha \eta^*-\alpha^* \eta)}e^{(\eta \beta^* - \eta^* \beta)/2} \\
&= \frac{4}{\pi^2 \hbar} \, e^{-6|\alpha|^{2}} e^{2(\alpha\lambda^*+\alpha^* \lambda)} \int \d^{2}\beta \; e^{-|\beta-\lambda|^{2}-|\beta+\lambda|^{2}} e^{-2|\beta|^{2}+4(\alpha \beta^* + \alpha^* \beta)}\\
&= \frac{4}{\pi^{2}\hbar} \, e^{-6|\alpha|^{2}}e^{-2|\lambda|^{2}}e^{2(\alpha \lambda^*+\alpha^* \lambda)} \int \d^{2}\beta \; e^{-4|\beta|^{2}+4(\alpha \beta^*+\alpha^* \beta)}=\frac{1}{\pi \hbar} \, e^{-2|\alpha-\lambda|^{2}} 
\end{split}
\label{coherent}
\end{equation}
which is another Gaussian but now displaced and centred on $\alpha=\lambda$. The shape of the vacuum or a coherent state is exactly the same.

\subsubsection{Energy eigenstates of a harmonic oscillator}

Another important example is the Wigner function of the Fock state 
\be
|n\rangle=\frac{\left(a^{\dagger}\right)^{n}}{\sqrt{n !}}|0\rangle \;.
\ee
%The overlap of the Fock state and coherent state can be trivially computed  from  \eqref{momen},
%\be
%\langle n | \alpha \rangle=\frac{\alpha^{n} \exp \left[-|\alpha|^{2} / 2\right]}{\sqrt{n !}}
%\ee
%Therefore
%\begin{align}
%W_{n}\left( \alpha,\alpha^{*}\right)&=\frac{1}{n ! \pi} \iint \d \eta \d \eta^{*}\left(\alpha^{*}-\frac{\eta^{*}}{2}\right)^{N}\left(\alpha+\frac{\eta}{2}\right)^{n} \mathrm{e}^{-|\alpha|^{2}-|\eta|^{2} / 4} \mathrm{e}^{\frac{1}{2}\left(\eta^{*} \alpha-\eta \alpha^{*}\right)}\\
%&=\frac{4}{n !\pi} \iint \d \tilde{\eta} \d \tilde{\eta}^{*}\left(2 \alpha^{*}-\tilde{\eta}^{*}\right)^{n}(2 \alpha+\tilde{\eta})^{n} \mathrm{e}^{-2|\alpha|^{2}-|\tilde{\eta}|^{2}}\\
%&=\frac4\pi \mathrm{e}^{-2|\alpha|^{2}} \sum_{m=0}^{n}(-1)^{n-m}|2 \alpha|^{2 m} \frac{n !}{(m !)^{2}((n-m) !)^{2}}
%\iint d \tilde{\eta} d \tilde{\eta}^{*}|\tilde{\eta}|^{2(n-m)} \mathrm{e}^{-|\tilde{\eta}|^{2}}\\
%&=\frac{4(-1)^{n}}{\pi} \mathrm{e}^{-2|\alpha|^{2}} \sum_{m=0}^{N}(-1)^{m}|2 \alpha|^{2 m} \frac{n !}{(m !)^{2}(n-m) !}=\frac{2(-1)^{n}}{\pi} \mathrm{e}^{-2|\alpha|^{2}} L_{n}\left(4|\alpha|^{2}\right)
%\end{align}
We will show in Section~\ref{CoherentRepRev}, when a more suitable expression for the Wigner function is given, that the result for the $n$-th Fock state in the coherent basis representation is
\be
W_{\ket{n}}(\alpha,\alpha^*)=\frac{(-1)^n}{\pi \hbar} \, e^{-2|\alpha|^{2}}L_{n}(4|\alpha|^{2}),
\label{WignercohFock}
\ee
where \(L_{n}(x)\) is the Laguerre polynomial,
\be\label{Laguerredef}
L_{n}(x)=\sum_{m=0}^{n}(-1)^{m} \frac{n !}{(m !)^{2}(n-m) !} x^{m}.
\ee
The Wigner function for the Fock state highly oscillates at \(|\alpha|^{2}<N\) and then rapidly decays at \(|\alpha|^{2}>N .\) It also takes on negative values, indicating that the first Fock excitation cannot be understood in terms of a classical theory.

\subsection{Single-mode squeezed states - The Squeezed vacuum}

Squeezed states are, same as coherent states, states with positive Wigner functions and minimal uncertainty. The difference is that the uncertainty is distributed unequally between independent quadratures. 

Let us introduce the squeezing operator
\be
\hat S(\xi)=\exp \left[\frac12\left(\xi^{*} \hat a^{2}- \xi \hat a^{\dagger 2}\right)\right],
\label{singsq}
\ee
where \(\xi=r \exp (\ii \theta)\) is an arbitrary complex number.  $r$ is called the squeezing parameter and the phase $\theta$ is the squeezing direction. It is easy to see that
$$
\hat S^{\dagger}(\xi)=\hat S^{-1}(\xi)=\hat S(-\xi).
$$
The squeezing operator is unitary (therefore symplectic and preserves the area of phase space) but squeezes the Wigner function in the direction of a quadrature. %You will show in general in an assignment that this is the case.

A straightforward application of the BCH formula
\be
e^{-\alpha \hat A} \hat B e^{\alpha \hat A}=\hat B-\alpha[\hat A, \hat B]+\frac{\alpha^{2}}{2 !}[\hat A,[\hat A, \hat B]]+\ldots, 
\ee
leads to the following useful unitary transformation properties of the squeezing operator:
\begin{align}
\hat S^{\dagger}(\xi) \hat a \hat S(\xi)&=\hat a \cosh r-\hat a^{\dagger} e^{i \theta} \sinh r,\\
\hat S^{\dagger}(\xi) \hat a^{\dagger} \hat S(\xi)&=\hat a^{\dagger} \cosh r-\hat a e^{-i \theta} \sinh r.
\end{align}
Let's compute the action of the squeezing operator on the ground state $\hat S(\xi)\ket0$. For that let us use a little trick, and the following analytic sum:
\begin{equation}
\label{sum}
\sum_{n=0}^\infty \frac{(2 n) !}{2^{2n} (n !)^2}\tanh^{2n}r=\cosh r.
\end{equation}
Now let's consider the following simple equation:
\begin{align}
\hat a\ket0=0\Rightarrow\hat a\hat S(-\xi) \hat S(\xi)\ket0=0&\Rightarrow \hat S^\dagger(-\xi)a\hat S(-\xi) \hat S(\xi)\ket0=0\\
&\Rightarrow (\hat a \cosh r+\hat a^{\dagger} e^{\ii \theta} \sinh r)\hat S\ket0=0.
\end{align}
If we write the squeezed ground state in terms of the Fock basis, $\hat S\ket0=\sum_n C_n \ket n$, we get
\be
(\hat a \cosh r+\hat a^{\dagger} e^{\ii \theta} \sinh r)\sum_n^\infty C_n \ket n=0\Rightarrow \cosh r\sum_{n=1}^\infty C_n \sqrt{n}\ket{n-1}+e^{\ii\theta}\sinh r\sum_{n=0}^\infty C_n\sqrt{n+1}\ket{n+1}=0,
\ee
since the Fock states are linearly independent, the cancellations need to happen term by term. Let's see them all one by one. First, the odd coefficients by imposing the cancellation of the even Fock terms:
\begin{itemize}
\item $\ket0\longrightarrow\cosh r\, C_1=0\Rightarrow C_1=0$.
\item $\ket{2k}\longrightarrow \cosh r\, C_{2k+1}\sqrt{2k+1}+e^{\ii\theta}\sinh r\, C_{2k-1}\sqrt{2k}=0\Rightarrow C_{2k+1}=-e^{\ii\theta}\tanh r\dfrac{\sqrt{2k}}{\sqrt{2k+1}}C_{2k-1}$.
\end{itemize}
This already allows us to determine all the odd coefficients of the Fock expansion of the squeezed states:
\begin{equation}
C_{2k+1}\propto\dots\propto C_5\propto C_3\propto C_1=0.
\end{equation}
We can now deal with the even coefficients by demanding the cancellation of the odd Fock state coefficients:
\begin{itemize}
\item $\ket1\longrightarrow\cosh r\, C_2\sqrt2+e^{\ii\theta}\sinh r\, C_0=0\Rightarrow C_2=-\frac{1}{\sqrt2}e^{\ii\theta}\tanh r\, C_0$.
\item $\ket3\longrightarrow\cosh r\, C_4\sqrt4+e^{\ii\theta}\sinh r\, C_2 \sqrt3=0\Rightarrow C_4=-\frac{\sqrt{3}}{\sqrt4}e^{\ii\theta}\tanh r\, C_2=\frac{\sqrt{3}}{\sqrt4\sqrt2}e^{\ii2\theta}\tanh^2 r\, C_0$.
\item $\ket5\longrightarrow\cosh r\, C_6\sqrt6+e^{\ii\theta}\sinh r\, C_4 \sqrt5=0\Rightarrow C_6=-\frac{\sqrt{5}}{\sqrt6}e^{\ii\theta}\tanh r\, C_4=-\frac{\sqrt5\sqrt{3}}{\sqrt6\sqrt4\sqrt2}e^{\ii3\theta}\tanh^3 r\, C_0$.
\item $\ket{2n-1}\longrightarrow C_{2n}=(-1)^{n}\frac{\sqrt{2n-1}\sqrt{2n-3}\dots}{\sqrt{2n}\sqrt{2n-2}\dots}e^{\ii n\theta}\tanh^n r\, C_0=(-1)^{n}\frac{\sqrt{(2n-1)!!}}{\sqrt{(2n)!!}}e^{\ii n\theta}\tanh^n r\, C_0$.
\end{itemize}
We can simplify the coefficients noting that
\be
\sqrt{\frac{(2 n-1) ! !}{(2 n) ! !}}=\sqrt{\frac{(2 n-1)(2 n-3) \cdots}{(2 n) !!}}  \sqrt{\frac{(2 n) ! !}{(2 n) ! !}}=\frac{\sqrt{(2 n) !}}{(2 n) ! !}=\frac{\sqrt{(2 n) !}}{2 n(2n-2)(2 n-4)\ldots}=\frac{\sqrt{(2n)!}}{2^nn!},
\ee
which means that
\be
C_{2n}=(-1)^{n}\frac{\sqrt{(2n)!}}{2^nn!}e^{\ii n\theta}\tanh^n r\, C_0.
\ee
So we know that the squeezed state in the Fock basis takes the expression
\be
\hat S(\xi)\ket0=C_0\sum_{n=0}^{\infty} \frac{\sqrt{(2n)!}}{2^n n!}(-e^{\ii \theta}\tanh r)^n\ket{2n}.
\ee
We can determine $C_0$ through the state normalization. Since the squeezing is a unitary operation:
\be
||{\hat S(\xi)}\ket0||^2=1\Rightarrow |C_0|^2 \sum_{n=0}^{\infty} \frac{(2n)!}{2^{2n} (n!)^2}\tanh^{2n} r =1 \Rightarrow |C_0|^2\cosh r=1\Rightarrow C_0=\frac{1}{\sqrt{\cosh r}},
\ee
finally yielding
\be
\hat S(\xi)\ket0=\frac{1}{\sqrt{\cosh r}}\sum_{n=0}^{\infty}\frac{\sqrt{(2n)!}}{2^n n!}(-e^{\ii \theta}\tanh r)^n\ket{2n}.
\ee
It is also easy to check that the squeezing the vacuum has an average energy cost:
\begin{align}
\bra 0 \hat S^\dagger \hat a^\dagger \hat a \hat S \ket 0&=\bra 0 \hat S^\dagger \hat a^\dagger \hat S \hat S^\dagger\hat a \hat S \ket 0= \bra 0(\hat a^{\dagger} \cosh r-\hat a e^{-i \theta} \sinh r)(\hat a \cosh r-\hat a^{\dagger} e^{i \theta} \sinh r) \ket 0\\
&=\bra 0(-\hat a e^{-i \theta} \sinh r)(-\hat a^{\dagger} e^{i \theta} \sinh r) \ket 0=\sinh^2 r.
\end{align}
And as we saw in the visualization, and as you will demonstrate in a tutorial, the effect of the squeezing operator is to conserve the area under the Wigner function squeezing one direction while expanding the perpendicular.

\subsection{Two-Mode Squeezed state}

Consider now a system of two harmonic oscillators with creation and annihilation operators $\hat a,\hat b$ such that $\hat q_1\propto (\hat a+\hat  a^\dagger)$, $\hat q_2\propto (\hat b+\hat b^\dagger)$. Now let us consider the operator
\bel{beler}
\hat S(\xi)=\exp\left[\frac12\xi^{*} \hat b \hat a-\frac12\xi \hat b^{\dagger} \hat a^{\dagger}\right].
\ee
The action on the annihilation and creation operators can easily be obtained from BCH formulas. Defining $\xi=r e^{\ii\theta}$ we get
\begin{align}
&\hat S^{\dagger}\hat  a \hat S =\hat a \cosh r-\hat b^{\dagger} e^{i \theta} \sinh r, \\
&\hat S^{\dagger}\hat  a^{\dagger} \hat S =\hat a^{\dagger} \cosh r-\hat b e^{-i \theta} \sinh r, \\
&\hat S^{\dagger}\hat  b \hat S =b \cosh r-\hat a^{\dagger} e^{i \theta} \sinh r, \\
&\hat S^{\dagger} \hat b^{\dagger} \hat S =\hat b^{\dagger} \cosh r-\hat a e^{-i \theta} \sinh r.
\end{align}
Let's compute the action of the two-mode squeezing operator on the ground state of the two oscillators $\hat S(\xi)\ket0\ket0$. For that, let's consider the following simple equation:
\begin{align}
\hat a\ket0\ket{0}=0\Rightarrow\hat a\hat S(-\xi) \hat S(\xi)\ket0\ket{0}=0&\Rightarrow \hat \nonumber S^\dagger(-\xi)a\hat S(-\xi) \hat S(\xi)\ket0\ket{0}=0\\
&\Rightarrow (\hat a \cosh r+\hat b^{\dagger} e^{\ii \theta} \sinh r)\hat S\ket0\ket0=0.\label{jerooo}
\end{align}
Now we write the squeezed ground state in terms of the Fock basis. To do that we first realize that \eqref{beler} can be expanded in a Taylor series and that the excitations (and deexcitations) always happen in pairs: a quantum of the first oscillator is always excited simultaneously with a quantum of the other oscillator, which means that  $\hat S\ket0\ket0=\sum_n C_n \ket n\ket{n}$. Knowing this we can write \eqref{jerooo} as
\be
\cosh r \sum_{n=1}^\infty C_{n} \sqrt{n} |n-1\rangle|n\rangle+ e^{i \phi} \sinh r \sum_{n=0}^\infty C_n \sqrt{n+1} \ket{n}\ket{n+1}= 0.
\ee
First, for the odd coefficients, by imposing the cancelation of the even Fock terms:
\begin{itemize}
\item $\ket0\ket1\longrightarrow\cosh r\, C_1+e^{\ii\theta}\sinh r\, C_{0}=0\Rightarrow C_1=-e^{\ii\theta}\tanh r\, C_0$.
\item $\ket{1}\ket{2}\longrightarrow \cosh r\, C_{2}\sqrt{2}+e^{\ii\theta}\sinh r\, C_{1}\sqrt{2}=0\Rightarrow C_2=-e^{\ii\theta}\tanh r\, C_1=(-e^{\ii\theta}\tanh r)^2 C_0$.
\item $\ket{n-1}\ket{n}\longrightarrow C_n=(-e^{\ii\theta}\tanh r)^n C_0$.
\end{itemize}
So we know that the squeezed state in the Fock basis takes the expression
\be
\hat S(\xi)\ket0=C_0\sum_{n=0}^{\infty}\left(-e^{i \phi} \tanh r\right)^{n}\ket{n}\ket{n}.
\ee
We can determine $C_0$ through the state normalization. Since the squeezing is a unitary operation:
\be
||{\hat S(\xi)}\ket0||^2=1\Rightarrow |C_0|^2 \sum_{n=0}^{\infty} \tanh ^{2 n} r=1 \Rightarrow |C_0|^2\frac{1}{1-\tanh^2 r}=1\Rightarrow C_0=\sqrt{1-\tanh^2 r}=\frac{1}{\cosh r},
\ee
finally yielding
\be
\hat S(\xi)\ket0=\frac{1}{{\cosh r}}\sum_{n=0}^{\infty}\left(-e^{i \phi} \tanh r\right)^{n}\ket{n}\ket{n}.
\ee

\subsubsection{Entanglement in the two-mode squeezed state}

A two-mode squeezed state is a pure bipartite state; let us see what the partial states are like. The density matrix of a two-mode squeezed state is
\begin{equation}
\hat S(\xi)\ket0\bra{0} \hat S^\dagger(\xi)=\frac{1}{\cosh^2 r}\sum_{n=0}^{\infty}\sum_{m=0}^{\infty}e^{\ii \phi(n-m)}\left(- \tanh r\right)^{n+m}\ket{n}\ket{n}\bra{m}\bra{m}.
\end{equation}
Therefore the partial state $\hat \rho_\textsc{a}$ of one of the oscillators is given by tracing out the second oscillator from the two-mode squeezed state 
\begin{equation}
\label{yoshimura}
\hat\rho_\textsc{a}=\tr_{\textsc{b}}\left(\hat S(\xi)\ket0\bra{0} \hat S^\dagger(\xi)\right)=\frac{1}{\cosh^2 r}\sum_{n=0}^{\infty} \tanh^{2n} r\, \ket{n}\bra{n}.
\end{equation}
This is an infinite rank diagonal state in the Fock basis, and thus a mixed state. But not just only a mixed state, in fact we can identify it by remembering that the Hamiltonian of the harmonic oscillator is (modulo a multiple of identity) $\hat H_\textsc{a}=\hbar\omega\,\hat N=\hbar\omega\,\hat a^\dagger \hat a $. Then, looking at the spectral decomposition of the number operator,
\be
\hat N=\hat a^\dagger \hat a=\sum_{n=0}^\infty n\, \ket{n}\bra{n}\Rightarrow \hat H_\textsc{a}=\sum_{n=0}^\infty \hbar\omega n\, \ket{n}\bra{n}
\ee
Now a little bit of algebra tells us that
\be
 f(\hat H)=\sum_{n=0}^\infty f(\hbar\omega n) \ket{n}\bra{n}.
\ee
Then, for the function $f(\hat H)=\exp\left(\frac{2}{\hbar\omega}\hat H\log\left[\tanh r\right]\right)$ we get that
\be
 f(\hat H)=\sum_{n=0}^\infty \exp\left(\frac{2}{\hbar\omega}\hbar \omega n\log\left[\tanh r\right]\right)\ket{n}\bra{n}=\sum_{n=0}^\infty \exp\left(\frac{2}{\hbar\omega}\hbar \omega n\log\left[\tanh r\right]\right)\ket{n}\bra{n}=\sum_{n=0}^{\infty} \tanh^{2n} r\, \ket{n}\bra{n}.
\ee
Therefore, the partial state $\hat \rho_\textsc{a}$ in \eqref{yoshimura} can be written as
\be
\hat \rho_\textsc{a}=\frac{1}{\cosh^2 r} e^{\frac{2}{\hbar\omega}\log\left(\tanh r\right)\hat H},
\ee
which is a thermal state (recall that a thermal state is $\rho_{T}=Z(T)^{-1}e^{-\hat H/(k_\textsc{b}T)}$) of temperature and partition function given by
\be
T=\frac{-\hbar\omega}{2k_\textsc{b}\log\left(\tanh r\right)},\qquad Z=\cosh^2 r,
\ee
where the temperature is zero for $r\to0$ and grows very fast for large $r$.

We recall that for a pure bipartite state, the more mixed the partial state of a subsystem is, the more entangled the state is. Thermal states of the harmonic oscillator can be thought of as one part of a bipartite two-mode squeezed vacuum. The hotter the state, the more entanglement present in the squeezed ground state.

Additionally, notice that the two-mode squeezed  vacuum is a Gaussian state. Its Wigner function is just a 4D Gaussian squeezed in the direction of cross-quadratures. So, a state can be Gaussian, and therefore `classical' and have entanglement? This is a mystery you will have to live with for a little longer until we get deeper into Gaussian quantum mechanics.

\section{Covariant formulation of the Wigner function for $n$ degrees of freedom:}

\subsection{Weyl quantization of $n$ degrees of freedom}\label{secWeyl}

It is time to go beyond one or two degrees of freedom (let's call them modes), and consider an arbitrary number of modes. Before we start, it is important to set the notation. There are different ways to build the canonical symplectic matrix (the components of the symplectic form in the canonical basis) corresponding to different orderings of the canonical basis in the matrix representation of the symplectic form. In particular for the canonical bases such that an arbitrary vector in phase space is represented respectively as $\boldsymbol\xi=(q^1...,q^n,p_1,\dots,p_n)$ and $\boldsymbol\xi=(q^1,p_1,\dots,q^n,p_n)$, we have, respectively,
\begin{align}
    (\tilde\Omega_{\alpha \beta})=
    \begin{pmatrix}
    0&-\openone\\
    \openone&0
    \end{pmatrix},\qquad     (\tilde\Omega_{\alpha \beta}) =\bigoplus_{a=1}^n\begin{pmatrix}
    0&-1\\
    1&0
    \end{pmatrix}.
\end{align}
We will call the second one the pairwise canonical basis and we will adopt this one from now on since it will be more practical to handle most of the calculations. Keep in mind that this is only a choice of basis by reordering the coordinates. The arbitrary basis expressions as well as the coordinate independent ones do not change. 

Consider the $2n$ dimensional symplectic vector space with canonical coordinates $\boldsymbol\xi=(q^1,p_1,\dots,q^n,p_n)$. It is endowed with the following symplectic form:
\begin{align}
    \Omega(\boldsymbol\xi_1,\boldsymbol\xi_2)=\tilde\Omega_{\alpha \beta}\xi^\alpha_1\xi^{\beta}_2,
\end{align}
where we recall the symplectic matrix and its inverse
\begin{align}
    (\tilde\Omega_{\alpha \beta}) =\bigoplus_{a=1}^n\begin{pmatrix}
    0&-1\\
    1&0
    \end{pmatrix},\qquad     (\tilde\Omega^{\alpha \beta})=\bigoplus_{a=1}^n\begin{pmatrix}
    0&1\\
    -1&0
    \end{pmatrix}.\end{align}

Let us study the canonical quantization of this system.
 Consider a set of $2n$ self-adjoint operators $\hat{\Xi}^\alpha$ defined over a Hilbert space,  each associated with a classical coordinate $\xi^a$ in such a way that for every vector in the symplectic space $\boldsymbol{\xi}=\xi^\alpha \mathsf{e}_\alpha$ there is an associated operator $\hat{\Xi}(\boldsymbol\xi)=\xi^\beta\tilde\Omega_{\alpha\beta}\hat{\Xi}^\alpha$.  According to the canonical quantization prescription, these operators fulfill the following commutation relations
 \begin{align}
     [\hat{\Xi}(\boldsymbol\xi_1),\hat{\Xi}(\boldsymbol\xi_2)]=-\ii\hbar\boldsymbol{\Omega}(\boldsymbol\xi_1,\boldsymbol\xi_2)\openone.
 \label{step1}
 \end{align}
Explicitly in terms of canonical coordinates, the l.h.s. and the r.h.s. of \eqref{step1} are
\be
 [\hat{\Xi}(\boldsymbol\xi_1),\hat{\Xi}(\boldsymbol\xi_2)]=\xi_1^\alpha\tilde\Omega_{\beta\alpha}  \xi_2^\gamma\tilde\Omega_{\delta\gamma} [\hat{\Xi}^\beta,\hat{\Xi}^\delta],\qquad \ii\hbar\boldsymbol{\Omega}(\boldsymbol\xi_1,\boldsymbol\xi_2)\openone=-\ii\hbar\tilde\Omega_{\alpha \gamma}\xi^\alpha_1\xi^{\gamma}_2\openone.
\ee
 For these two expressions to be equal we need that
 \be
 [\hat{\Xi}^\beta,\hat{\Xi}^\delta]=\ii\hbar\tilde\Omega^{\beta\delta}\openone.
 \ee
 This means that $[\hat \Xi^1,\hat\Xi^2]=[\hat \Xi^3,\hat\Xi^4]=\dots=[\hat \Xi^{n-1},\hat\Xi^n]=\ii\hbar\openone$, with all other commutators identically zero (except the antisymmetrized versions of the previously listed ones of course). Given these commutation relations, it is common convention that the odd indices correspond to generalized position operators \mbox{$\hat\Xi^{2k-1}\equiv\hat q^{k}$}, and that the even indices correspond to generalized momenta \mbox{$\hat\Xi^{2k}\equiv\hat p_{k}$}. We can define generalized creation and annihilation operators as well and apply the same ordering convention for mixed powers of generalized positions and momenta as for the Weyl quantization of a single degree of freedom.
 
%For example, if we particularize for a single mode/harmonic oscillator (phase space of dimension 2) we obtain that for two arbitrary phase space vector
%\begin{align}
%&\boldsymbol{\xi_1}=(q_1,p_1)\Rightarrow \hat{\Xi}(\boldsymbol\xi_1)=\xi_1^\alpha\tilde\Omega_{\alpha\beta}\hat{\Xi}^\beta=\xi^1_1\tilde\Omega_{12}\hat{\Xi}^2+\xi_1^2\tilde\Omega_{21}\hat{\Xi}^1=-q_1\hat{\Xi}^2+p
%_1\hat{\Xi}^1\\
%&\boldsymbol{\xi_2}=(q_2,p_2)\Rightarrow \hat{\Xi}(\boldsymbol\xi_1)=\xi_2^\alpha\tilde\Omega_{\alpha\beta}\hat{\Xi}^\beta=\xi_2^1\tilde\Omega_{12}\hat{\Xi}^2+\xi_2^2\tilde\Omega_{21}\hat{\Xi}^1=-q_2\hat{\Xi}^2+p
%_2\hat{\Xi}^1
%\end{align}
%and the canonical commutation relation becomes
%\be
 %[\hat{\Xi}(\boldsymbol\xi_1),\hat{\Xi}(\boldsymbol\xi_2)]=\xi_1^\alpha\tilde\Omega_{\alpha\beta}  \xi_2^\gamma\tilde\Omega_{\gamma\delta} [\hat{\Xi}^\beta,\hat{\Xi}^\delta]
%\ee

\subsection{The generalization of the Wigner function}

%\hat \Delta_{_{\boldsymbol{\alpha}(\boldsymbol{\xi})}}

Given a density matrix $\hat\rho$, we define its Wigner function over phase space as 
    \begin{align}
        W_{\hat{\rho}}(\boldsymbol\xi)\coloneqq\frac{1}{(2\hbar\pi)^{2n}}\int \d^{2n}\boldsymbol\xi' e^{\frac{\ii}{\hbar}\boldsymbol{\Omega}(\boldsymbol\xi,\boldsymbol\xi')}\langle e^{-\frac{\ii}{\hbar}\hat{\Xi}(\boldsymbol\xi')}\rangle_{\hat{\rho}}=\frac{1}{(2\hbar\pi)^{2n}}\int \d^{2n}\boldsymbol\xi' e^{\frac{\ii}{\hbar}\tilde\Omega_{\alpha \beta}\xi^\alpha \xi'^{\beta}}\langle e^{\frac{\ii}{\hbar}\xi'^{\alpha}\tilde\Omega_{\alpha\beta}\hat{\Xi}^\beta} \rangle_{\hat{\rho}}.
        \label{Wigeneral}
    \end{align}
 %   where 
%\begin{align}
  %  \hat
    %    \Delta_{_{\boldsymbol{\alpha}(\boldsymbol{\xi})}}\coloneqq\delta^{(2n)}(\boldsymbol{\alpha}(\boldsymbol{\xi})\openone-\hat{\boldsymbol{a}}),
%    \end{align}
%and the delta  is such that
 %  \begin{align}
 %      \delta^{(2n)}(\boldsymbol{\alpha}(\boldsymbol{\xi})\openone-\hat{\boldsymbol{a}})\ket{\boldsymbol{\beta}}= \delta^{(2n)}(\boldsymbol{\alpha}(\boldsymbol{\xi})\openone-\boldsymbol{\beta}\openone)\ket{\boldsymbol{\beta}}=\delta^{(2n)}(\boldsymbol{\alpha}(\boldsymbol{\xi})-\boldsymbol{\beta})\ket{\boldsymbol{\beta}}
 %   \end{align}
% and $\boldsymbol{\alpha}(\boldsymbol{\xi})$ is the complex phase space variable associated with the vector $\boldsymbol{\xi}$.
In this subsection we are going to prove that the covariant form of the Wigner function \eqref{Wigeneral} generalizes the Wigner function we have introduced in the previous sections to an arbitrary number of degrees of freedom. For a single particle/mode we have that $\boldsymbol{\xi}=(q,p)\Rightarrow \xi^1=q$, $\xi^2=p$, and therefore 
\be
\boldsymbol{\Omega}(\boldsymbol\xi,\boldsymbol\xi')=\tilde\Omega_{\alpha \beta}\xi^\alpha \xi^{'\beta}= \tilde\Omega_{12}\xi^1\xi^{'2}+\tilde\Omega_{21}\xi^2\xi^{'1}=-q p' +pq'. 
\label{resultadejo2}
\ee
 In the same fashion,
  \be
 e^{\frac{\ii}{\hbar}\xi^{'\alpha}\tilde\Omega_{\alpha\beta}\hat{\Xi}^\beta}=e^{\frac{\ii}{\hbar}\xi^{'1}\tilde\Omega_{12}\hat{\Xi}^2+\frac{\ii}{\hbar}\xi^{'2}\tilde\Omega_{21}\hat{\Xi}^1}=e^{-\frac{\ii}{\hbar} q'\hat p+\frac{\ii}{\hbar}p'\hat q}=e^{-\frac{\ii}{\hbar}q'\hat p}e^{\frac{\ii}{\hbar}p'\hat q}e^{\frac{\ii}{2\hbar}q'p'\openone},
 \label{tola2}
 \ee
 where in the last step we have used the Zassenhaus formula (another BCH identity)
 \be
 e^{t(\hat X+\hat Y)}=e^{t \hat X} e^{t \hat Y} e^{-\frac{t^{2}}{2}[\hat X, \hat Y]} e^{\frac{t^{3}}{6}(2[\hat Y, [ \hat X, \hat Y] |+[\hat X,[\hat X, \hat Y]])}\ldots
 \ee
Now, substituting \eqref{resultadejo2} and \eqref{tola2} in the definition of the Wigner function \eqref{Wigeneral}, we get
    \begin{align}
          \label{Wigeneralstep2}
        W_{\hat{\rho}}(\boldsymbol\xi)=W_{\hat{\rho}}(p,q)&=\frac{1}{(2\hbar\pi)^{2}}\int \d  q' \int \d p'\, e^{\frac{-\ii}{\hbar}(qp'-pq')}\langle e^{-\frac{\ii}{\hbar}q'\hat p}e^{\frac{\ii}{\hbar}p'\hat q}e^{\frac{\ii}{2\hbar}q'p'\openone}\rangle_{\hat{\rho}}\\
 \notag &=\frac{1}{(2\hbar\pi)^{2}}\int \d  q' \int \d p'\, e^{\frac{-\ii}{\hbar}(qp'-pq')}\tr\left(\hat{\rho}\, e^{-\frac{\ii}{\hbar}q'\hat p}e^{\frac{\ii}{\hbar}p'\hat q}e^{\frac{\ii}{2\hbar}q'p'\openone}\right)\\
\notag  &=\frac{1}{(2\hbar\pi)^{2}}\int \d  q' \int \d p'\, e^{\frac{-\ii}{\hbar}(qp'-pq')}\tr\left(e^{-\frac{\ii}{2\hbar}q'\hat p}e^{\frac{\ii}{\hbar}p'\hat q}e^{\frac{\ii}{2\hbar}q'p'\openone}\hat{\rho}\,e^{-\frac{\ii}{2\hbar}q'\hat p} \right)\\
\notag  &=\frac{1}{(2\hbar\pi)^{2}}\int \d x\int\d y \int \d  q' \int \d p'\, e^{\frac{-\ii}{\hbar}(qp'-pq')}\bra{x}e^{-\frac{\ii}{2\hbar}q'\hat p}e^{\frac{\ii}{\hbar}p'\hat q}e^{\frac{\ii}{2\hbar}q'p'\openone}\proj{y}{y}\hat{\rho}e^{-\frac{\ii}{2\hbar}q'\hat p} \ket{x}   \\
\notag  &=\frac{1}{(2\hbar\pi)^{2}}\int \d x\int\d y \int \d  q' \int \d p'\, e^{\frac{-\ii}{\hbar}(qp'-pq')}e^{\frac{\ii}{\hbar}p'(x-\tfrac{q'}{2})}\underbrace{\braket{x-\tfrac{q'}{2}}{y}}_{\delta[y-(x-q'/2)]}\!\!\!\bra{y}\hat{\rho} \ket{x+\tfrac{q'}{2}}   e^{\frac{\ii}{2\hbar}q'p'}\\
\notag  &=\frac{1}{(2\hbar\pi)^{2}}\int \d x \int \d  q' \int \d p'\, e^{\frac{-\ii}{\hbar}(qp'-pq')}e^{\frac{\ii}{\hbar}p'(x-\frac{q'}{2})}e^{\frac{\ii}{2\hbar}q'p'}\bra{x-\tfrac{q'}{2}}\hat{\rho} \ket{x+\tfrac{q'}{2}}  \\
\notag  &=\frac{1}{(2\hbar\pi)^{2}}\int \d x \int \d  q' \int \d p'\, e^{\frac{-\ii}{\hbar}p'[q-(x-\frac{q'}{2}+\frac{q'}{2})]}e^{\frac{\ii}{\hbar}pq'}\bra{x-\tfrac{q'}{2}}\hat{\rho} \ket{x+\tfrac{q'}{2}}   \\
\notag  &=\frac{1}{(2\hbar\pi)^{2}}\int \d x \int \d  q' \underbrace{\int \d p'\, e^{\frac{-\ii}{\hbar}p'(q-x)}}_{2\pi\hbar\,\delta(q-x)}e^{\frac{\ii}{\hbar}pq'}\bra{x-\tfrac{q'}{2}}\hat{\rho} \ket{x+\tfrac{q'}{2}} \\ 
\notag  &= \frac{1}{2\hbar\pi}\int \d  q' e^{\frac{\ii}{\hbar}pq'}\bra{q-\tfrac{q'}{2}}\hat{\rho} \ket{q+\tfrac{q'}{2}} \quad \stackrel{\mathclap{q'\to-x}}{=}\quad  \frac{1}{2\hbar\pi}\int \d  x\, e^{\frac{-\ii}{\hbar}px}\bra{q+\tfrac{x}{2}}\hat{\rho} \ket{q-\tfrac{x}{2}},  
    \end{align}
which is the Wigner function introduced in \eqref{Wigneryes}.  This equation can also be trivially vectorized to obtain the Wigner function in the case of $n$ degrees of freedom.
 
\subsection{Revisiting the coherent state representation of the Wigner function}\label{CoherentRepRev}
 
We can now obtain a much more powerful version  the coherent state representation for the Wigner function than the one we saw in Section~\ref{CoherentRep}. For a single degree of freedom/mode we recall that $\boldsymbol{\xi}=(q,p)\Rightarrow \xi^1=q$, $\xi^2=p$, and therefore 
\be
\boldsymbol{\Omega}(\boldsymbol\xi,\boldsymbol\xi')=\tilde\Omega_{\alpha \beta}\xi^\alpha \xi^{'\beta}= \tilde\Omega_{12}\xi^1\xi^{'2}+\tilde\Omega_{21}\xi^2\xi^{'1}=-q p' +pq' =\ii \hbar (\eta\alpha^*-\alpha\eta^*),
\label{resultadejo}
\ee
where in the last step we used that
\begin{equation}
q=\sqrt{\frac{\hbar}{2} \frac{1}{m \omega}}\left(\alpha^*+\alpha\right),\qquad {p}=\ii \sqrt{\frac{\hbar}{2} m \omega}\left(\alpha^*-\alpha\right),\qquad q'=\sqrt{\frac{\hbar}{2} \frac{1}{m \omega}}\left(\eta^*+\eta\right),\qquad {p'}=\ii \sqrt{\frac{\hbar}{2} m \omega}\left(\eta^*-\eta\right).
\label{chngvar2}
\end{equation}
In the same fashion
 \be
 e^{\frac{\ii}{\hbar}\xi^{'\alpha}\tilde\Omega_{\alpha\beta}\hat{\Xi}^\beta}=e^{\frac{\ii}{\hbar}\xi^{'1}\tilde\Omega_{12}\hat{\Xi}^2+\frac{\ii}{\hbar}\xi^{'2}\tilde\Omega_{21}\hat{\Xi}^1}=e^{-\frac{\ii}{\hbar} q'\hat p+\frac{\ii}{\hbar}p'\hat q}=e^{\eta \hat a^\dagger - \eta^*\hat a},
 \label{tola}
 \ee
where in the last step we have used the fact that 
\begin{equation}\hat{q}=\sqrt{\frac{\hbar}{2} \frac{1}{m \omega}}\left(\hat a^{\dagger}+\hat a\right),\qquad \hat{p}=\ii \sqrt{\frac{\hbar}{2} m \omega}\left(\hat a^{\dagger}-\hat a\right),\end{equation}
which upon comparison with \eqref{chngvar2} and realization that \eqref{resultadejo} and the exponent of \eqref{tola} have the same algebraic structure, it allows us to perform the substitution  $\eta\to\hat a,\eta^*\to\hat a^\dagger$, $\alpha\to\eta$ in \eqref{resultadejo} to obtain the last step, that is
\be
-q p' +pq' =\ii \hbar (\eta\alpha^*-\alpha\eta^*)\Rightarrow \ii( -q'\hat p+ p'\hat q)=\ii\big[\ii\hbar(\eta^*\hat a-\eta\hat a^\dagger)\big] =\hbar(\eta \hat a^\dagger - \eta^*\hat a).
\ee
Substituting \eqref{resultadejo} and \eqref{tola} in \eqref{Wigeneral} we get
        \begin{align}
        W_{\hat{\rho}}(\boldsymbol\xi)=W_{\hat{\rho}}(\alpha,\alpha^*)=\frac{\ii}{4\pi^2\hbar}\int \d  \eta^*\d \eta\, e^{(\eta^*\alpha-\eta\alpha^*)}\langle e^{\eta \hat a^\dagger - \eta^*\hat a}\rangle_{\hat{\rho}},
        \label{Wigeneralparallel}
    \end{align}
where we have also used that $\d^2\boldsymbol{\xi}=\d q\d p=\ii\hbar\, \d  \eta^*\d \eta$. Finally, we need to evaluate the expectation of $e^{\eta \hat a^\dagger - \eta^*\hat a}$. First we notice from \eqref{firstdisp} that the exponential is exactly a displacement operator $\hat D(\eta)=e^{\eta \hat a^\dagger - \eta^*\hat a}$, then
\begin{align}
\langle e^{\eta \hat a^\dagger - \eta^*\hat a}\rangle_{\hat{\rho}}&=\tr\left(\hat\rho\hat D(\eta)\right)=\tr\left(\hat D(\tfrac{\eta}{2})\hat\rho\hat D(\tfrac{\eta}{2})\right)=\frac{1}{\pi}\int\d^2\beta \tr\left(\hat D(\tfrac{\eta}{2})\hat\rho\hat D(\tfrac{\eta}{2})\proj{\beta}{\beta}\right)\\
&=\frac{1}{\pi}\int\d^2\beta \bra{\beta}\hat D(\tfrac{\eta}{2})\hat\rho\hat D(\tfrac{\eta}{2})\ket{\beta}=\frac{1}{\pi}\int\d^2\beta\, e^{(\eta \beta^{*}-\eta^* \beta)/2} \, \bra{\beta-\tfrac{\eta}{2}}\hat\rho\ket{\beta+\tfrac{\eta}{2}} ,
\end{align}
where we used \eqref{useful}. Finally, substituting this into \eqref{Wigeneralparallel} we get
\be
        W_{\hat{\rho}}(\boldsymbol\xi)=W_{\hat{\rho}}(\alpha,\alpha^*)=\frac{\ii}{4\pi^3\hbar}\int \d\eta\d\eta^*\int\d^2\beta \, e^{\eta^*\big(\alpha-\frac{\beta}{2}\big)-\eta\big(\alpha^{*}-\frac{\beta^{*}}{2}\big)}\bra{\beta-\tfrac{\eta}{2}}\hat\rho\ket{\beta+\tfrac{\eta}{2}}.
        \label{Wigeneralparallelf}
\ee
%To compare with previous expressions one has to realize that this Wigner function still has to be integrated against the element of volume of phase space (which is $\d^2\boldsymbol{\xi}=\d q\d p=\hbar\, \d  \eta^*\d \eta$), and that will cancel the $\hbar$ in practical calculations. 
This expression certainly resembles \eqref{Wignercoherent}, but it is not exactly the same. First of all, notice that these expressions are actually comparable. We deduced \eqref{Wignercoherent} in the context of the harmonic oscillator, but the result is not restricted to it. In particular, the change \eqref{changvar} and the definition \eqref{defcreationannihilation} can be done in general for any one-dimensional system and for arbitrary $\omega$. Finally, it is easy (albeit wearisome) to check that for the general matrix element in the coherent basis, $\ket{\lambda}\bra{\mu}$, both expressions coincide (noting that $2\text d^2\eta=\ii\,\text{d}\eta\text{d}\eta^*$), i.e.
\be
\begin{split}
\frac{\ii}{2\pi^3\hbar}\iint \d \eta \d\eta^* \, \d^2\beta \; & e^{\eta^*(\alpha-\frac{\beta}{2})-\eta(\alpha^{*}-\frac{\beta^{*}}{2})} \braket{\beta-\tfrac{\eta}{2}}{\lambda}\braket{\mu}{\beta+\tfrac{\eta}{2}}=\\
& =\frac{1}{\pi^3 \hbar} \iint \d^2\eta \,\d^2\beta \; e^{-2|\alpha-\beta|^{2}}e^{(\alpha\eta^*-\alpha^*\eta)}e^{(\eta \beta^*-\eta^* \beta)/2}\braket{\beta-\tfrac{\eta}{2}}{\lambda}\braket{\mu}{\beta+\tfrac{\eta}{2}}\\
\end{split}
\ee
for every $\lambda,\mu \in \mathbb{C}$. As the coherent states form a basis of the Hilbert space of a one-dimensional system, we conclude that \eqref{Wignercoherent} and \eqref{Wigeneralparallelf} give the same result for any $\hat{\rho}$.
We can now revisit an example with this powerful expression for the coherent state representation of the Wigner function:
        \begin{align}
        W_{\hat{\rho}}(\boldsymbol\xi)=W_{\hat{\rho}}(\alpha,\alpha^*)=\frac{\ii}{4\pi^2\hbar}\int \d  \eta^*\d \eta\, e^{(\eta^*\alpha-\eta\alpha^*)}\langle \hat D(\eta)\rangle_{\hat{\rho}}.
        \label{Wigenercohcalc}
    \end{align}

 Finally,  we can generalize the displacement and the squeezing operators to the case of multiple degrees of freedom. Let us define
    \begin{align}
        \hat{D}(\boldsymbol{\xi})=e^{-\ii\hat{\Xi}(\boldsymbol\xi)}.
    \end{align}
    One can check that  this definition generalizes the single mode one, and it satisfies
    \begin{align}
         \hat{D}^{\dagger}({\boldsymbol\xi})\hat{\Xi}(\boldsymbol\xi') \hat{D}({\boldsymbol\xi})=\hat{\Xi}(\boldsymbol\xi')+\Omega(\boldsymbol\xi,\boldsymbol\xi')\openone.
         \end{align}
Similarly, consider the unitary operator
    \begin{align}
        \hat{S}_{\boldsymbol F}=e^{ \frac\ii2F_{ab}\hat{\Xi}^a\hat{\Xi}^b},
    \end{align}
    where $F_{ab}$ is a symmetric matrix. Using BCH formulas one can show that 
    \begin{align}
         \hat{S}_{\boldsymbol{F}}^{\dagger}\hat{\Xi}(\boldsymbol\xi) \hat{S}_{\boldsymbol F}=\hat{\Xi}(\boldsymbol{C}\boldsymbol\xi)=\hat{\Xi}(\tensor{C}{^\alpha_\beta}\xi^\beta),
    \end{align}
    where $C$ is the matrix given by
    \begin{align}
        \tensor{C}{^\alpha_\beta}=\tensor{(e^{\Omega F})}{^\alpha_\beta}.
    \end{align}

\subsection{Wigner function of all the eigenstates of a harmonic oscillator}

Now that we found an expression for the coherent state representation of the Wigner function, it is possible to easily calculate it for both the ground state and all the Fock excitations in a much more straightforward manner than using the formulations we saw in Section~\ref{CoherentRep}.

First, let us quickly recompute the Wigner function of the ground state of the harmonic oscillator. We know that
\be
\langle \hat D(\eta)\rangle_{\proj{0}{0}}=\bra{0}{\hat D(\eta)}\ket{0}=\braket{0}{\eta}=e^{-|\eta|^2/2}.
\ee
Inserting this into \eqref{Wigenercohcalc} we get
        \begin{align}
        W_{\ket{0}}(\alpha,\alpha^*)&=\frac{\ii}{4\pi^2\hbar}\int \d  \eta^*\d \eta\, e^{(\eta^*\alpha-\eta\alpha^*)}e^{-|\eta|^2/2}=\frac{1}{2\pi^2\hbar}\int \d x \d y \, e^{(x-\ii y)\alpha-(x+\ii y)\alpha^*}e^{-\frac12x^2}e^{-\frac12y^2}\\&=\frac{1}{2\pi^2\hbar}\int \d x \d y \, e^{2\ii x\,\text{Im}\alpha}e^{-2\ii y\,\text{Re}\alpha}e^{-\frac12x^2}e^{-\frac12y^2}=\frac{1}{2\pi^2\hbar}2\pi e^{-2|\alpha|^2}=\frac{1}{\pi\hbar}e^{-2|\alpha|^2},
        \label{Wigenercohcalc2}
    \end{align}
which coincides with the previously obtained result in \eqref{coherentground}. 

More importantly: expression \eqref{Wigenercohcalc} finally allows us to calculate with a reasonable amount of work the Wigner function of the Fock state \eqref{WignercohFock} which we postponed before. To do so, we will rewrite \eqref{Wigenercohcalc} in a more convenient form. Note, first, that we can introduce the integral in the expectation value,
\be
W_{\hat{\rho}}(\alpha,\alpha^*)=\frac{\ii}{4\pi^{2}\hbar}\left\langle \int \d \eta^* \,\d \eta \; e^{(\eta^* \alpha - \eta \alpha^*)}\hat{D}(\eta) \right\rangle_{\hat{\rho}}
\ee
Now, the integrand
\be
e^{(\eta^*\alpha-\eta\alpha^*)}\hat{D}(\eta)=e^{\eta(\hat{a}^{\dagger}-\alpha^*)-\eta^*(\hat{a}-\alpha)}=\hat{D}(\alpha) \hat{D}(\eta) \hat{D}(\alpha)^{\dagger},
\ee
where we used \eqref{displacementopa}. Thus,
\be
W_{\hat{\rho}}(\alpha,\alpha^*)=\frac{\ii}{4\pi^{2}\hbar}\left\langle \hat{D}(\alpha) \left(\int \d\eta^* \,\d\eta \; \hat{D}(\eta) \right) \; \hat{D}(\alpha)^{\dagger} \right\rangle_{\hat{\rho}}=\frac{1}{2\pi^2\hbar}\left\langle \hat{D}(\alpha)\hat{T}\hat{D}(\alpha)^{\dagger} \right\rangle_{\hat{\rho}},
\label{WignerdeductFock1}
\ee
where we have noted
\be
\hat{T}=\int \d^2 \eta\; \hat{D}(\eta) \;.
\ee
Now, it turns out that
\be
\hat{T}=2\pi e^{i\pi \hat{a}^{\dagger}\hat{a}} \;.
\label{integdisplacement}
\ee
To check this result, we first note that, assuming $n \geq m$ with $l:=n-m$,
\be
\begin{split}\label{braketDFock}
\bra{n}\hat{D}(\eta)\ket{m}&=e^{-|\eta|^{2}/2}\bra{n}e^{\eta \hat{a}^{\dagger}}e^{-\eta^* \hat{a}} \ket{m}=e^{-|\eta|^{2}/2} \sum_{k=0}^{m}{\frac{\eta^{k+l}(-\eta^*)^k}{(k+l)!\,k!}\bra{n}(\hat{a}^{\dagger})^{k+l}\hat{a}^k \ket{m}} \\
&=e^{-|\eta|^{2}/2} \sum_{k=0}^{m}\frac{(-1)^{k}\sqrt{n!\,m!}}{(k+l)!\,k!\,(m-k)!}\eta^{k+l}(\eta^*)^k.
\end{split}
\ee
Now,
\be
\begin{split}
\bra{n}\hat{T}\ket{m}&= \int \d^{2}\eta \bra{n}\hat{D}(\eta) \ket{m}=\sum_{k=0}^{m}\frac{(-1)^{k}\sqrt{n!\,m!}}{(k+l)!\,k!\,(m-k)!} \int \d^{2}\eta \;e^{-|\eta|^{2}/2}\eta^{k+l}(\eta^*)k \\
&=\sum_{k=0}^{m}\frac{(-1)^{k}\sqrt{n!\,m!\,}}{(k+l)!\,k!\,(m-k)!} \, 2^{k+1}(\sqrt{2})^l \underbrace{\int \d^{2}\eta \;e^{-|\xi|^{2}}\xi^{k+l}(\xi^*)^k}_{\pi \, k! \, \delta_{l0}} \;,\\
\end{split}
\ee
where we used \eqref{orthogonalcohconj}. Recalling $l=n-m$, we can rewrite
\be
\bra{n}\hat{T}\ket{m}=\sum_{k=0}^{n}\frac{n!}{(k!)^{2}(n-k)!}(-1)^{k}2^{k+2}\pi\,k!\,\delta_{nm}=2\pi \sum_{k=0}^{n} {n \choose k} (-2)^{k} \delta_{nm}=2\pi (-1)^{n} \delta_{nm} \;.
\ee
As the Fock states form a basis of the Hilbert space, the relation
\be
\bra{n}\hat{T}\ket{m}=2\pi\bra{n}e^{i\pi \hat{a}^{\dagger}\hat{a}} \ket{m}
\ee
completes the proof \eqref{integdisplacement}. Now, returning to \eqref{WignerdeductFock1},
\be
W_{\hat{\rho}}(\alpha,\alpha^*)=\frac{1}{\pi \hbar}\left\langle \hat{D}(\alpha) e^{i\pi \hat{a}^{\dagger}\hat{a}} \hat{D}(\alpha)^{\dagger} \right\rangle_{\hat{\rho}}.
\label{WignerdeductFock2}
\ee
Finally, $e^{i\pi \hat{a}^{\dagger}\hat{a}}$ is clearly a unitary operator, which satisfies
\be
e^{i\pi \hat{a}^{\dagger}\hat{a}}\hat{a}e^{-i\pi \hat{a}^{\dagger}\hat{a}}=-\hat{a} \qquad \textrm{and} \qquad e^{i\pi \hat{a}^{\dagger}\hat{a}}\hat{a}^{\dagger}e^{-i\pi \hat{a}^{\dagger}\hat{a}}=-\hat{a}^{\dagger} 
\ee
as can be straightforwardly checked looking at the matrix elements in the Fock basis. In particular, this implies that
\be
e^{i\pi \hat{a}^{\dagger}\hat{a}} \hat{D}(\alpha)^{\dagger}e^{-i\pi \hat{a}^{\dagger}\hat{a}}=e^{i\pi \hat{a}^{\dagger}\hat{a}} \hat{D}(-\alpha) e^{-i\pi \hat{a}^{\dagger}\hat{a}}=\hat{D}(\alpha) \;.
\ee
Applying this to \eqref{WignerdeductFock2}, we get
\be
W_{\hat{\rho}}(\alpha,\alpha^*)=\frac{1}{\pi \hbar} \left\langle \hat{D}(2\alpha) e^{i\pi \hat{a}^{\dagger}\hat{a}} \right\rangle_{\hat{\rho}}\;. \ee
This is the ``suitable'' form we wanted in order to calculate the Wigner function of a Fock state, which is now straightforward:
\be
\begin{split}
W_{\ket{n}}(\alpha,\alpha^*)&=\frac{1}{\pi \hbar}\bra{n}\hat{D}(2\alpha) e^{i\pi \hat{a}^{\dagger}\hat{a}} \ket{n}=\frac{(-1)^{n}}{\pi \hbar}e^{-2|\alpha|^{2}}\sum_{k=0}(-1)^{k}\frac{n!}{(k!)^2(n-k)!}|2\alpha|^{2k}\\
&=\frac{(-1)^{n}}{\pi \hbar}e^{-2|\alpha|^{2}}L_{n}(4|\alpha|^{2})
\end{split}
\ee
where we have used \eqref{braketDFock} and the definition of Laguerre polynomials given in \eqref{Laguerredef}. We have thus obtained the result anticipated in \eqref{WignercohFock}.

\section{Gaussian Quantum Mechanics}\label{Gstatessec}

\subsection{The covariance matrix and the vector of means}

We are now going to focus on those states that have a Gaussian Wigner function\footnote{Note that some references may use ``Gaussian states'' to refer to a larger set, including also all the possible convex combinations of the Gaussian states we define here. Notice that those states' Wigner functions will not be Gaussian in general, although they are positive and therefore they also correspond to  classical states (that can be thought of as classical probability distributions over the Gaussian states we define in \eqref{GaussDef}). However, the most common definition for Gaussian state is the one we give in these notes.}. Those states are families of squeezed, coherent thermal states. A Gaussian state's Wigner function can be written as
\bel{GaussDef}
W(\boldsymbol{\xi})=\frac{1}{\pi^n \sqrt{\operatorname{det}(\boldsymbol{\sigma})}} \exp \left(-(\boldsymbol{\xi}-\boldsymbol{\xi}_0)^{\mu}(\boldsymbol{\xi}-\boldsymbol{\xi}_0)^{\nu} ({\boldsymbol{\sigma}}^{-1})_{\mu\nu}\right),
\ee
where the vector $\boldsymbol{\xi}_0$ captures the state's first statistical moments (i.e., the mean, the peak of the Gaussian distribution),
\be
{\xi}_0^\mu\coloneqq\langle\hat\Xi^\mu\rangle_{\hat \rho}.
\ee
The matrix \(\boldsymbol{\sigma}\) captures the canonical vairable's statistical second moments, i.e., the covariance between each pair of canonical observables,
\be
\sigma^{\mu\nu}\coloneqq\left\langle\hat{\Xi}^{\mu} \hat{\Xi}^{\nu}+\hat{\Xi}^{\nu} \hat{\Xi}^{\mu}\right\rangle_{\hat \rho}- 2\left\langle\hat{\Xi}^{\mu}\right\rangle_{\hat \rho}\left\langle\hat{\Xi}^{\nu}\right\rangle_{\hat \rho}.
\ee
Note that there are other conventions for the covariance matrix (e.g., dividing by 2, etc.). This one is one of the most convenient notations.

To characterize any Gaussian state it is enough to know its first moments and the covariance matrix, for a total of $n(2n+3)$ real numbers. There is no need to deal with infinite dimensional Hilbert spaces for these states.

The condition of positive definiteness together with the canonical commutation relations  implies that $\boldsymbol{\sigma}\ge \ii\boldsymbol \Omega^{-1}$ where $\boldsymbol{\Omega}^{-1}$ is the matrix inverse of the symplectic matrix (Exercise!).

Note that the condition $\boldsymbol{\sigma}\ge \ii\boldsymbol \Omega^{-1} \Rightarrow \boldsymbol{\sigma}>\boldsymbol{0}$. To prove that the condition $\boldsymbol{\sigma}\ge \ii\boldsymbol \Omega^{-1}$  implies that the covariance matrix is positive semidefinite, we use that conjugating a positive semidefinite form keeps it positive semidefinite, and we have
$$
\boldsymbol{\sigma} \geq \mathrm{i} \boldsymbol{\Omega}^{-1} \Rightarrow \boldsymbol{\sigma}=\boldsymbol{\sigma}^{*} \geq(\mathrm{i} \boldsymbol \Omega^{-1})^{*}=-\mathrm{i} \boldsymbol \Omega^{-1}
$$
where we have used the fact that \(\boldsymbol{\sigma}\) and \(\boldsymbol \Omega^{-1}\) are both real valued. Since \(\boldsymbol{\sigma}\) is greater than both $\ii\boldsymbol \Omega^{-1}$ and \(-\mathrm{i} \boldsymbol \Omega^{-1},\) it is greater than their average as well, thus \(\boldsymbol{\sigma} \geq 0 .\) This implication cannot be reversed.

Note that in addition to keeping the covariance matrix positive semidefinite, the condition $\boldsymbol{\sigma}\ge \ii\boldsymbol \Omega^{-1}$ also enforces the uncertainty principle by preventing the covariances from being arbitrarily small.

\subsubsection{Example 1: Thermal state of a single Harmonic oscillator}

For a system with one degree of freedom, the covariance matrix is a $2\times2$ matrix and the first moment vector is a 2-dimensional vector:
 \bel{cother}
(\sigma^{\mu\nu})= \begin{pmatrix}
2 \langle \hat q^2 \rangle_{\hat \rho}-2\langle \hat q \rangle_{\hat \rho}^2& \langle \hat q \hat p +\hat p \hat q \rangle_{\hat \rho}-2\langle \hat q \rangle_{\hat \rho}\langle \hat p \rangle_{\hat \rho}\\
 \langle \hat p\hat q +\hat q\hat p\rangle_{\hat \rho}-2\langle \hat p \rangle_{\hat \rho}\langle \hat q \rangle_{\hat \rho}&2 \langle \hat p^2 \rangle_{\hat \rho}-2\langle \hat p \rangle_{\hat \rho}^2
 \end{pmatrix},\qquad (\xi_0^\mu)=\begin{pmatrix} \langle \hat q \rangle_{\hat \rho} \\ \langle \hat p \rangle_{\hat \rho}\end{pmatrix}.
\ee
For a single harmonic oscillator, the thermal state covariance matrix is easy to compute. Let us consider the thermal state of temperature $T$:
\be
\hat \rho =\frac{1}{Z}e^{-\beta \hat H}=\frac{1}{Z}e^{-\frac{\hbar \omega}{k_\textsc{b}T}\hat a^\dagger \hat a},
\ee
where
\be
Z=\tr\left(e^{-\frac{\hbar \omega}{k_\textsc{b}T}\hat a^\dagger \hat a}\right)=\sum_{n=0}^\infty\langle n |e^{-\frac{\hbar \omega}{k_\textsc{b}T}\hat a^\dagger \hat a}\ket n = \sum_{n=0}^\infty e^{-\frac{\hbar \omega}{k_\textsc{b}T} n}= \frac{e^{\frac{\hbar \omega}{k_\textsc{b}T}}}{e^{\frac{\hbar \omega}{k_\textsc{b}T} }-1}=\frac{1}{1-e^{-\frac{\hbar \omega}{k_\textsc{b}T}}}.
\label{partitionw}
\ee
The expectation of $\hat q$ and $\hat p$ in a thermal state are zero, which we can quickly see by considering that
\be
\tr\left[e^{-C \hat{a}^{\dagger} \hat{a}}\left(\hat{a}\pm\hat{a}^\dagger\right)\right]=\sum_{n}\langle n |e^{-C \hat{a}^{\dagger} \hat a}(\hat{a}\pm\hat{a}^\dagger)| n\rangle=\sum_{n} e^{-C n}\langle n |(\hat{a}\pm\hat{a}^\dagger) | n\rangle= 0.
\ee
There are four covariances to compute. First, the diagonal ones:
\begin{align}
\nonumber\langle \hat q^2 \rangle_{\hat \rho}&=\frac{\hbar}{2Z}\frac{1}{m\omega}\tr\left(e^{-\frac{\hbar \omega}{k_\textsc{b}T}\hat a^\dagger \hat a}(\hat a^\dagger+\hat a)^2\right)=\frac{\hbar}{2Z}\frac{1}{m\omega}\sum_{n=0}^\infty\langle n |e^{-\frac{\hbar \omega}{k_\textsc{b}T}\hat a^\dagger \hat a}(\hat a^\dagger+\hat a)^2\ket n \\
\nonumber&= \frac{\hbar}{2Z}\frac{1}{m\omega}\sum_{n=0}^\infty e^{-\frac{\hbar \omega}{k_\textsc{b}T} n} \bra{n}(\hat a^\dagger \hat a^\dagger +\hat a \hat a +\hat a^\dagger \hat a+\underbrace{\hat a\hat a^\dagger}_{\hat a^\dagger \hat a+\openone})\ket n\\
\nonumber&=\frac{\hbar}{2Z}\frac{1}{m\omega}\sum_{n=0}^\infty e^{-\frac{\hbar \omega}{k_\textsc{b}T} n} (2n+1)=\frac{\hbar}{2Z}\frac{1}{m\omega}\frac{e^{\frac{\hbar \omega}{k_\textsc{b}T} }(1+e^{\frac{\hbar \omega}{k_\textsc{b}T} })}{(e^{\frac{\hbar \omega}{k_\textsc{b}T} }-1)^2}\\
&=\frac{\hbar}{2m\omega}\operatorname{cotanh}\left(\frac{\hbar\omega}{2k_\textsc{b}T}\right),
\end{align}
where in the last step we substituted $Z$ by  \eqref{partitionw} and simplified. The variance of $\hat p$ can be analogously computed to yield a very similar value: 
\begin{align}
\nonumber\langle \hat p^2 \rangle_{\hat \rho}&=-\frac{\hbar m\omega}{2Z}\tr\left(e^{-\frac{\hbar \omega}{k_\textsc{b}T}\hat a^\dagger \hat a}(\hat a^\dagger-\hat a)^2\right)=-\frac{\hbar m\omega}{2Z}\sum_{n=0}^\infty\langle n |e^{-\frac{\hbar \omega}{k_\textsc{b}T}\hat a^\dagger \hat a}(\hat a^\dagger-\hat a)^2\ket n \\
\nonumber&= -\frac{\hbar m\omega}{2Z}\sum_{n=0}^\infty e^{-\frac{\hbar \omega}{k_\textsc{b}T} n} \bra{n}(\hat a^\dagger \hat a^\dagger +\hat a \hat a -\hat a^\dagger \hat a-\underbrace{\hat a\hat a^\dagger}_{\hat a^\dagger \hat a+\openone})\ket n\\
\nonumber&=\frac{\hbar m\omega}{2Z}\sum_{n=0}^\infty e^{-\frac{\hbar \omega}{k_\textsc{b}T} n} (2n+1)=\frac{\hbar m\omega}{2Z}\frac{e^{\frac{\hbar \omega}{k_\textsc{b}T} }(1+e^{\frac{\hbar \omega}{k_\textsc{b}T} })}{(e^{\frac{\hbar \omega}{k_\textsc{b}T} }-1)^2}\\
&=\frac{\hbar m\omega}{2}\operatorname{cotanh}\left(\frac{\hbar\omega}{2k_\textsc{b}T}\right),
\end{align}
where in the last step we substituted $Z$ by  \eqref{partitionw} and simplified. The non diagonal components can be easily proven to be zero since $(\hat a^\dagger \pm\hat a)(\hat a \mp \hat a^\dagger)=\hat a \hat a+\hat a^{\dagger}\hat a^\dagger$,  which have zero trace. Therefore, from \eqref{cother}, the covariance matrix for the thermal state is
 \bel{fulldime}
(\sigma^{\mu\nu})= \begin{pmatrix}
 \frac{\hbar}{m\omega} \nu & 0\\
0&{\hbar m \omega}\, \nu
 \end{pmatrix},\qquad \nu=\operatorname{cotanh}\left(\frac{\hbar\omega}{2k_\textsc{b}T}\right)>1,
\ee
where $\nu\to 1$ in the limit $T\to0$.

Oftentimes (more often than not, actually) the covariance matrix is represented in a dimensionless form by defining alternate dimensionless position and momentum operators 
\be
\hat{\tilde q}\coloneqq \sqrt{\frac{m\omega}{\hbar}}\hat q=\frac{1}{\sqrt{2}}(\hat a^\dagger +\hat a),\qquad \hat{\tilde p}\coloneqq\sqrt{\frac{1}{\hbar m\omega}}\hat p=\frac{\ii}{\sqrt{2}}(\hat a^\dagger -\hat a).
\label{dimensionless}
\ee
In terms of these dimensionless operators, that we will call quadratures, the covariance matrix takes a very simple form since we restore a symmetry between all the directions in phase space:
 \bel{thermalityyy}
\boldsymbol{\sigma}\coloneqq(\sigma^{\mu\nu})= \begin{pmatrix}
 \nu & 0\\
0&  \nu
 \end{pmatrix}=\nu\openone,\qquad \nu=\operatorname{cotanh}\left(\frac{\hbar\omega}{2k_\textsc{b}T}\right)>1.
\ee
This also means that the covariance matrix of the ground state is the identity (in the dimensionless quadrature basis).

From now on we will use the much more convenient dimensionless quadrature operators to calculate the covariance matrices and vectors of first moments, that is
 \be
\boldsymbol{\xi}_0^\mu\coloneqq\langle\q^\mu\rangle_{\hat \rho},\qquad \sigma^{\mu\nu}\coloneqq\left\langle\q^{\mu} \q^{\nu}+\q^{\nu} \q^{\mu}\right\rangle_{\hat \rho}- 2\left\langle\q^{\mu}\right\rangle_{\hat \rho}\left\langle\q^{\nu}\right\rangle_{\hat \rho},
\label{covariancebueno}
\ee
 where the dimensionless quadrature operators here $ \q^\alpha\propto \hat\Xi^\alpha$ are for simplicity taken to be the  operators in the way defined in \eqref{dimensionless} (that is, generalizing the  $\tq$ and $\tp$  defined before) so that they satisfy the canonical commutation relations
 \be
 [\q^\alpha,\q^\beta]=\ii\tilde\Omega^{\alpha\beta}\openone.
 \ee

\subsubsection{Example 2: Single mode squeezed vacuum}

It is easy to see that the expectation of $\tq$ and $\tp$ are zero in a squeezed vacuum. Knowing that the squeezing operator just squeezes the Wigner function in some direction preserving the centre and the area, we know that a state whose Wigner function is centred at zero will still be centred at zero after squeezing. In fact, it is easy to prove algebraically that any state that is diagonal in the Fock basis (Fock states, thermal states, etc.) will have zero expectation value of the quadratures after squeezing.

Consider an arbitrary single mode state $\hat\rho$. Consider now its squeezed version $\hat\rho_S=\hat S\hat\rho\hat S^\dagger$. The expectation of the $\tq$ quadrature is
\begin{align}
\nonumber\left\langle\tq\right\rangle_{\hat \rho_S}&=\frac{1}{\sqrt2}\left[\tr(\hat S\hat\rho \nonumber S^\dagger\hat a^\dagger)+\tr(\hat S\hat\rho\hat S^\dagger\hat a)\right]=\frac{1}{\sqrt2}\left[\tr(\hat\rho\hat S^\dagger\hat a^\dagger\hat S)+\tr(\hat\rho\hat S^\dagger\hat a\hat S)\right]\\
&=\frac{1}{\sqrt2}\left[\tr\left(\hat\rho (\hat a^{\dagger} \cosh r-\hat a e^{-\ii \theta} \sinh r)\right)+\tr\left(\hat\rho(\hat a \cosh r-\hat a^{\dagger} e^{\ii \theta} \sinh r)\right)\right]\\
&=\frac{1}{\sqrt2}\left[(\cosh r-e^{\ii\theta}\sinh r )\tr(\hat \rho \hat a^\dagger)+ (\cosh r-e^{-\ii\theta}\sinh r )\tr(\hat \rho \hat a)\right].
\end{align}
A completely analogous calculation for $\tp$ yields 
\begin{align}
\nonumber\left\langle\tp\right\rangle_{\hat \rho_S}&=\frac{\ii}{\sqrt2}\left[\tr(\hat S\hat\rho\hat S^\dagger\hat a^\dagger)-\tr(\hat S\hat\rho\hat S^\dagger\hat a)\right]=\\
&=\frac{\ii}{\sqrt2}\left[(\cosh r-e^{\ii\theta}\sinh r )\tr(\hat \rho \hat a^\dagger)- (\cosh r-e^{-\ii\theta}\sinh r )\tr(\hat \rho \hat a)\right].
\end{align}
 If the state before squeezing is diagonal in the Fock basis, that is, $\hat\rho=\sum_n C_n \proj{n}{n}$, then \mbox{$\tr(\hat \rho \hat a^\dagger)=\tr(\hat \rho \hat a)=0$}, and therefore for any such state the expectation of the quadratures for its squeezed version is zero. This includes the case of the squeezed vacuum.
 
 To evaluate the covariances we need to consider 
 \begin{align}
\nonumber& {\left(\tq\right)}^2=\frac12(\hat a^\dagger \hat a^\dagger +\hat a \hat a +2\hat a^\dagger \hat a+\openone),\\
\nonumber &  {\left(\tp\right)}^2=-\frac12(\hat a^\dagger \hat a^\dagger +\hat a \hat a -2\hat a^\dagger \hat a-\openone),\\
 &   {\left(\tq\tp+\tp\tq\right)}=\ii(\hat a^\dagger \hat a^\dagger -\hat a \hat a),
\label{panader}
 \end{align}
and that the squeezed monomials of ladder operators are
\begin{align}
\nonumber \hat S^\dagger\hat a^\dagger \hat a\hat S&=\hat S^\dagger\hat a^\dagger\hat S\hat S^\dagger\hat a\hat S= (\hat a^{\dagger} \cosh r-\hat a e^{-\ii \theta} \sinh r)(\hat a \cosh r-\hat a^{\dagger} e^{\ii \theta} \sinh r)\\
&=-\frac12\left(e^{\ii\theta}\sinh 2r\right)\hat a^\dagger \hat a^\dagger -\frac12\left(e^{-\ii\theta}\sinh 2r\right)\hat a \hat a +\cosh 2r\, \hat a^\dagger \hat a +\sinh^2 r\,\openone,\\
\nonumber \hat S^\dagger\hat a \hat a^\dagger\hat S&=\hat S^\dagger\hat a\hat S\hat S^\dagger\hat a^\dagger\hat S= (\hat a \cosh r-\hat a^{\dagger} e^{\ii \theta} \sinh r)(\hat a^{\dagger} \cosh r-\hat a e^{-\ii \theta} \sinh r)\\
&=-\frac12\left(e^{\ii\theta}\sinh 2r\right)\hat a^\dagger \hat a^\dagger -\frac12\left(e^{-\ii\theta}\sinh 2r\right)\hat a \hat a +\cosh 2r\, \hat a^\dagger \hat a +\cosh^2 r\,\openone,\\
\nonumber \hat S^\dagger\hat a^\dagger \hat a^\dagger\hat S&=\hat S^\dagger\hat a^\dagger\hat S\hat S^\dagger\hat a^\dagger\hat S= (\hat a^{\dagger} \cosh r-\hat a e^{-\ii \theta} \sinh r)(\hat a^{\dagger} \cosh r-\hat a e^{-\ii \theta} \sinh r)\\
&=(\cosh^2 r)  \hat a^\dagger \hat a^\dagger +\left(e^{-2\ii\theta}\sinh^2 r\right)\hat a \hat a -\left(e^{-\ii\theta}\sinh 2r\right)\hat a^\dagger \hat a -\frac12\left(e^{-\ii\theta}\sinh 2r\right)\openone,\\
\nonumber \hat S^\dagger\hat a \hat a\hat S&=\hat S^\dagger\hat a\hat S\hat S^\dagger\hat a\hat S= (\hat a \cosh r-\hat a^{\dagger} e^{\ii \theta} \sinh r)(\hat a \cosh r-\hat a^{\dagger} e^{\ii \theta} \sinh r)\\
&=\left(e^{2\ii\theta}\sinh^2 r\right)  \hat a^\dagger \hat a^\dagger +(\cosh^2 r)\hat a \hat a -\left(e^{\ii\theta}\sinh 2r\right)\hat a^\dagger \hat a -\frac12\left(e^{\ii\theta}\sinh 2r\right)\openone,
\end{align}
where we have used that $\hat a\hat a^\dagger=\openone +\hat a^\dagger \hat a$, and the identities $\sinh(2r)=2\sinh r\,\cosh r$, $\cosh(2r)=\cosh^2r+\sinh^2 r$.

The respective vacuum expectations (which give the expectation of the quadratures for the squeezed vacuum) are therefore
\begin{align}
&\bra0\hat S^\dagger\hat a^\dagger \hat a\hat S\ket0=\sinh^2 r,\\
&\bra0\hat S^\dagger\hat a \hat a^\dagger\hat S\ket0=\cosh^2 r,\\
&\bra0\hat S^\dagger\hat a^\dagger \hat a^\dagger\hat S\ket0=-\frac12e^{-\ii\theta}\sinh 2 r,\\
&\bra0\hat S^\dagger\hat a \hat a\hat S\ket0=-\frac12e^{\ii\theta}\sinh 2 r.
\end{align}
With this and \eqref{panader} we can quickly obtain that
\begin{align}
&\bra0\hat S^\dagger  \tq\,^2 \hat S\ket0=\frac12\left(-\frac{e^{-\ii\theta}+e^{\ii\theta}}{2}\sinh 2 r +2\sinh^2 r +1\right)=\frac12\left(\cosh 2r-\cos\theta\sinh 2r\right),\\
&\bra0\hat S^\dagger\tp\,^2\hat S\ket0=-\frac12\left(-\frac{e^{-\ii\theta}+e^{\ii\theta}}{2}\sinh 2 r -2\sinh^2 r -1\right)=\frac12\left(\cosh 2r+\cos\theta\sinh 2r\right),\\
&\bra0\hat S^\dagger\left(\tq\tp+\tp\tq\right)\hat S\ket0=\ii\left(-\frac{e^{-\ii\theta}-e^{\ii\theta}}{2}\sinh 2 r\right)=-\sin\theta\sinh2r,
\end{align}
where for the first two rightmost equalities we used that $\cosh 2r=1+2\sinh^2r$. Therefore  the covariance matrix is
 \be
(\sigma^{\mu\nu})= \begin{pmatrix}
\cosh 2r-\cos\theta\sinh 2r & -\sin\theta\sinh2r\\
-\sin\theta\sinh2r&  \cosh 2r+\cos\theta\sinh 2r
 \end{pmatrix}
\ee
Exercise! Compute the first moments and the covariance matrix of a squeezed thermal state. 

Note that this is the long and tedious way to compute the covariance matrix, and for Gaussian states certainly not the way we will use most in practice, as we will see in the following sections. Since Gaussian states are classical, we will see how to fully characterize them and their evolution under quadratic Hamiltonians without having to work with operators on Hilbert spaces.
 
 \subsection{Multipartite covariance matrix}
 
 One of the nicest features of the covariance matrix description of Gaussian quantum mechanics is that the covariance matrix is directly the correlators between the canonical observables of the system. That means that the covariance matrix of a composite system made of two completely uncorrelated subsystems with covariance matrices $\boldsymbol{\sigma}_\textsc{a}$ and $\boldsymbol{\sigma}_\textsc{b}$ is just
 \begin{equation}
 \boldsymbol{\sigma}_\textsc{ab}=\boldsymbol{\sigma}_\textsc{a}\oplus\boldsymbol{\sigma}_\textsc{b},  
 \end{equation}
 to be compared with the tensor product of density matrices that we have in the Hilbert space description of the system. 
 
 In fact, in a composite system it is very easy to identify where the correlations are in the covariance matrices and very easy to take partial traces, since in general, for a bipartite system we have
  \begin{equation}
 \boldsymbol{\sigma}_\textsc{ab}=\begin{pmatrix}
 \boldsymbol{\sigma}_\textsc{a} & \boldsymbol{\gamma_{\textsc{ab}}}\\
\boldsymbol{\gamma_{\textsc{ab}}}^\intercal&  \boldsymbol{\sigma}_\textsc{b}
 \end{pmatrix},
 \label{multisig}
 \end{equation}
 where $\boldsymbol{\gamma_{\textsc{ab}}}$ is a matrix of correlators between the observables of the two systems. Tracing out one of the systems is simply extracting the diagonal covariance matrices $ \boldsymbol{\sigma}_\textsc{a}$ and $ \boldsymbol{\sigma}_\textsc{b}$ in the joint covariance matrix: no need to perform traces of any kind.
 
 \subsection{Time evolution in the symplectic formulation of quantum mechanics}

With our algebraic structure and Gaussian states rewritten in terms of phase space objects, we now turn our attention toward the unitary transformations which preserve the Gaussianity of the states they act on. Such transformations are called Gaussian unitary transformations. 

Within all Gaussian transformations we will focus on unitary Gaussian transformations. A unitary transformation preserves the commutation relations \eqref{step1}, and therefore they preserve the symplectic form, which in turn means they preserve the volume of phase space as we discussed in previous sections. This means that unitary  transformations correspond to symplectic transformations in phase space (that preserve the `volume/area' under the Wigner function).

Gaussian unitary transformations map Gaussian Wigner functions to Gaussian Wigner functions preserving the area under the function, which means that they can be phase space rotations, squeezing and displacement operators. Both squeezing and rotations are generated by quadratic Hamiltonians, and displacements by linear ones. We can write the most general Hamiltonian that generates a Gaussian unitary transformation as
\be
\hat H =\frac12 \hat\Theta^{\alpha}\hat\Theta^\beta F_{\alpha\beta}+\alpha_\nu\hat\Theta^\nu,
\label{Hamo}
\ee
 where $\boldsymbol{F}$ is a Hermitian matrix, that is $F_{\alpha\beta}=F_{\beta\alpha}^*$, and the quadrature operators here $ \q^\alpha\propto \hat\Xi^\alpha$ such that $ [\q^\alpha,\q^\beta]=\ii\tilde\Omega^{\alpha\beta}\openone$ are the ones defined in \eqref{covariancebueno}.
 
Notice that for the most general Hamiltonian that is a quadratic form of the operators we can write Heisenberg equations for the quadrature operators:
  \begin{align}
 \nonumber \frac{\mathrm{d}}{\mathrm{d} t} \q^\mu&=\mathrm{i}[\hat{H}, \q^\mu]=\frac\ii2F_{\alpha\beta}[\hat\Theta^{\alpha}\hat\Theta^\beta,\q^\mu]+\ii\alpha_\nu [\q^\nu,\q^\mu]\\
 \nonumber &=\frac\ii2F_{\alpha\beta}\left(\q^\alpha[\q^\beta,\q^\mu]+[\q^\alpha,\q^\mu]\q^\beta\right)+\ii\alpha_\nu[\q^\nu,\q^\mu]\\
 \nonumber &=-\frac12F_{\alpha\beta}\left(\tilde\Omega^{\beta\mu}\q^\alpha+\tilde\Omega^{\alpha\mu}\q^\beta\right)-\tilde\Omega^{\nu\mu}\alpha_\nu\openone\\
 \nonumber &=\frac12\left(\tilde\Omega^{\mu\beta}F_{\alpha\beta}\q^\alpha+\tilde\Omega^{\mu\alpha}F_{\alpha\beta}\q^\beta\right) +\tilde\Omega^{\mu\nu}\alpha_\nu\openone\\
  \nonumber &=\frac12\left(\tilde\Omega^{\mu\beta}F_{\beta\alpha}^*\q^\alpha+\tilde\Omega^{\mu\alpha}F_{\alpha\beta}\q^\beta\right) +\tilde\Omega^{\mu\nu}\alpha_\nu\openone\\
 & =\tilde\Omega^{\mu\nu}\left(\bar F_{\nu\alpha}\q^\alpha +\alpha_\nu\openone\right),\label{pooooos}
  \end{align}
where $2\bar F_{\alpha\beta}\coloneqq F_{\alpha\beta}+(F^{\text t})_{\alpha\beta}=F_{\alpha\beta}+F^*_{\alpha\beta}$ (in matrix form, $\bar{\boldsymbol{F}}=(\boldsymbol{F}+\boldsymbol{F}^\text{t})/2$) is a symmetric matrix.  We could write \eqref{pooooos} in matrix form as 
  \be
  \frac{\mathrm{d}}{\mathrm{d} t} \boldsymbol{\q}=\boldsymbol{\Omega}^{-1}\left(\bar{\boldsymbol{F}}\boldsymbol{\q}+\boldsymbol{\alpha}\openone\right).
  \label{Heisen2}
  \ee
  Note that in the Gaussian quantum mechanics literature, the symplectic form is often defined with opposite signature, and therefore they would get the matrix $\boldsymbol{\Omega}$ in the equation above instead of the inverse. Technically, that is not correct, as vectors in phase space are contravariant objects and therefore the commutator of two vectors is a twice contravariant tensor, like the inverse of the symplectic form is. But it is important to have this in mind when reading the literature on the topic since this is the most common choice out there. We will keep the more rigorous notation in this course.
  
Let us recap. We have of course the usual quantum mechanical evolution, which starts from the Heisenberg equation
\be
\frac{\mathrm{d}}{\mathrm{d} t} \boldsymbol{\q}=\mathrm{i}[\hat{H}, \boldsymbol{\q}]
\ee
 whose solution is the unitary evolution given by
 \be
 \boldsymbol{\q}(t)=\hat U^\dagger(t)\boldsymbol{\q}(0)\hat U(t),
 \ee
where $\hat U(t)=\exp\big(-\frac{\ii}{\hbar} \hat H t\big)$ provided the Hamiltonian is time-independent with itself at all times (if not we would have to substitute the exponential by the time-ordered exponential of the integral of the Hamiltonian over time). However, equivalently, we have found completely different equations [Eq. \eqref{Heisen2}] for $ \boldsymbol{\q}(t)$  that tell us that
  \be
  \frac{\mathrm{d}}{\mathrm{d} t} \boldsymbol{\q}=\boldsymbol{\Omega}^{-1}\left(\bar{\boldsymbol{F}}\boldsymbol{\q}+\boldsymbol{\alpha}\openone\right),
  \label{Heisen3}
  \ee
  and therefore
 \be
 \boldsymbol{\q}(t)=\hat U^\dagger(t)\boldsymbol{\q}(0)\hat U(t)=\boldsymbol{S}(t) \boldsymbol{\q}(0) + \boldsymbol{d}(t)\openone,
 \ee
 where\footnote{This is a result that we will not dwell much on as we will quickly move to covariance matrix and vector of means, so I am not including the full steps of solving \eqref{Heisen3}, but solving this differential equation is elementary and a proof that this is the solution to \eqref{Heisen3} can be seen in Appendix A of Phys. Rev. A 97, 052120 (2018). } 
 \be
\boldsymbol{S}(t) =\exp (\boldsymbol{\Omega}^{-1} \bar{\boldsymbol{F}} t), \qquad \boldsymbol{d}(t) =\frac{\exp (\boldsymbol{\Omega}^{-1} \bar{\boldsymbol{F}} t)-\openone_{2 N}}{\boldsymbol{\Omega}^{-1} \bar{\boldsymbol{F}}} \boldsymbol{\Omega}^{-1} \boldsymbol{\alpha},
\label{minisol}
 \ee
again taking into account that if the matrix of Hamiltonian coefficients $\bF$ is time-dependent we need to write the time-ordered exponential $\boldsymbol{S}(t) =\mathcal{T}\exp (\int \d t\, \boldsymbol{\Omega}^{-1} \bar{\boldsymbol{F}}(t) )$.

 Importantly, note that \(\hat{U}(t)\) is a linear map on the system's Hilbert space and acts on \({\boldsymbol{\q}}\) componentwise. On the other hand, \(\boldsymbol{S}(t)\) is a linear map on the system's phase space and acts on \({\boldsymbol{\q}}\) as it would on a phase space vector, yielding linear combinations of its (operator-valued) components.
 
 Finally note that \(\boldsymbol{\Omega}^{-1} \boldsymbol{F}\) does not need to be invertible to make sense of \eqref{minisol} if one understands it in terms of the definition
$$
\frac{\exp (\boldsymbol{M} t)-\openone}{\boldsymbol{M}}:=\sum_{m=0}^{\infty} \frac{t^{m+1}}{(m+1) !} \boldsymbol{M}^{m},
$$
for a general square matrix \(\boldsymbol{M}\).

For now we have seen that the time evolution of the quadrature operators is just a linear affine  transformation in phase space, but it is not just any symplectic transformation: Unitary evolution preserves the commutation relations, so the linear affine transformation has to preserve them as well, which implies that $\boldsymbol{S}$  has to preserve the symplectic form, i.e.,
\be
\bS \bO{\bS}^\intercal=\bO\Rightarrow \tensor{S}{^\alpha_\beta}\Omega^{\beta\gamma}\tensor{S}{_\gamma^\delta}=\Omega^{\alpha\delta}.
\ee
Therefore, $\bS$ is a symplectic transformation.

\subsection{Evolution of Gaussian states}

Now that we have seen how a simple linear affine symplectic transformation on phase space of dimension $2N$ can do the job of an infinite dimensional unitary transformation on a Hilbert space (and this is true for all quantum states), we can focus on the `classical' Gaussian states, and for them, there is no need of invoking Hilbert spaces or operators whatsoever.

 Notice that we can use \eqref{Heisen2} to compute the evolution of the covariance matrix $\boldsymbol{\sigma}$ and the vector of first moments $\boldsymbol{\xi_0}$. From \eqref{covariancebueno}:
  \be
\dot{\xi}_0^\mu=\left\langle\frac{\mathrm{d}}{\mathrm{d} t}\q^\mu\right\rangle_{\hat \rho}=\tilde\Omega^{\mu\nu}\left(\bar F_{\nu\alpha}\xi_0^\alpha +\alpha_\nu\right),
\ee
or in matrix form, 
\be\dot{\boldsymbol{\xi}}_0=\boldsymbol{\Omega}^{-1}\!\left(\bar{\boldsymbol{F}}\boldsymbol{\xi}_0+\boldsymbol{\alpha}\right). 
\label{mean}
\ee
Analogously, the time derivative of the covariance matrix takes the form
\begin{align}
\nonumber\dot \sigma^{\mu\nu}\coloneqq&\left\langle \frac{\mathrm{d}\q^\mu}{\mathrm{d} t}\q^{\nu}+\q^{\mu}\frac{\mathrm{d}\q^\nu}{\mathrm{d} t}+\frac{\mathrm{d}\q^\nu}{\mathrm{d} t}\q^{\mu}+\q^{\nu}\frac{\mathrm{d}\q^\mu}{\mathrm{d} t}\right\rangle_{\hat \rho}- 2\left\langle\frac{\mathrm{d}\q^\mu}{\mathrm{d} t}\right\rangle_{\hat \rho}\left\langle\q^\nu\right\rangle_{\hat \rho}-2\left\langle\q^\mu\right\rangle\left\langle\frac{\mathrm{d}\q^\nu}{\mathrm{d} t}\right\rangle_{\hat \rho}\\
&= \left\langle \hat A^{\mu\nu}+ \hat B^{\mu\nu}\right\rangle-2\left(\dot{\xi}_0^\mu {\xi}_0^\nu+\dot{\xi}_0^\nu {\xi}_0^\mu\right),
\label{porcohu}
\end{align}
where the partial terms $\hat A^{\mu\nu}$ and $\hat B^{\mu\nu}$ are
\begin{align}
\nonumber \hat A^{\mu\nu}&\coloneqq\frac{\mathrm{d}\q^\mu}{\mathrm{d} t}\q^{\nu}+\q^{\nu}\frac{\mathrm{d}\q^\mu}{\mathrm{d} t}\\
\nonumber&=\tilde\Omega^{\mu\sigma}\left(\bar F_{\sigma\alpha}\q^\alpha +\alpha_\sigma\openone\right)\q^{\nu}+\q^{\nu}\tilde\Omega^{\mu\sigma}\left(\bar F_{\sigma\alpha}\q^\alpha +\alpha_\sigma\openone\right)\\
&=\tilde\Omega^{\mu\sigma}\bar F_{\sigma\alpha}\left(\q^\alpha\q^\nu+\q^\nu\q^\alpha\right) +2\tilde\Omega^{\mu\sigma}\alpha_\sigma \q^\nu ,
\end{align}
\begin{align}
\nonumber \hat B^{\mu\nu}&\coloneqq\frac{\mathrm{d}\q^\nu}{\mathrm{d} t}\q^{\mu}+\q^{\mu}\frac{\mathrm{d}\q^\nu}{\mathrm{d} t}\\
\nonumber&=\tilde\Omega^{\nu\sigma}\left(\bar F_{\sigma\alpha}\q^\alpha +\alpha_\sigma\openone\right)\q^{\mu}+\q^{\mu}\tilde\Omega^{\nu\sigma}\left(\bar F_{\sigma\alpha}\q^\alpha +\alpha_\sigma\openone\right)\\
&=\tilde\Omega^{\nu\sigma}\bar F_{\sigma\alpha}\left(\q^\alpha\q^\mu+\q^\mu\q^\alpha\right) +2\Omega^{\nu\sigma}\alpha_\sigma \q^\mu.
\end{align}
Therefore, 
\begin{align}
\left\langle \hat A^{\mu\nu}+ \hat B^{\mu\nu}\right\rangle=\bar \Omega^{\mu\sigma}\bar F_{\sigma\alpha}\left\langle\q^\alpha\q^\nu+\q^\nu\q^\alpha\right\rangle  +\Omega^{\nu\sigma}\bar F_{\sigma\alpha}\left\langle\q^\alpha\q^\mu+\q^\mu\q^\alpha\right\rangle +2\alpha_\sigma\left( \Omega^{\nu\sigma} \xi_0^\mu+\Omega^{\mu\sigma} \xi_0^\nu\right).
\end{align}
Evaluating now the second term of \eqref{porcohu}, we get
\begin{align}
\nonumber 2\left(\dot{\xi}_0^\mu {\xi}_0^\nu+\dot{\xi}_0^\nu {\xi}_0^\mu\right)&=2\tilde\Omega^{\mu\sigma}\left(\bar F_{\sigma\alpha}\xi_0^\alpha +\alpha_\sigma\right)\xi_0^\nu+  2\tilde\Omega^{\nu\sigma}\left(\bar F_{\sigma\alpha}\xi_0^\alpha +\alpha_\sigma\right)\xi_0^\mu\\
&=2\bar F_{\sigma\alpha}\tilde\Omega^{\mu\sigma} \xi_0^\alpha\xi_0^\nu+2\bar F_{\sigma\alpha} \tilde\Omega^{\nu\sigma}\xi_0^\alpha\xi_0^\mu+2\alpha_\sigma\left(\Omega^{\mu\sigma}\xi_0^\nu+\Omega^{\nu\sigma}\xi_0^\mu\right),
\end{align}
and therefore 
\begin{align}
\nonumber\dot \sigma^{\mu\nu}&=\left\langle \hat A^{\mu\nu}+ \hat B^{\mu\nu}\right\rangle-2\left(\dot{\xi}_0^\mu {\xi}_0^\nu+\dot{\xi}_0^\nu {\xi}_0^\mu\right)\\
\nonumber &= \tilde\Omega^{\mu\sigma}\bar F_{\sigma\alpha}\left(\left\langle\q^\alpha\q^\nu+\q^\nu\q^\alpha\right\rangle-2\xi^\alpha_0\xi_0^\nu\right)  +\tilde\Omega^{\nu\sigma}\bar F_{\sigma\alpha}\left(\left\langle\q^\alpha\q^\mu+\q^\mu\q^\alpha\right\rangle-2\xi_0^\alpha\xi_0^\mu\right)\\ 
&= \tilde\Omega^{\mu\sigma}\bar F_{\sigma\alpha}\sigma^{\alpha\nu}  +\tilde\Omega^{\nu\sigma}\bar F_{\sigma\alpha}\nonumber \sigma^{\alpha\mu}= \tilde\Omega^{\mu\sigma}\bar F_{\sigma\alpha}\sigma^{\alpha\nu}  +\sigma^{\mu\alpha}\bar F_{\alpha\sigma}\tilde\Omega^{\nu\sigma}\\
&=\tensor{(\boldsymbol{\Omega}^{-1} \bar{\boldsymbol{F}})}{^\mu_\alpha}\sigma^{\alpha\nu}+\sigma^{\mu\alpha}\tensor{\left[(\boldsymbol{\Omega}^{-1} \bar{\boldsymbol{F}})^\intercal\right]}{_\alpha^\nu},
\end{align}
or, in full matrix form,
\be
\dot{\boldsymbol{\sigma}}=(\boldsymbol{\Omega}^{-1} \bar{\boldsymbol{F}})\boldsymbol{\sigma}+\boldsymbol{\sigma}(\boldsymbol{\Omega}^{-1} \bar{\boldsymbol{F}})^\intercal.
\label{covariant}
\ee

Let us write together the equations of motion for the covariance matrix and for the vector of first moments:
\begin{align}
\dot{\boldsymbol{\xi}}_0&=\boldsymbol{\Omega}^{-1}\!\left(\bar{\boldsymbol{F}}\boldsymbol{\xi}_0+\boldsymbol{\alpha}\right),\\ 
\dot{\boldsymbol{\sigma}}&=(\boldsymbol{\Omega}^{-1} \bar{\boldsymbol{F}})\boldsymbol{\sigma}+\boldsymbol{\sigma}(\boldsymbol{\Omega}^{-1} \bar{\boldsymbol{F}})^\intercal.
\label{motion}
\end{align}
These two equations can readily be integrated to yield
\begin{equation}
\boldsymbol{\xi}_0(t)=\bS(t) \boldsymbol{\xi}_0(0)+\boldsymbol{d},\qquad \boldsymbol{\sigma}(t)=\bS(t)\boldsymbol\sigma(0)\bS^\intercal(t),
\end{equation}
where we recall that
\be
\bS(t)=\mathcal{T}\exp\left(\int_{t_0}^t\d t\,\bO\bar{\bF}\right).
\ee
As a convenient tool, it is useful to notice that a Hamiltonian of the form \eqref{Hamo} can always be written as a quadratic form of the creation and annihilation operators, and it is actually very common to do so (think of the squeezing operators, for instance). Consider the $n$-dimensional vectors of creation and annihilation operators respectively:
\begin{align}
\hat{\mathbf{a}} \coloneqq\big(\hat{a}_{1}, \ldots \hat{a}_{N}\big)^{\intercal}, \qquad \hat{\mathbf{a}}^{\dagger} \coloneqq\big(\hat{a}_{1}^{\dagger}, \ldots \hat{a}_{N}^{\dagger}\big)^{\intercal}.
\label{annihil}
\end{align}
Then the quadratic part of the Hamiltonian \eqref{Hamo} can be rewritten\footnote{Note that I didn't include the linear term here. In fact a linear term can be also put in this form except for multiples of the identity. However let us cosnider for now oepratons that do not translate the Gaussian states.} as
\be
\hat{H}=\big(\hat{\mathbf{a}}^{\dagger}\big)^{\intercal} \mathbf{W}\, \hat{\mathbf{a}}+\big(\hat{\mathbf{a}}^{\dagger}\big)^{\intercal} \mathbf{G}\, \hat{\mathbf{a}}^{\dagger}+\hat{\mathbf{a}}^{\intercal} \mathbf{G}^{*} \hat{\mathbf{a}},
\label{annihilHa}
\ee
where $\mathbf{G}^{*}$ denotes here the Hermitian conjugate matrix (Conjugate transpose) of $ \mathbf{G}$. Knowing the matrices $ \mathbf{G}$ and $\mathbf{W}$ one can readily obtain the matrix $\bF$. By equating \eqref{Hamo} and \eqref{annihilHa} we get that 
\begin{equation}
\mathbf{F}=\left(\begin{array}{cc}
\mathbf{A} & \mathbf{X} \\
\mathbf{X}^{*} & \mathbf{B}
\end{array}\right),
\end{equation}
where
\begin{align}
\mathbf{A}=\mathbf{W}+\mathbf{G}+\mathbf{G}^{*}, \qquad\mathbf{B}=\mathbf{W}-\mathbf{G}-\mathbf{G}^{*},\qquad\mathbf{X}=\ii\left(\mathbf{W}-\mathbf{G}+\mathbf{G}^{*}\right).
\end{align}

Notice that our choice to lump all creators and annihilators together in Eq.~\eqref{annihil} means that for this calculation we are using the ordering $\hat{\boldsymbol\Xi}=(\hat q_1...,\hat q_n,\hat p_1,\dots,\hat p_n)$ rather than $\hat{\boldsymbol\Xi}=(\hat q_1,\hat p_1,\dots,\hat q_n,\hat p_n)$ employed in the rest of the notes. This is so because it is the easiest choice to obtain simple expressions for the matrix $\boldsymbol{F}$ for the problem at hand. For a single mode, the two conventions are obviously equivalent, but the moment we have multipartite system it is good to remember which convention we use. If we compute in this convention we will need to (trivially) transform to the other one to recover the graphical multipartite block-matrix structure of Eq.~\eqref{multisig}. We will see that with an example later on with the two-mode squeezing. 

 After developing all this machinery, we see that for Gaussian states we can completely forget about the infinite-dimensional Hilbert space representation of the observables and instead work with the classical phase-space distributions based objects $\boldsymbol{\xi}_0$ and $\boldsymbol{\sigma}$. There is nothing quantum about this and Lagrange would happily declare this a classical theory of mechanics (of states distributed Gaussianly on phase space).
 
 \subsubsection{Revisiting squeezing and the covariance matrix of the squeezed vacuum}

 With what we have learned, calculating the covariance matrix for the single mode vacuum becomes pretty easy. First, let's consider the single mode squeezing operator that is given by equation \eqref{singsq}:
 \be
\hat S(re^{\ii\theta})=\exp \left[\frac{r}{2}  \left(e^{-\ii\theta} \hat a^{2}- e^{\ii\theta}\hat a^{\dagger 2}\right)\right]=e^{-\ii \hat H},
 \ee
where $\hat H$ is the generator of the squeezing. Therefore the squeezing can be seen as an unitary generated by a quadratic Hamiltonian of the form \eqref{annihilHa} with
 \be
 W=0,\qquad G=-\ii \frac{r}{2} e^{\ii\theta},
 \ee
 which means that the matrix $\bF$ is given by
 \be
\bF=r\begin{pmatrix}
 \sin\theta & -\cos\theta \\
-\cos\theta & -\sin\theta
\end{pmatrix}\Rightarrow \bar{\bF}=\frac{\bF+\bF^\intercal}{2}=r\begin{pmatrix}
 \sin\theta & -\cos\theta \\
-\cos\theta & -\sin\theta
\end{pmatrix}.
 \ee
 Therefore, 
 \be
 \bO\bar{\bF}=r\begin{pmatrix}
    0&1\\
    -1&0
    \end{pmatrix}\begin{pmatrix}
 \sin\theta & -\cos\theta \\
-\cos\theta & -\sin\theta
\end{pmatrix}=r\begin{pmatrix}
-\cos\theta & - \sin\theta  \\
-\sin\theta & \cos\theta 
\end{pmatrix},
 \ee
 and the symplectic matrix implementing this squeezing in phase space is
  \be
\bS=\exp(\bO\bar{\bF})=\exp\left[r\begin{pmatrix}
-\cos\theta & - \sin\theta  \\
-\sin\theta & \cos\theta 
\end{pmatrix}\right]=\begin{pmatrix}
 \cosh r-\cos \theta  \sinh r & -\sin \theta  \sinh r \\
 -\sin \theta  \sinh r & \cos \theta  \sinh r+\cosh r \\
\end{pmatrix}.
 \ee
And this is the symplectic operation that applies squeezing to any state. Applied to the ground state,  recalling from previous sections that the covariance matrix for the ground state of a single harmonic oscillator is the identity, we get
 \be
\boldsymbol{\sigma}_\text{sq}= \bS\sigma_0\bS^\intercal=\bS\begin{pmatrix}1 &0\\0&1\end{pmatrix}\bS^\intercal= \begin{pmatrix}
\cosh 2r-\cos\theta\sinh 2r & -\sin\theta\sinh2r\\
-\sin\theta\sinh2r&  \cosh 2r+\cos\theta\sinh 2r
 \end{pmatrix},
\ee
 which is the covariance matrix of the squeezed vacuum. This is much easier to compute than using the quadrature operators. Notice that no quantum mechanics was used to obtain this. 
 
 \subsubsection{Symplectic representation of the two mode squeezing operator and two-mode squeezed vacuum}

Let us now consider a system of two harmonic oscillators and a two-mode squeezing operator that we can write as
\bel{sqor}
\hat S(r e^{\ii\theta})=\exp\left[\frac{r}{2}\left(e^{-\ii\theta} \hat a_1 \hat a_2-e^{\ii\theta}\hat a_1^{\dagger} \hat a_2^{\dagger}\right)\right]=e^{-\ii \hat H}.
\ee
The squeezing can be seen as an unitary generated by a quadratic Hamiltonian of the form \eqref{annihilHa} with
 \begin{equation}
 \boldsymbol{W}=\boldsymbol{0},\qquad
\mathbf{G}=-\ii\frac{re^{\ii\theta}}{4}\begin{pmatrix}
0 & 1 \\
1 & 0
\end{pmatrix},
\end{equation}
which yields
 \begin{equation}
\mathbf{A}=\frac{r \sin\theta}{2}\begin{pmatrix}
0 & 1 \\
1 & 0
\end{pmatrix}=-\mathbf{B},\qquad \mathbf{X}=-\frac{r \cos\theta}{2}\begin{pmatrix}
0 & 1 \\
1 & 0
\end{pmatrix}.
\end{equation}
This in turn means that
\begin{equation}
 \boldsymbol{F}=\frac{r}{2}\begin{pmatrix}
0 & \sin\theta & 0 & -\cos\theta \\
\sin\theta & 0 &-\cos\theta & 0 \\
0 & -\cos\theta  & 0 & -\sin\theta\\
 -\cos\theta & 0 &-\sin\theta & 0 \end{pmatrix}\Rightarrow \bar{\bF}=\frac{\bF+\bF^\intercal}{2}=\frac{r}{2}\begin{pmatrix}
0 & \sin\theta & 0 & -\cos\theta \\
\sin\theta & 0 &-\cos\theta & 0 \\
0 & -\cos\theta  & 0 & -\sin\theta\\
 -\cos\theta & 0 &-\sin\theta & 0 \end{pmatrix}.
\end{equation}
Therefore, 
 \begin{align}
 \bO\bar{\bF}&=\frac{r}{2}\begin{pmatrix}
0 & 0 & 1 & 0 \\
0 & 0 &0 & 1 \\
-1 & 0  & 0 & 0\\
 0 & -1 &0 & 0 \end{pmatrix}\begin{pmatrix}
0 & \sin\theta & 0 & -\cos\theta \\
\sin\theta & 0 &-\cos\theta & 0 \\
0 & -\cos\theta  & 0 & -\sin\theta\\
 -\cos\theta & 0 &-\sin\theta & 0 \end{pmatrix}=\frac{r}{2}\begin{pmatrix}
0 & -\cos\theta & 0 &  -\sin\theta  \\
 -\cos\theta &0 & -\sin\theta& 0 \\
0  &-\sin\theta& 0 &  \cos\theta \\
 -\sin\theta & 0  & \cos\theta & 0 \end{pmatrix}
 \end{align}
 and the symplectic matrix implementing this squeezing in phase space is
  \be
\bS=\exp(\bO\bar{\bF})=\begin{pmatrix}
 \cosh \frac{r}{2} & -\cos \theta  \sinh \frac{r}{2} & 0 & -\sin \theta  \sinh \frac{r}{2} \\
 -\cos \theta  \sinh \frac{r}{2}&  \cosh \frac{r}{2}& -\sin \theta  \sinh \frac{r}{2} &0 \\
0  & -\sin \theta  \sinh \frac{r}{2} &  \cosh \frac{r}{2}& \cos \theta  \sinh \frac{r}{2} \\
-\sin \theta  \sinh \frac{r}{2}& 0 & \cos \theta  \sinh \frac{r}{2} &\cosh \frac{r}{2}\\
\end{pmatrix}.\ee
This is the symplectic operation that applies squeezing to any two-mode state. Let us now apply it to the ground state of the two oscillators. The covariance matrix in that case is
\be
 \boldsymbol{\sigma}_\textsc{ab}=\begin{pmatrix} 1 & 0\\ 0& 1\end{pmatrix}\oplus\begin{pmatrix} 1 & 0\\ 0& 1\end{pmatrix}=\begin{pmatrix} 1 & 0 & 0 & 0\\
  0& 1 & 0& 0\\
  0& 0 & 1& 0\\
    0& 0 & 0& 1\\
  \end{pmatrix},  
\ee
and thus the two mode squeezed vacuum covariance matrix is
 \begin{align}
\boldsymbol{\sigma}_\text{sq}&= \bS\sigma_0\bS^\intercal=\bS\begin{pmatrix} 1 & 0 & 0 & 0\\
  0& 1 & 0& 0\\
  0& 0 & 1& 0\\
    0& 0 & 0& 1\\
  \end{pmatrix}\bS^\intercal= 
 \begin{pmatrix}
 \cosh r & -\cos \theta  \sinh r & 0 & -\sin \theta  \sinh r \\
 -\cos \theta  \sinh r&  \cosh r& -\sin \theta  \sinh r&0 \\
0  & -\sin \theta  \sinh r &  \cosh r& \cos \theta  \sinh r \\
-\sin \theta  \sinh r& 0 & \cos \theta  \sinh r &\cosh r\\
\end{pmatrix}.
\label{covtwomodsqv}
\end{align}
Notice how the covariance matrix becomes the identity when $r=0$ (the vacuum is not squeezed). Additionally, it is trivial to trace out one of the systems and check indeed that the partial states are thermal. To do so is also convenient to switch from the ordering of quadratures  $\hat{\boldsymbol\Xi}=(\hat q_1...,\hat q_n,\hat p_1,\dots,\hat p_n)\to\hat{\boldsymbol\Xi}=(\hat q_1,\hat p_1,\dots,\hat q_n,\hat p_n)$ so that the covariance matrix goes from the convention  $\hat{\boldsymbol\Xi}=(\hat q_1...,\hat q_n,\hat p_1,\dots,\hat p_n)$ employed in this calculation,
\be
\boldsymbol\sigma\!=\!\begin{pmatrix}
2\langle \hat q_1^2\rangle -2\langle \hat q_1\rangle^2  &  2\langle \hat q_1 \hat q_2  \rangle-2\langle \hat q_1 \rangle\langle \hat q_2 \rangle &  \langle \hat q_1 \hat p_1 +\hat p_1 \hat q_1 \rangle-2\langle \hat q_1 \rangle\langle \hat p_1 \rangle &  2\langle \hat q_1 \hat p_2  \rangle-2\langle \hat q_1 \rangle\langle \hat p_2 \rangle \\
 2\langle \hat q_2 \hat q_1  \rangle-2\langle \hat q_2 \rangle\langle \hat q_1 \rangle&  \langle \hat q_2^2\rangle -2\langle \hat q_2\rangle^2& 2\langle \hat q_2 \hat p_1  \rangle-2\langle \hat q_2 \rangle\langle \hat p_1 \rangle& \langle \hat q_2 \hat p_2 +\hat p_2 \hat q_2 \rangle-2\langle \hat q_2 \rangle\langle \hat p_2 \rangle \\
 \langle \hat p_1 \hat q_1 +\hat q_1 \hat p_1 \rangle-2\langle \hat p_1 \rangle\langle \hat q_1 \rangle  &  2\langle \hat p_1 \hat q_2  \rangle-2\langle \hat p_1 \rangle\langle \hat q_2 \rangle  & \langle \hat p_1^2\rangle -2\langle \hat p_1\rangle^2& 2\langle \hat p_1 \hat p_2  \rangle-2\langle \hat p_1 \rangle\langle \hat p_2 \rangle\\
2\langle \hat p_2 \hat q_1 \rangle-2\langle \hat p_2 \rangle\langle \hat q_1 \rangle&  \langle \hat p_2 \hat q_2 +\hat q_2 \hat p_2 \rangle-2\langle \hat p_2 \rangle\langle \hat q_2 \rangle &2\langle \hat p_2 \hat p_1  \rangle-2\langle \hat p_2 \rangle\langle \hat p_1 \rangle &\langle \hat p_2^2\rangle -2\langle \hat p_2\rangle^2\\
\end{pmatrix},
\ee
 to  the ordering $\hat{\boldsymbol\Xi}=(\hat q_1,\hat p_1,\dots,\hat q_n,\hat p_n)$ which is much more convenient to work with multipartite systems,
 \be
\boldsymbol\sigma\!=\!\begin{pmatrix}
2\langle \hat q_1^2\rangle -2\langle \hat q_1\rangle^2  &  \langle \hat q_1 \hat p_1 +\hat p_1 \hat q_1 \rangle-2\langle \hat q_1 \rangle\langle \hat p_1 \rangle  & 2\langle \hat q_1 \hat q_2  \rangle-2\langle \hat q_1 \rangle\langle \hat q_2 \rangle  &  2\langle \hat q_1 \hat p_2  \rangle-2\langle \hat q_1 \rangle\langle \hat p_2 \rangle \\
 \langle \hat p_1 \hat q_1 +\hat q_1 \hat p_1 \rangle-2\langle \hat p_1 \rangle\langle \hat q_1 \rangle &  \langle \hat p_1^2\rangle -2\langle \hat p_1\rangle^2&2\langle \hat p_1 \hat q_2  \rangle-2\langle \hat p_1 \rangle\langle \hat q_2 \rangle & 2\langle \hat p_1 \hat p_2  \rangle-2\langle \hat p_1 \rangle\langle \hat p_2 \rangle\\
2\langle \hat q_2 \hat q_1  \rangle-2\langle \hat q_2 \rangle\langle \hat q_1 \rangle  & 2\langle \hat q_2 \hat p_1  \rangle-2\langle \hat q_2 \rangle\langle \hat p_1 \rangle   & \langle \hat p_1^2\rangle -2\langle \hat p_1\rangle^2&  \langle \hat q_2 \hat p_2 +\hat p_2 \hat q_2 \rangle-2\langle \hat q_2 \rangle\langle \hat p_2 \rangle\\
2\langle \hat p_2 \hat q_1 \rangle-2\langle \hat p_2 \rangle\langle \hat q_1 \rangle&    2\langle \hat p_2 \hat p_1  \rangle-2\langle \hat p_2 \rangle\langle \hat p_1 \rangle&  \langle \hat p_2 \hat q_2 +\hat q_2 \hat p_2 \rangle-2\langle \hat p_2 \rangle\langle \hat q_2 \rangle &\langle \hat p_2^2\rangle -2\langle \hat p_2\rangle^2,\\
\end{pmatrix},
\ee
so that now we have,  as in Eq.~\eqref{multisig}, that
\begin{equation}
 \boldsymbol{\sigma}_{12}=\begin{pmatrix}
 \boldsymbol{\sigma}_1 & \boldsymbol{\gamma}_{12}\\
\boldsymbol{\gamma}_{12}^\intercal&   \boldsymbol{\sigma}_2
 \end{pmatrix}
 \end{equation}
 where $\boldsymbol{\gamma}_{12}$ is a matrix of correlators between the observables of the two systems. Again, tracing out one of the systems reduces to extracting the covariance matrices $\boldsymbol{\sigma}_1$ and $ \boldsymbol{\sigma}_2$ from the diagonal blocks of the joint covariance matrix: no need to perform traces of any kind. 
 
 For our particular case, the entries of the two mode squeezed vacuum covariance matrix are reshuffled from \eqref{covtwomodsqv}  [with he quadrature ordering $\hat{\boldsymbol\Xi}=(\hat q_1...,\hat q_n,\hat p_1,\dots,\hat p_n)$]   into 
 \begin{align}
\boldsymbol{\sigma}_\text{sq}&= 
 \begin{pmatrix}
 \cosh r & 0 &  -\cos \theta  \sinh r & -\sin \theta  \sinh r \\
 0&  \cosh r& -\sin \theta  \sinh r&\cos \theta  \sinh r \\
-\cos \theta  \sinh r & -\sin \theta  \sinh r &  \cosh r& 0 \\
-\sin \theta  \sinh r&  \cos \theta  \sinh r & 0 &\cosh r\\
\end{pmatrix}.
\label{covtwomodsqv}
\end{align}
with the quadrature ordering  $\hat{\boldsymbol\Xi}=(\hat q_1,\hat p_1,\dots,\hat q_n,\hat p_n)$. It is now obvious that the two partial covariance matrices are
\be
\boldsymbol{\sigma}_1=\boldsymbol{\sigma}_2  =\begin{pmatrix}
\cosh r & 0\\
0& \cosh r 
 \end{pmatrix}=\cosh r \openone,
\ee
which is indeed the covariance matrix of a thermal state that we saw in Eq.~\eqref{thermalityyy} with $\nu=\cosh r$.

\section{Entropy and Entanglement of Gaussian states}

\subsection{Symplectic diagonalization of the covariance matrix}

Williamson's theorem says that any symmetric positive definite real matrix of even dimensions can be diagonalized by a symplectic transformation:
\be
\boldsymbol{\sigma}_\textsc{d}=\bS_\textsc{d}\boldsymbol{\sigma}\bS^\intercal_\textsc{d},
\ee
where the  matrix $\boldsymbol{\sigma}_\textsc{d}$ is diagonal and of the form
\bel{williamsonf}
\boldsymbol{\sigma}_\textsc{d}=\bigoplus_{i=1}^{n}\begin{pmatrix}
\nu_i & 0\\
0 & \nu_i
\end{pmatrix}=
\begin{pmatrix}
\nu_1 & & & & & \\
&\nu_1 & & & &   &\\
& &\nu_2 &  & &  &\\
& &  &\nu_2 &  & \\
& & & &\ddots & &\\
& & & & &\nu_n & \\
& & & & & & \nu_n 
\end{pmatrix}.
\ee
Notice that a symplectic transformation defines a change of canonical coordinates, and therefore Williamson's theorem is saying that there always exist a set of canonical coordinates $\tilde\xi^\mu$ such that the symplectic matrix is the direct sum of harmonic oscillators in thermal states, one per mode in the system! This is equivalent to finding a normal mode decomposition. We call the set $\{\nu_1,\dots,\nu_n\}$ the \textit{symplectic spectrum} of the covariance matrix $\boldsymbol{\sigma}$, and we call \textit{symplectic eigenvalues} to every element of the symplectic spectrum.

It is important to note that the symplectic eigenvalues are not the eigenvalues of the covariance matrix. We can find a relationship between the eigenvalues of $\bSi$ and its symplectic eigenvalues. To do this let us first define a matrix $\boldsymbol{M}\coloneqq \bSi\bO$. Let us evaluate how $\boldsymbol{M}$ transforms under a symplectic transformation of the covariance matrix:
\be
\boldsymbol{M}'= \bS \bSi   \bS^{\intercal}\bO=\bS \bSi  \bS^{\intercal} (\bS^{\intercal})^{-1}\bO \bS^{-1}=\bS \bSi \bO \bS^{-1}=\bS \boldsymbol M \bS^{-1}.
\ee
Here we have used that $\bS\bOr\bS^{\intercal}=\bOr\Rightarrow (\bS^{\intercal})^{-1}\bO \bS^{-1}=\bO$. So we see that the symplectic transformation of the covariance matrix induces a similarity transformation on $\boldsymbol{M}$ that leaves the eigenvalues of $\boldsymbol{M}$ invariant. 

If we particularize now to the symplectic transformation that symplectically diagonalizes $\boldsymbol\sigma$ to the form \eqref{williamsonf}, then the explicit form of $\boldsymbol{M}'= \bSi_{\textsc{d}}\bO$ is
\be
\boldsymbol{M}'= \bigoplus_{i=1}^{n}\begin{pmatrix}
0& \nu_i\\
- \nu_i & 0
\end{pmatrix},
\ee
and therefore the eigenvalues of $M$ are $\{\pm\ii\nu_i\}$. We conclude that the symplectic eigenvalues of $\bS$ are the modulus of the eigenvalues of $\bSi\bO$. This shows as well that the symplectic eigenvalues are invariant under any symplectic transformation.

The symplectic eigenvalues, as we will see, contain all the information about any entanglement present in a Gaussian state. Finding the symplectic eigenvalues of a covariance matrix is a much easier endeavour than finding the eigenvalues of the corresponding density matrix.

\subsection{How to perform a symplectic diagonalization}\label{Symplectic diagonalization}

Before seeing how we evaluate the entanglement of Gaussian states, let us stop for a moment to see how to perform a general symplectic diagonalization\footnote{We follow the method described in Appendix B of \href{https://journals.aps.org/pra/abstract/10.1103/PhysRevA.79.052327}{S. Pirandola, A. Serafini, and S. Lloyd, Phys. Rev. A \textbf{79}, 052327 (2009)}.}. 

Given a real symmetric positive definite matrix $\bm{F}$ of even dimension $2n$, we want to find the symplectic transformation $\bm{\Sigma}$ that satisfies
\begin{equation}\label{symplectic diagonalized}
\bm{\Sigma} \bm{F} \bm{\Sigma}^\intercal= \bigoplus_{i=1}^{n}\begin{pmatrix}
\nu_i & 0\\
0 & \nu_i
\end{pmatrix}  \equiv \bm{\mathcal{{F}}},
\end{equation}
where $\{\nu_1,\hdots,\nu_n\}$ are the symplectic eigenvalues of $\bm{F}$, which can be calculated easily since they coincide with the (doubly degenerate) eigenvalues of $|\ii\bm{F}\bm{\Omega}^{-1}|$. In other words, $\{\pm\ii\nu_1,\hdots,\pm\ii\nu_n\}$ is the eigenspectrum of $\bm{F}\bm{\Omega}^{-1}$, which we know how to calculate with basic linear algebra techniques. 

The algorithm to perform the symplectic diagonalization goes as follows: we look for a symplectic transformation with the form
\begin{equation}\label{sixsix}
\bm{\Sigma}=\bm{\mathcal{F}}^{1/2} \bm{O} \bm{F}^{-1/2},
\end{equation}
where $\bm{O}$ is an orthogonal matrix. Notice that $\bm{\mathcal{F}}^{1/2}$ can be calculated straightforwardly, since $\bm{\mathcal{F}}$ is diagonal. To calculate $\bm{F}^{-1/2}$, we need to first diagonalize it (by similarity, in the usual sense), then perform the inverse and the square root, and then reverse the diagonalization. With a $\bm{\Sigma}$ of the form~\eqref{sixsix}, we get that
\begin{equation}
\bm{\Sigma} \bm{F} \bm{\Sigma}^{\intercal}=\bm{\mathcal{F}}^{1/2} \bm{O} \bm{F}^{-1/2} \bm{F} \bm{F}^{-1/2} \bm{O}^\intercal \bm{\mathcal{F}}^{1/2} =\bm{\mathcal{F}}^{1/2} \bm{O} \bm{O}^\intercal \bm{\mathcal{F}}^{1/2}=\bm{\mathcal{F}},
\end{equation}
where we used that $\bm{F}$ and $\bm{\mathcal{F}}$ are symmetric, and the orthogonality of $\bm{O}$. We have thus checked that $\bm{\Sigma}$ diagonalizes $\bm{F}$. We need to fulfill the condition that it is a symplectic transformation:
\begin{equation}
\bm{\Sigma}\bm{\Omega} \bm{\Sigma}^\intercal=\bm{\mathcal{F}}^{1/2} \bm{O} \bm{F}^{-1/2} \bm{\Omega} \bm{F}^{-1/2} \bm{O}^\intercal \bm{\mathcal{F}}^{1/2}=\bm{\Omega} \; \Leftrightarrow \; \bm{O} \bm{Y} \bm{O}^\intercal=\bm{Z},
\end{equation}
where we have defined 
\begin{equation}\label{aux matrices}
\bm{Y}=\bm{F}^{-1/2} \bm{\Omega} \bm{F}^{-1/2} \quad \textrm{and} \quad \bm{Z}=\bm{\mathcal{F}}^{-1/2} \bm{\Omega} \bm{\mathcal{F}}^{-1/2} .
\end{equation}
Thus, we need our orthogonal matrix $\bm{O}$ to transform $\bm{Y}$ into $\bm{Z}$ via congruence. This will determine which $\bm{O}$ we need to use. In order to find it, first notice that $\bm{Z}$'s form is very simple:
\begin{equation}
\bm{Z}= \bigoplus_{i=1}^{n} \begin{pmatrix}
0 & -\nu_i^{-1} \\
\nu_i^{-1} & 0 \\
\end{pmatrix}
\end{equation}
It is straightforward to check that with the unitary
\begin{equation}\label{eq: matrix K}
\bm{K}=\frac{1}{\sqrt{2}} \bigoplus_{i=1}^{n} \begin{pmatrix}
\ii & -\ii\\
1 & 1 \\
\end{pmatrix}
\end{equation}
we get
\begin{equation}
\bm{K}^{\dagger}\bm{Z}\bm{K}= \bigoplus_{i=1}^{n} \begin{pmatrix}
\ii \nu_i^{-1} & 0\\
0 & -\ii\nu_i^{-1}\\
\end{pmatrix}.
\end{equation}
Now, $\bm{Z}$ and $\bm{Y}$ have the same eigenspectrum\footnote{It can be seen from Eqs.~\eqref{symplectic diagonalized} and~\eqref{aux matrices} that $\bm{Y}$ and $\bm{Z}$ are related by a similarity transformation.}. Since $\bm{Y}$ is antisymmetric (and therefore normal), we can diagonalize it with a unitary matrix $\bm{U}$, such that
\begin{equation}
\bm{U}^\dagger \bm{Y} \bm{U}=\bigoplus_{i=1}^{n} \begin{pmatrix}
\ii \nu_i^{-1} & 0\\
0 & -\ii\nu_i^{-1}\\
\end{pmatrix}.
\end{equation}
Thus, 
\begin{equation}
\bm{K}\bm{U}^{\dagger} \bm{Y} \bm{U}\bm{K}^\dagger=\bm{Z}.
\end{equation}
Finally, it can be checked that taking $\bm{O}=\bm{K}\bm{U}^\dagger$, $\bm{O}$ is always a \textit{real} matrix, so that $\bm{O}^\dagger=\bm{O}^\intercal$. And that completes the diagonalization, since
\begin{equation}
\bm{\Sigma}=\bm{\mathcal{F}}^{1/2}\bm{K}\bm{U}^\dagger \bm{F}^{-1/2}
\end{equation}
does the job. 

\subsection{Computing the von Neumann entropy of a Gaussian state}

    For the this subsection we are going to follow closely the pedagogical derivation developed by Tommaso F. Demarie in \href{https://arxiv.org/abs/1209.2748}{arXiv:1209.2748}. Let us consider a system of $N$ continuous variable degrees of freedom. The state of the system is the arbitrary Gaussian state with density operator $\hat\rho$ and associated covariance matrix $\bSi$. One can always find a (symplectic) transformation $\bS$ that symplectically diagonalizes $\bSi$ so that $\bSi^{\prime}=\bS \bSi \bS^{\intercal}$ is its Williamson form. Such a symplectic transformation maps the quadrature operators to their primed versions $\bXi^{\prime}=\bS \bXi$. This symplectic transformation  is associated to an unitary transformation in the Hilbert space $\hat U$ such that $\hat\rho^{\prime}=\hat{U} \hat\rho \hat{U}^{\dagger}$ corresponds to the state of coveraince matrix $\bSi'$. Thus, we can write $\hat\rho^{\prime}$ as
\be
\hat\rho^{\prime}=\hat\rho_{1}^{\prime} \otimes \hat\rho_{2}^{\prime} \ldots \otimes \hat\rho_{N}^{\prime},
\ee
where $\hat\rho_{i}^{\prime}$ is the density operator of a single quantum harmonic oscillator in a thermal state. We know this because, as we saw in Sec. \ref{Gstatessec}, a diagonal covariance matrix can always be read as a thermal state of a set of uncoupled harmonic oscillators. Note that the primed oscillators are now uncoupled, although each oscillator is non-local, as each mode is a linear combination of possibly many of the original modes.
Recall that the thermal equilibrium state of a harmonic oscillator at temperature $T$ is represented by the Gibbs state we saw in previous sections. Concretely, in the Fock basis $\left\{\left| n \right\rangle\right\}$, the state is given by
\be
\hat\rho= Z^{-1}\sum_{n=1}^\infty e^{-E_{n} / k_{B} T}\left| n \right\rangle\!\left\langle n \right|,
\ee
where $E_{n}$ is the energy of each Fock state: $\hat{H}\left| n \right\rangle=E_{n}\left| n \right\rangle,$ and the partition function \mbox{$Z=\operatorname{Tr}\left(e^{-\hat{H} / k_{B} T}\right)$}. 

We can now define primed creation and annihilation operators for the modes on which the covariance matrix of the system of oscillators takes the normal form:
\be
\hat{a}_{i}^{\prime}=\sqrt{\frac{m_{i}^{\prime} \omega_{i}^{\prime}}{2 \hbar}}\left(\hat{q}_{i}^{\prime}+\frac{\ii}{m_{i}^{\prime} \omega_{i}^{\prime}} \hat{p}_{i}^{\prime}\right), \qquad \hat{a}_{i}^{\dagger \prime}=\sqrt{\frac{m_{i}^{\prime} \omega_{i}^{\prime}}{2 \hbar}}\left(\hat{q}_{i}^{\prime}-\frac{\ii}{m_{i}^{\prime} \omega_{i}^{\prime}} \hat{p}_{i}^{\prime}\right),
\ee
which can be expressed as a symplectic transformation of the unprimed operators: if $\hat{\boldsymbol{\alpha}}\coloneqq(\hat a_1,\hat a_1^\dagger \cdots,\hat a_N,\hat a_N^\dagger) $, $\hat{\boldsymbol{\alpha}}'\coloneqq(\hat a_1',{{\hat a_1}^{'\dagger}}\cdots,\hat a_N',{{\hat a_N}^{'\dagger}}) $ then
\be
\hat{\boldsymbol{\alpha}}'=\bS\hat{\boldsymbol{\alpha}},
\ee
and again the state $\hat\rho_i'$ for each transformed oscillator corresponds to the thermal state associated with the Hamiltonian 
\be
\hat{H}_{i}^{\prime}=\hbar \omega_{i}^{\prime}\left(\hat{a}_{i}^{\dagger \prime} \hat{a}_{i}^{\prime}+\frac{1}{2}\right).
\ee
Therefore, the density operator of each uncoupled (non-local) mode is given by
\be
\hat\rho_{i}^{\prime} =\sum_{n} Z_{i}^{-1} e^{-E_{n, i}^{\prime} / k_{B} T_{1}}|\varphi_{n^{\prime}_{i}}\rangle\langle\varphi_{n^{\prime}_{i}}|=Z_{i}^{-1} e^{-\hat{H}_{i}^{\prime} / k_{B} T_{i}},
\ee
where $\hat{H}'\left|\varphi_{n'}\right\rangle=E_{n'}\left|\varphi_{n'}\right\rangle,$ and each partition function is
$$
Z_{i}=\operatorname{Tr}\left(e^{-\hat{H}_{i}^{\prime} / k_{B} T_{i}}\right)=\sum_{n_{i}^{\prime}=0}^{\infty}\langle\varphi_{n_{i}^{\prime}}|e^{-\left(\hat{a}_{i}^{\dagger\prime} \hat{a}_{i}^{\prime}+\frac{1}{2}\right) \hbar \omega_{i}^{\prime} / k_{B} T_{i}}| \varphi_{n_{i}^{\prime}}\rangle=\sum_{n_{i}^{\prime}=0}^{\infty} e^{-\left(n_{i}^{\prime}+\frac{1}{2}\right) \tilde\beta_{i}},
$$
where the $\{n_{i}^{\prime}\}$ are the eigenvalues of the primed number operator $\hat{a}_{i}^{\dagger\prime} \hat{a}_{i}^{\prime}|\varphi_{n_{i}^{\prime}}\rangle=n_{i}^{\prime}|\varphi_{n_{i}^{\prime}}\rangle$. We also define $\tilde\beta_{i} \coloneqq \hbar \omega_{i}^{\prime} / k_{B} T_{i}$. We can simplify the partition function summing the geometric series
$$
Z_{i}=e^{-\tilde\beta_{i} / 2} \sum_{n_{i}^{\prime}=0}^{\infty} e^{-n_{i}^{\prime} \tilde\beta_{i}}=e^{-\tilde\beta_{i} / 2}\left[1+e^{-\tilde\beta_{i}}+e^{-2 \tilde\beta_{i}}+\ldots\right] \Rightarrow Z_{i}=\frac{e^{-\tilde\beta_{i} / 2}}{1-e^{-\tilde\beta_{i}}},
$$
thus we get that
\be
\hat\rho_{i}^{\prime}=\left(1-e^{-\tilde\beta_{i}}\right) e^{-\hat{a}_{i}^{\dagger\prime} \hat a_{i}^{\prime} \tilde\beta_{i}}.
\ee
In order to simplify we can express the density operator as a function of the expectation value of the number operator of the primed modes  $\bar{n}_{i}^{\prime}\coloneqq\langle \hat{a}_{i}^{\dagger\prime} \hat a_{i}^{\prime}\rangle_{\hat\rho_i^\prime}$  so that
\bel{number0}
\begin{aligned}
\bar{n}_{i}^{\prime} &=\operatorname{Tr}\left(\rho_{i}^{\prime} \hat{a}_{i}^{\dagger} \hat{a}_{i}^{\prime}\right)=\sum_{n_{i}^{\prime}=0}^{\infty}\langle\varphi_{n_{i}^{\prime}}|\left(1-e^{-\tilde\beta_{i}}\right) e^{-\hat{a}_{i}^{\dagger} \hat{a}_{i}^{\prime} \tilde\beta_{i}} \hat{a}_{i}^{\dagger} \hat{a}_{i}^{\prime}| \varphi_{n_{i}^{\prime}}\rangle=\left(1-e^{-\tilde\beta_{i}}\right) \sum_{n_{i}^{\prime}=0}^{\infty} n_{i}^{\prime} e^{-n_{i}^{\prime} \tilde\beta_{i}}=\frac{1}{e^{\tilde\beta_{i}}-1} 
\end{aligned}
\ee
which means that
\bel{number}
 e^{\tilde\beta_{i}}=\frac{1+\bar{n}_{i}^{\prime}}{\bar{n}_{i}^{\prime}}.
\ee
Using \eqref{number} we can rewrite the primed states' density operators as:
$$
\rho_{i}^{\prime}=\frac{1}{1+\bar{n}_{i}^{\prime}}\left(\frac{\bar{n}_{i}^{\prime}}{1+\bar{n}_{i}^{\prime}}\right)^{\hat{a}_{1}^{\dagger\prime} \hat{a}_{i}^{\prime}}.
$$
The von Neumann entropy for a single harmonic oscillator $S\left(\hat\rho_{i}^{\prime}\right)=-\operatorname{Tr}\left(\hat\rho_{i}^{\prime} \log \hat\rho_{i}^{\prime}\right)$ is moderately long to compute but relatively straightforward. Let us first evaluate
\be
\begin{aligned}
\operatorname{Tr}\left(\hat\rho_{i}^{\prime} \log \hat\rho_{i}^{\prime}\right) &=\sum_{n_{i}^{\prime}=0}^{\infty}\langle\varphi_{n_{i}^{\prime}}|\left(\frac{1}{1+\bar{n}_{i}^{\prime}}\left(\frac{\bar{n}_{i}^{\prime}}{1+\bar{n}_{i}^{\prime}}\right)^{\hat{a}_{i}^{\dagger} \hat{a}_{i}^{\prime}} \log \left[\frac{1}{1+\bar{n}_{i}^{\prime}}\left(\frac{\bar{n}_{i}^{\prime}}{1+\bar{n}_{i}^{\prime}}\right)^{\hat{a}_{i}^{\dagger} \hat{a}_{i}^{\prime}}\right]\right)| \varphi_{n_{i}^{\prime}}\rangle\\
&=\frac{1}{1+\bar{n}_{i}^{\prime}} \sum_{n_{i}^{\prime}=0}^{\infty}\langle\varphi_{n_{i}^{\prime}}|\left(\left(\frac{\bar{n}_{i}^{\prime}}{1+\bar{n}_{i}^{\prime}}\right)^{n_{i}^{\prime}}\left[-\log \left(1+\bar{n}_{i}^{\prime}\right)+n_{i}^{\prime} \log \left(\frac{\bar{n}_{i}^{\prime}}{1+\bar{n}_{i}^{\prime}}\right)\right]\right)| \varphi_{n_{i}^{\prime}}\rangle\\
&=\frac{1}{1+\bar{n}_{i}^{\prime}}\left[-\log \left(1+\bar{n}_{i}^{\prime}\right) \sum_{n_{i}^{\prime}=0}^{\infty}\left(\frac{\bar{n}_{i}^{\prime}}{1+\bar{n}_{i}^{\prime}}\right)^{n_{i}^{\prime}}+\log \left(\frac{\bar{n}_{i}^{\prime}}{1+\bar{n}_{i}^{\prime}}\right) \sum_{n_{i}^{\prime}=0}^{\infty} n_{i}^{\prime}\left(\frac{\bar{n}_{i}^{\prime}}{1+\bar{n}_{i}^{\prime}}\right)^{n_{i}^{\prime}}\right]\\
&=\bar{n}_{i}^{\prime} \log \bar{n}_{i}^{\prime}-\left(1+\bar{n}_{i}^{\prime}\right) \log \left(1+\bar{n}_{i}^{\prime}\right).
\end{aligned}
\ee
where we have used that
$$
\sum_{n_{i}=0}^{\infty}\left(\frac{\bar{n}_{i}^{\prime}}{1+\bar{n}_{i}^{\prime}}\right)^{n_{i}}\!\!\!=1+\bar{n}_{i}^{\prime},  \qquad \sum_{n_{i}=0}^{\infty} n_{i}\left(\frac{\bar{n}_{i}^{\prime}}{1+\bar{n}_{i}^{\prime}}\right)^{n_{i}}\!\!\!=\bar{n}_{i}^{\prime}\left(1+\bar{n}_{i}^{\prime}\right).
$$
This means that the von Neumann entropy for a single harmonic oscillator in a thermal state
in terms of its expected occupation number is
\bel{entropp}
S\left(\hat\rho_{i}^{\prime}\right)=\left(1+\bar{n}_{i}^{\prime}\right) \log \left(1+\bar{n}_{i}^{\prime}\right)-\bar{n}_{i}^{\prime} \log \bar{n}_{i}^{\prime}.
\ee

As we will see next, it is easy to write the mean occupation number in terms of the symplectic eigenvalues of the covariance matrix. We know from \eqref{thermalityyy} that in its normal form, for the $i$-th oscillator,
 \be
\boldsymbol{\sigma}'_i= \begin{pmatrix}
 \nu_i & 0\\
0&  \nu_i
 \end{pmatrix}=\nu\openone,\qquad \nu=\operatorname{cotanh}\frac{\tilde\beta_i}{2},
\ee
and therefore the symplectic eigenvalues of $\bSi'$ can be written as
\be
\nu_i=\operatorname{cotanh}\frac{\tilde\beta_i}{2} =\frac{e^{\frac{\tilde\beta_i}{2}}+e^{-\frac{\tilde\beta_i}{2}}}{e^{\frac{\tilde\beta_i}{2}}-e^{-\frac{\tilde\beta_i}{2}}}=\frac{e^{\tilde\beta_i}+1}{e^{\tilde\beta_i}-1}=\frac{\frac{1+\bar{n}_{i}^{\prime}}{\bar{n}_{i}^{\prime}}+1}{\frac{1+\bar{n}_{i}^{\prime}}{\bar{n}_{i}^{\prime}}-1}=2\bar{n}_{i}^{\prime}+1,
\ee
where in the last step we used \eqref{number}. This in turn means that
\be
\bar{n}_{i}^{\prime}=\frac{\nu_i-1}{2}.
\ee
Substituting this into \eqref{entropp}, we get finally that the von Neumann entropy of the $i$-th non-local mode is
\bel{entropp}
S\!\left(\hat\rho_{i}^{\prime}\right)=\frac{\nu_i+1}{2} \log \left(\frac{\nu_i+1}{2}\right)-\frac{\nu_i-1}{2} \log\left(\frac{\nu_i-1}{2}\right).
\ee
Finally, since the total state $\hat\rho'$ is a tensor product of all the individual thermal non-local modes (found by the symplectic diagonalization of the covariance matrix of the state $\hat\rho$), and the entropy is additive under tensor product, we find that the total entropy is
\bel{totentropp}
S\!\left(\hat\rho\right)=\sum_{i=1}^N\left[\frac{\nu_i+1}{2} \log \left(\frac{\nu_i+1}{2}\right)-\frac{\nu_i-1}{2} \log\left(\frac{\nu_i-1}{2}\right)\right],
\ee
where $\nu_i$ are the symplectic eigenvalues of the covariance matrix $\bSi$.

Now, if the total state is pure, computing the partial state entropies will give us a measure of entanglement, the entanglement entropy. Computing the partial state of a Gaussian state for which we know its covariance matrix is very easy, as it is just a cropping of the total covariance matrix. Let us see it with an example.

\subsection{Entanglement entropy of the ground state of a system of two coupled quantum harmonic oscillators}\label{Example two oscillators}

Consider a system of two coupled harmonic oscillators with mass $m$ and frequency $\omega$ with Hamiltonian\footnote{This example is also analyzed in a slightly different way in  \href{https://arxiv.org/abs/1209.2748}{arXiv:1209.2748}}:
\be
\hat{H}=\frac{\hat{p}_{1}^{2}+\hat{p}_{2}^{2}}{2 m}+\frac{m \omega^{2}}{2}\left(\hat{q}_{1}^{2}+\hat{q}_{2}^{2}\right)+\lambda\left(\hat{q}_{1}-\hat{q}_{2}\right)^{2}.
\ee
This Hamiltonian can be written as
\be
\hat H =\frac12 \hat\Xi^{\alpha}\hat\Xi^\beta F_{\alpha\beta},
\label{Hamo2}
\ee
where $\bXi=(\hat q_1,\hat p_1,\hat q_2,\hat p_2)$ and
\be
\bF=\bar\bF=\begin{pmatrix}
m\omega^2\! +\!2\lambda & 0 & -2\lambda & 0\\
0& m^{-1} & 0 & 0\\
-2\lambda&  0 & {m\omega^2}\! +\!2\lambda & 0\\
0 & 0& 0& m^{-1}
\end{pmatrix}.
\ee
Notice that for this example we have recovered full dimensional units. Since this is a quadratic Hamiltonian, the ground state of the coupled harmonic oscillators is Gaussian. 

We would like to evaluate the entanglement between the two oscillators when the full system is in the ground state. To find the ground state we first symplectically diagonalize the Hamiltonian to find its normal modes. We carry out the following symplectic transformation inserting the identity twice
\be
\hat H=\frac12 \underbrace{\tensor{(S^{-1})}{^\alpha_\rho}\tensor{S}{^\rho_\mu}}_{\delta^\alpha_\mu}\underbrace{\tensor{(S^{-1})}{^\beta_\sigma}\tensor{S}{^\sigma_\nu}}_{\delta^\beta_\nu}\hat\Xi^{\mu}\hat\Xi^{\nu} F_{\alpha\beta}=\frac12 \underbrace{(\tensor{S}{^\rho_\mu}\hat\Xi^\mu)}_{\Xi^{'\rho}} \underbrace{(\tensor{S}{^\sigma_\nu}\hat\Xi^{\nu})}_{\Xi^{'\sigma}}\underbrace{[ \tensor{(S^{-1})}{^\alpha_\rho}F_{\alpha\beta}\tensor{(S^{-1})}{^\beta_\sigma}]}_{F'_{\rho\sigma}}\;,
\ee
where the new modes $\bXi'=(\hat q_1',\hat p_1',\hat q_2',\hat p_2')$ and the new Hamiltonian matrix $\bF'$ are defined in matrix form as
\be
\bXi'=\bS\bXi,\qquad \bF'=(\bS^{-1})^{\intercal}\bF\bS^{-1} .
\ee
Notice the new modes are linear combinations of the old modes. In order to find $\bm{S}$, let us first calculate the symplectic eigenvalues: define
\begin{equation}
\bm{M}=\bm{F}\bm{\Omega}^{-1}=\begin{pmatrix}
0 & m\omega^2+2\lambda & 0 & -2\lambda \\
-m^{-1} & 0 & 0 & 0 \\
0 & -2\lambda & 0 & 2\lambda+m\omega^2 \\
0 & 0 & -m^{-1} & 0 \\
\end{pmatrix},
\end{equation}
which has eigenvalues $\{\pm\ii\omega,\pm\ii\omega\sqrt{1+4\lambda/m\omega^2}\}$. These are the frequencies of the two decoupled oscillators that we will obtain upon diagonalization of the Hamiltonian:
\be
\omega_{1}^{\prime}=\omega, \qquad \omega_{2}^{\prime}=\omega \sqrt{1+\frac{4 \lambda}{m \omega^{2}}} \eqqcolon \omega \alpha.
\ee
Thus, $\omega$ and $\omega\alpha$ are the symplectic eigenvalues of $F$. Since we are working with units, upon diagonalization the value associated to the position variable of the $i$-th mode has to be $m\omega_i$ times the one associated to the momentum variable, with the product of both giving $\omega^2$ and $\omega^2\alpha^2$, respectively. We conclude that the diagonal form we are looking for is
\begin{equation}
\bm{\mathcal{F}}=\begin{pmatrix}
m\omega^2 & 0 & 0 & 0\\
0& m^{-1} & 0 & 0\\
0&  0 &m\omega^2\alpha^2 & 0\\
0 & 0& 0& m^{-1}
\end{pmatrix}=\begin{pmatrix}
m\omega^2 & 0 & 0 & 0\\
0& m^{-1} & 0 & 0\\
0&  0 &4\lambda+m\omega^2 & 0\\
0 & 0& 0& m^{-1}
\end{pmatrix}.
\end{equation}
We now calculate
\begin{equation}
\bm{F}^{-1/2}=\bm{P} \bm{D}^{-1/2} \bm{P}^{-1} =\begin{pmatrix}
\frac{1}{\sqrt{4\lambda+m\omega^2}} & 0 & 0 & 0 \\
0 & \sqrt{m} & 0 & 0 \\
0 & 0 & \frac{\sqrt{m}}{2}\big( 1 + \frac{1}{m\omega} \big) & \frac{\sqrt{m}}{2}\big( -1 + \frac{1}{m\omega} \big) \\
0 & 0 & \frac{\sqrt{m}}{2}\big( -1 + \frac{1}{m\omega} \big) & \frac{\sqrt{m}}{2}\big( 1 + \frac{1}{m\omega} \big) \\
\end{pmatrix}
\end{equation}
where $\bm{P}$ is the transformation that diagonalizes $\bm{F}$ to $\bm{D}$,
\begin{equation}
\bm{P}=\begin{pmatrix}
0 & 0 & 1 & -1 \\
0 & 1 & 0 & 0 \\
0 & 0 & 1 & 1 \\
1 & 0 & 0 & 0 \\
\end{pmatrix} \qquad \textrm{and} \qquad \bm{D}=\begin{pmatrix}
m^{-1} & 0 & 0 & 0 \\
0 & m^{-1} & 0 & 0 \\
0 & 0 & m\omega^2 &  \\
0 & 0 & 0 & 4\lambda+m\omega^2 \\
\end{pmatrix}.
\end{equation}
If we build $\bm{Y}$ and $\bm{Z}$ as in Section~\ref{Symplectic diagonalization}, we get that they are diagonalized by the unitaries 
\begin{equation}
\bm{K}=\frac{1}{\sqrt{2}}\begin{pmatrix}
\ii & -\ii & 0 & 0 \\
1 & 1 & 0 & 0 \\
0 & 0 & \ii & -\ii \\
0 & 0 & 1 & 1 \\
\end{pmatrix} \qquad \textrm{and} \qquad \bm{U}=\frac{1}{2}\begin{pmatrix}
\ii & -\ii & -\ii & \ii \\
1 & 1 & -1 & -1 \\
\ii & -\ii & \ii & -\ii \\
1 & 1 & 1 & 1 \\
\end{pmatrix}.
\end{equation}
Thus, we can define
\begin{equation}
\bm{O}=\bm{K}\bm{U}^\dagger=\frac{1}{\sqrt{2}}\begin{pmatrix}
1 & 0 & 1 & 0 \\
0 & 1 & 0 & 1 \\
-1 & 0 & 1 & 0 \\
0 & -1 & 0 & 1 \\
\end{pmatrix},
\end{equation}
and the symplectic transformation that diagonalizes $\bm{F}$ is then
\begin{equation}
(\bm{S}^{-1})^\intercal=\bm{\mathcal{F}}^{1/2}\bm{O}\bm{F}^{-1/2}=\frac{1}{\sqrt{2}}\begin{pmatrix}
1 & 0 & 1 & 0 \\
0 & 1 & 0 & 1 \\
-1 & 0 & 1 & 0 \\
0 & -1 & 0 & 1 \\
\end{pmatrix}.
\end{equation}
Equivalently
\be
\bS^{-1}=\frac{1}{\sqrt2}\begin{pmatrix}
1 & 0 & -1 & 0\\
0& 1 & 0 & -1\\
1&  0 &1 & 0\\
0 & 1& 0& 1
\end{pmatrix},
\ee
such that the normal mode Hamiltonian matrix is
\begin{align}
\bF'=\frac12\begin{pmatrix}
1 & 0 & 1 & 0\\
0& 1 & 0 & 1\\
-1&  0 &1 & 0\\
0 & -1& 0& 1
\end{pmatrix}&\begin{pmatrix}
m\omega^2\! +\!2\lambda & 0 & -2\lambda & 0\\
0& m^{-1} & 0 & 0\\
-2\lambda&  0 & {m\omega^2}\! +\!2\lambda & 0\\
0 & 0& 0& m^{-1}
\end{pmatrix}\begin{pmatrix}
1 & 0 & -1 & 0\\
0& 1 & 0 & -1\\
1&  0 &1 & 0\\
0 & 1& 0& 1
\end{pmatrix}\\
&=\begin{pmatrix}
m\omega^2 & 0 & 0 & 0\\
0& m^{-1} & 0 & 0\\
0&  0 &4\lambda+m\omega^2 & 0\\
0 & 0& 0& m^{-1}
\end{pmatrix}.
\end{align}
Notice that the oscillators defined by $\bXi'=\bS\bXi$ are non-local:  $\hat q'_1$ and $p'_1$  are combinations of the positions and momenta of the two oscillators, therefore they don't belong to a single one of them.

Since we have the Hamiltonian in terms of normal modes, finding the ground state of the coupled oscillators is trivial: we can write the covariance matrix of the ground state for the system of decoupled oscillators directly from \eqref{fulldime} and then transform back with the inverse of the symplectic transformation to recover the covariance matrix of the ground state in terms of the local modes.

The ground state covariance matrix in terms of the normal modes is the direct sum of the covariance matrices of the individual ground states as given by \eqref{fulldime}:
\be
\bSi'=\hbar\begin{pmatrix}
 \frac{1}{m\omega}  & 0 & 0 & 0\\
0&{ m \omega}\, & 0 & 0\\
0 & 0&  \frac{1}{m\omega\alpha}  & 0\\
0& 0 &0&{ m \omega\alpha}\, 
\end{pmatrix}.
\ee
we can now obtain the covariance matrix for the local oscillator modes through the symplectic transformation
\begin{align}
\bSi=\bS^{-1}\bSi'(\bS^{-1})^\intercal=&\frac\hbar2\begin{pmatrix}
1 & 0 & -1 & 0\\
0& 1 & 0 & -1\\
1&  0 &1 & 0\\
0 & 1& 0& 1
\end{pmatrix}\begin{pmatrix}
 \frac{1}{m\omega}  & 0 & 0 & 0\\
0&{ m \omega}\, & 0 & 0\\
0 & 0&  \frac{1}{m\omega\alpha}  & 0\\
0& 0 &0&{ m \omega\alpha}\, 
\end{pmatrix}\begin{pmatrix}
1 & 0 & 1 & 0\\
0& 1 & 0 & 1\\
-1&  0 &1 & 0\\
0 & -1& 0& 1
\end{pmatrix}\\
&=\hbar\left(
\begin{array}{cccc}
 \frac{\alpha +1}{2 \alpha  m \omega } & 0 & \frac{\alpha -1}{2 \alpha  m \omega } & 0 \\
 0 & \frac{1}{2} (\alpha +1) m \omega  & 0 & -\frac{1}{2} (\alpha -1) m \omega  \\
 \frac{\alpha -1}{2 \alpha  m \omega } & 0 & \frac{\alpha +1}{2 \alpha  m \omega } & 0 \\
 0 & -\frac{1}{2} (\alpha -1) m \omega  & 0 & \frac{1}{2} (\alpha +1) m \omega  \\
\end{array}
\right),
\end{align}
where it is obvious that this ground state contains correlations between observables of the two coupled oscillators. Notice that the ground state of the system of two coupled oscillators is a pure state, therefore the entanglement entropy is well defined. Computing the entanglement entropy of the two oscillators is very easy since the partial state for each oscillator is just given by the following covariance matrices
\be
\bSi_1=\bSi_2=\hbar\begin{pmatrix}
 \frac{\alpha +1}{2 \alpha  m \omega } & 0\\
  0 & \frac{1}{2} (\alpha +1) m \omega  
\end{pmatrix}.
\ee
We just need to compute the symplectic eigenvalues of this covariance matrix and plug them in \eqref{totentropp} (which has only one summand in this case). We recall that the symplectic eigenvalues are given by the modulus of the (purely imaginary)  eigenvalues of  $\bSi\bO$. Those symplectic eigenvalues are the same for the two oscillators and are (taking $\hbar=1$ for simplicity applying the derived formulas)
\be
\nu=\frac{1+\alpha}{2\sqrt{\alpha}}.
\ee
We can compute the entanglement entropy, which is non-zero and increasing with the value of the coupling strength:
\bel{entropp}
S_E=\frac{\nu+1}{2} \log \left(\frac{\nu+1}{2}\right)-\frac{\nu-1}{2} \log\left(\frac{\nu-1}{2}\right).
\ee

\section{Acknowledgements}

The author is indebted to Jos\'e Polo-G\'omez for his invaluable help reviewing these notes. It is because of him that I feel comfortable posting these course notes on arXiv so that perhaps they may be helpful to someone learning the basics of phase space quantum mechanics and Gaussian QM. I also would like to acknowledge Jos\'e de Ram\'on, who was the teaching assistant for this course for two years and who has reviewed and detected typos in previous versions of these notes. 

\end{document}